\documentclass[11pt,letterpaper]{article}
\usepackage[colorlinks=true,linkcolor=blue,urlcolor=blue,citecolor=blue,anchorcolor=blue,
pdftex,breaklinks,pdfencoding=auto,psdextra,
bookmarksopenlevel=1,bookmarksopen=true]{hyperref}
\usepackage{xcolor}
\definecolor{darkblue}{rgb}{0,0,.6}
\usepackage{fontenc,fontaxes,amsmath,amsfonts,mathtools,arydshln,setspace,bbm,tikz,multirow,longtable,subcaption,graphicx,booktabs,bm,dsfont,booktabs,longtable,subcaption,ragged2e,paralist}

\usepackage[authoryear]{natbib}
\usepackage{orcidlink,float}
\usepackage[inline,shortlabels]{enumitem}

\usepackage[margin=1.1in]{geometry}

\graphicspath{{plots/}}
\mathtoolsset{mathic=true}
\DeclareMathOperator{\ran}{ran}

\DeclareMathOperator{\spn}{span}

\DeclareMathOperator{\HS}{HS}

\DeclareMathOperator{\sgn}{sgn}

\newcommand{\YY}{\mathfrak Y}
\newcommand{\YYY}{\mathcal Y}
\newcommand{\YYYY}{\mathcal G}

\newcommand{\varv}{\bm{\mathsf{v}}}
\newcommand{\opnorm}{\textrm{op}}
\newcommand{\PPi}{\bm{{\Pi}}}
\newcommand{\ttau}{\bm{{\tau}}}
\newcommand{\pp}{p}

\newcommand{\KK}{{\mathsf K}}
\newcommand{\PP}{{\rm P}}
\newcommand{\QQ}{{\rm Q}}
\renewcommand{\hat}{\widehat}
\newcommand{\llambda}{\bm{\mathsf{\lambda}}}
\newcommand{\edex}[2]{_{#1}[#2]}

  {\end{figure}} 

\newenvironment{proofs}[1][Proof]
  {\begin{trivlist}
   \item[\hskip \labelsep {\bfseries #1.}] 
   \itshape} 
  {\hfill $\square$ \end{trivlist}} 

\newtheorem{theorems}{Theorem}[section]
\newtheorem{lemmas}{Lemma}[section]
\newtheorem{corollarys}{Corollary}[section]
\newtheorem{assumptions}{Assumption}

\newtheorem{propositions}{Proposition}[section]
\newtheorem{remarks}{Remark}[section]

\graphicspath{{plots/}}

\newtheorem{assumpE}{Assumption}

\begin{document}

\def\spacingset#1{\renewcommand{\baselinestretch}%
{#1}\small\normalsize} \spacingset{1}


  \title{\bf Functional Regression with Nonstationarity and Error Contamination: Application to the Economic Impact of Climate Change \\ \vspace{1em} }  
  \author{ Kyungsik Nam\hspace{.2cm}\\
     Division of Climate Change, Hankuk University of Foreign Studies\\
\\
     Won-Ki Seo\thanks{Data and computing code used in this paper are available at \url{https://github.com/wonkiseo86/FRNE}.
} \\
     School of Economics, University of Sydney}
\date{}
  \maketitle

\bigskip
\begin{abstract}
This paper studies a regression model with functional dependent and explanatory variables, both of which exhibit nonstationary dynamics. The model assumes that the nonstationary stochastic trends of the dependent variable are explained by those of the explanatory variables, and hence that there exists a stable long-run relationship between the two variables despite their nonstationary behavior. We also assume that the functional observations may be error-contaminated. We develop novel autocovariance-based estimation and inference methods for this model. The methodology is broadly applicable to economic and statistical functional time series with nonstationary dynamics. To illustrate our methodology and its usefulness, we apply it to evaluating the global economic impact of climate change, an issue of intrinsic importance.
\end{abstract}

\noindent%
\vfill

\newpage
\onehalfspacing 
\newpage
	\section{Introduction}\label{Sec_intro}
	In data-rich environments, practitioners often need to deal with non-traditional observations, such as curves, probability density functions, or images. Accordingly, recent literature on functional data analysis, which provides statistical methods for handling such complex data, has gained popularity. For a comprehensive and broad review of this topic, readers are referred to \cite{Ramsay2005} and \cite{HK2012}. Practitioners in various fields have benefited from advances in this area. In particular, functional linear regression models have become a central tool for those interested in analyzing the relationships between two or more such variables. Some early contributions to this topic include \cite{Yao2005}, \cite{Hall2007}, \cite{Park2012397}, \cite{Florence2015}, \cite{Benatia2017}  and  \cite{imaizumi2018}, and, more recently, \cite{Chen_et_al_2020}, \cite{Babii2022} and \cite{seong2021functional} study the issue of endogeneity. A common feature of all these papers is that they all consider functional regression models with iid or stationary sequence of random functions.
	
	Only recently has the literature begun to consider nonstationary dependent observations, even if many economic and statistical functional time series tend to be nonstationary, as noted in recent papers (e.g., \citealp{Chang2016152, BSS2017,Franchi2017b, LRS, NSS, NSS2, seo2020functional, seoshang22}). As a result, statistical methods developed for such time series are currently limited to analyzing their essential properties, such as cointegration, stochastic trends, and the dominant subspace. Despite its empirical relevance, articles developing inferential methods for functional (auto-)regression models involving nonstationary functional time series are scarce; to the best of the authors' knowledge, there are currently only a few preprints  (e.g., \citealp{changshocks,Hu2016}). We fill this gap by developing novel statistical methods for functional regression models where both regressand and regressor exhibit unit-root-type nonstationary behavior allowing cointegration, a feature particularly important for economic and financial applications.
	
	In addition to incorporating nonstationarity into functional regression models, we aim to enhance the real-world applicability of our methods by addressing a typical and practical aspect of functional data that has recently been discussed in the literature: incomplete and partially observed data (see, e.g., \citealp{Chen_et_al_2020}; \citealp{seong2021functional}). In the majority of real data analyses, (i) each functional observation, say $x_t(u)$ for $u\in [a_1,a_2]$, is not directly observed, and often constructed from its partial and discrete realizations $(x_t(u_1),\ldots,x_t(u_n))'$ for $u_1,\ldots,u_n \in [a_1,a_2]$ and (ii) the number of discrete observations $n$ is often not large enough. In fact, (i) and (ii) are pointed out by \cite{seong2021functional} in the context of functional linear models, and they argued that ``endogeneity'' caused by measurement errors need to be properly addressed for estimation and inference (see also \citealp{Chen_et_al_2020}). As a specific example, consider a case where functional observations are probability density-valued (as in Section \ref{Sec_empirics} to appear). This case has gained significant interest in the literature; for the stationary case, see e.g., \cite{kneip2001inference} and \cite{Park2012397}, while for the nonstationary case refer to \cite{Chang2016152} and \cite{seo2019cointegrated}. In this scenario, the true probability density is not observable, and thus, it needs to be replaced by a proper nonparametric estimate. This naturally introduces small or large measurement errors in practice. In this paper, we explicitly consider cases where the variables of interest, which are nonstationary, are also error-contaminated, and then develop statistical methods that are robust to error contamination. This not only distinguishes the present paper significantly from existing works 
	(cf., e.g., \citealp{Benatia2017, Park2012397, Chen_et_al_2020, Babii2022,seong2021functional})
	but also makes our proposed methods  more appealing to applied researchers. We also believe that our methodology can be applied to various economic and financial time series.  

	More technically and specifically, we assume that the variables of interest are cointegrated functional time series, following the framework of \cite{Chang2016152} and \cite{BSS2017}. This assumption has been widely used in the recent literature on nonstationary functional time series, especially in economic applications (see, e.g., \citealp{NSS, seo2020functional}). We then assume that these variables can only be observed with additive measurement errors. As noted by \cite{seong2021functional}, the problem of neglected error contamination generally results in  inconsistency of standard estimators constructed from the sample covariance operator $\widehat{C}_0$ of the regressor (this is also true in our model, as will be discussed in Section \ref{Sec_empirics} in greater detail). This inconsistency arises primarily because $\widehat{C}_0$ is inherently contaminated by measurement errors and, consequently, becomes a distorted estimator of its population counterpart. To address this issue of error contamination, we consider autocovariance-based inference, avoiding the direct use of the covariance operator of an error-contaminated variable for statistical inference, as in some recent articles on functional regression models (see, e.g., \citealp{Chen_et_al_2020}). More specifically, we construct our proposed estimator based on the lag-$\kappa$ sample autocovariance $\widehat{C}_{\kappa}$ for some positive $\kappa$. This approach is grounded in the observation that, as long as the measurement errors are not strongly correlated and satisfy certain mild regularity conditions (to be detailed), (i) the sample autocovariance $\widehat{C}_{\kappa}$ will be less affected by measurement errors, and (ii) the assumption that measurement errors are not strongly correlated does not seem overly restrictive, given that such errors in functional data analysis commonly arise from constructing each individual functional observation based on its discrete realizations; as will be detailed in Section \ref{Sec_empirics}, our asymptotic analysis requires a much weaker condition on the serial correlation of the measurement errors rather than complete serial uncorrelatedness. It should be noted that, in this paper, we also consider the case $\kappa=0$, which yields the standard covariance-based estimator in the functional linear model, and study its detailed asymptotic properties, a contribution that is, to the best of the authors' knowledge, also novel. 
	
We develop autocovariance-based inferential methods that are robust to the potential presence of measurement errors. This includes a novel dimension-reduction method, our proposed estimator of the slope parameter in the functional regression model, and their asymptotic properties. The proposed estimator bears some resemblance to the conventional two-step estimator of \cite{engle1987} in that both use residuals computed from the estimated relationship between the nonstationary components of the model; however, beyond this superficial similarity, the two approaches differ substantially in structure and purpose.
We also provide numerical studies with real-world data and simulation experiments to examine the performance of our proposed estimator. As an application, we illustrate the empirical relevance of our proposed methodology by applying it to an empirical model for studying the global economic impact of climate change. Specifically, we show that the proposed framework effectively estimates the distributional relationship between land temperature anomalies--often considered a measure of climate change--and regional economic growth rates in the possible presence of measurement errors, thereby offering a robust basis for assessing heterogeneous climate–economy relationships across the globe.
	

	The rest of the paper is organized as follows. Section \ref{Sec_prelim} reviews essential preliminaries on nonstationary cointegrated functional time series. Section \ref{Sec_econometrics} details the regression model, and Section \ref{Sec_econometrics2} develops inferential methods for it. In Section \ref{Sec_empirics}, we apply the proposed method to examine the global economic impact of climate change. Section \ref{Sec_conclude} concludes.

	\bigskip
	\section{Preliminaries}\label{Sec_prelim}
	\subsection{Notation and simplification}
	
	We let \(\mathcal{H}\) be a real separable Hilbert space of functions on the interval $[a_1, a_2]$, and let $\langle \cdot, \cdot \rangle$ (resp.\ $\|\cdot\|$) denote the associated inner product (resp.\ norm). We let ${\mathcal H}_y$ denote another Hilbert space, which will be set to $\mathbb{R}$ (when the dependent variable $y_t$ is real-valued) or $\mathcal H$ (when $y_t$ is function-valued). Throughout, regardless of whether $\mathcal{H}_y = \mathbb{R}$ or $\mathcal{H}$, we adopt a slight abuse of notation by using $\langle \cdot, \cdot \rangle$ and $\|\cdot\|$ to denote the inner product and norm associated with $\mathcal{H}_y$, respectively. This notational simplification facilitates the exposition and poses minimal risk of confusion, as the meaning of each operation is readily inferred from the context. For the same reason, we use $I$ to denote the identity map on any Hilbert space under consideration. As a further simplification, we henceforth write $\int F$ to denote $\int_{0}^1 F(s)ds$ for any operator- or vector-valued function $F$ defined on $[0,1]$.

	Section \ref{Sec_mathprelim} of the Supplementary Material reviews basic concepts of bounded linear operators and random elements associated with two (possibly different) Hilbert spaces. Accordingly, we let $\mathcal L_{\mathcal H}$ denote the space of bounded linear operators on $\mathcal H$ with the usual operator norm $\|\cdot\|_{\opnorm}$, and let $\otimes$ denote the tensor product associated with $\mathcal H$, $\mathcal H_y$, or both (see \eqref{eqtensor}). Section \ref{Sec_mathprelim} also reviews $\mathcal H$-valued random elements $X$, their expectation (denoted $\mathbb{E}[X]$), covariance operator (denoted $C_X:=\mathbb{E}[(X-\mathbb{E}(X)) \otimes (X-\mathbb{E}(X))]$), and cross-covariance with an $\mathcal H_y$-valued random element $Y$ (denoted $C_{XY}:=\mathbb{E}[(X-\mathbb{E}(X)) \otimes (Y-\mathbb{E}(Y))]$). For $A \in \mathcal L_{\mathcal H}$, concepts such as the adjoint (denoted $A^\ast$), range (denoted $\ran A$), and kernel (denoted $\ker A$), as well as properties such as self-adjointness, compactness, Hilbert–Schmidtness, and nonnegativity are introduced in that section, and they will be useful subsequently.

	We will consider sequences of random linear operators, constructed from random elements in $\mathcal H$ and $\mathcal H_y$ (for a more detailed discussion on general random  linear operators, see \citealp{skorohod2001}). For any such operator-valued random sequence $\{A_j\}_{j\geq1}$, we write $A_j \to_{\pp} A$ to denote convergence in probability with respect to the operator norm (i.e., $\|A_j - A\|_{\opnorm} \to_p 0$). In the subsequent discussion, convergence in probability sometimes occurs for $\mathcal H$- or $\mathcal H_y$-valued elements (in the appropriate norm), but for convenience we use the same notation $\to_p$ to denote such convergence throughout, as distinguishing between the two would add notational complexity with little benefit. Moreover, as is common in the literature (see, e.g., \citealp{seo2020functional}), we write $A_j = A + O_p(a_T)$ (resp.\ $A_j = A + o_p(a_T)$) if $\|A_j - A\|_{\opnorm} = O_p(a_T)$ (resp.\ $\|A_j - A\|_{\opnorm} = o_p(a_T)$) for some sequence $a_T$. For any two operators $A$ and $B$, we write $A=_d B$ to denote equivalence in their finite-dimensional distributions as in \cite{seo2020functional}, i.e., $A=_dB$ if, for any $n>0$,  $\{v_j\}_{j=1}^n$ ($\subset  \mathcal H \text{ or } \mathcal H_y$)  and  $\{w_j\}_{j=1}^n$ ($\subset  \mathcal H \text{ or } \mathcal H_y$), the distribution of $(\langle A v_1,w_1 \rangle,\ldots,\langle A v_n,w_n \rangle)'$ equals that of $(\langle B v_1,w_1 \rangle,\ldots,\langle B v_n,w_n \rangle)'$.

	\subsection{Cointegrated $\mathcal H$-valued time series}\label{AP_FTS}
	We review cointegrated linear processes in $\mathcal H$, which have been used to model the persistent nonstationary behavior of many economic functional time series (see, e.g., \citealp{Chang2016152, NSS, NSS2, seo2020functional}).
	Suppose that $\Delta x_t = x_{t}-x_{t-1} = \sum_{j=0}^\infty \psi_j \varepsilon_{t-j}$ for some sequence of bounded linear operators $\{\psi_j\}_{j\geq 0}$  and an iid sequence $\{\varepsilon_t\}_{t \in \mathbb{Z}}$ satisfying $\mathbb{E}[\varepsilon_t] = 0$ and $\mathbb{E}[\|\varepsilon_t\|^4] < \infty$ and having a positive definite covariance $C_{\varepsilon}$. If $\sum_{j=0}^\infty j \|\psi_j\|_{\opnorm}<\infty$ holds, 
	we know from the Phillips-Solo decomposition of \cite{Phillips1992} and its extension to a function space (see e.g., \citealp{seo_2022}), $x_t$ allows the following representation, ignoring the initial values that are negligible in our asymptotic analysis:
	\begin{equation} \label{psdecom}
		x_t = \psi(1) \sum_{s=1}^t \varepsilon_s + \eta_t, 
	\end{equation}
	where $\psi(1) =\sum_{j=0}^\infty \psi_j$, $\eta_t = \sum_{j=0}^\infty \widetilde{\psi}_j \varepsilon_{t-j}$ and $\widetilde{\psi}_j = -\sum_{k=j+1}^\infty \psi_k$. Let $\PP^S$ be the orthogonal projection onto $[\ran \psi(1)]^\perp$ and let $\PP^N = I-\PP^S$. Then $\langle x_t,v \rangle$ is stationary if and only if $v \in \mathcal H^S$ (see \citealp{BSS2017}). Thus the entire Hilbert space $\mathcal H$ can be orthogonally decomposed into $\mathcal H^N = \ran \PP^N$ and $\mathcal H^S = \ran \PP^S$. We call $\mathcal H^N$ (resp.\ $\mathcal H^S$) the nonstationary (resp.\ stationary) subspace induced by $\{x_t\}_{t\geq1}$.   

Subsequently, we will consider cointegrated time series introduced in this section, but some additional restrictions will be imposed for our asymptotic analysis in Section \ref{Sec_econometrics}.

	\section{Proposed model}\label{Sec_econometrics}
	Let \(\{x_t\}_{t \geq 1}\) be a cointegrated \(\mathcal{H}\)-valued time series, as detailed in Section~\ref{AP_FTS}, which induces a bipartite partition of \(\mathcal{H}\) into a nonstationary subspace \(\mathcal{H}^N\) and a stationary subspace \(\mathcal{H}^S\). We consider the following data-generating mechanism: for $f:\mathcal H \to {\mathcal H}_y$,
	\begin{equation}
		y_{t} = f(x_t)  + u_t, \quad \,\,u_t^N = \PP^N \Delta x_t = \sum_{j=0}^{\infty} \psi_j^N \epsilon_{t-j}, \quad  \text{and} \quad  u_t^S  = \PP^S x_t =  \sum_{j=0}^{\infty} \psi_j^S \epsilon_{t-j}.  \label{eqreg1}
	\end{equation}
Note that the above model includes no deterministic terms. We first develop inferential methods for this case and then discuss extending our methods to a model with deterministic terms in Section \ref{sec_det}; as may be expected, this extension requires only modest and non-substantial modifications of the results developed for the case without deterministic terms.
	
	Throughout this paper, we assume that $x_t$ cannot be directly observed but that only $\tilde{x}_t$, observed with measurement errors, is available. As highlighted by \cite{seong2021functional}, this assumption is empirically relevant because functional observations used in practice are often incompletely observed, with only finitely many discrete realizations available to practitioners. Consequently, it is common to construct a functional observation $z_t$ in advance by smoothing its $n$ discrete data points $z_t(s_1), \ldots, z_t(s_n)$, with $s_j$ included in the entire interval  $[a_1, a_2]$, before computing estimators or test statistics. While one may disregard measurement errors for simplicity if $n$ is large enough and the data points are densely observed over $[a_1, a_2]$, this is often not the case in practice (our empirical application in Section \ref{Sec_empirics} is an example). We develop theoretical results under the presence of measurement errors in functional variables, while also discussing how these results simplify in the absence of such errors. Accordingly, the subsequent theoretical developments remain applicable to the error-free case, which has been more commonly considered in the functional data analysis literature.
	
Particularly, the issue of measurement errors is prominent when considering probability density–valued functional observations, say $\{z_t\}_{t\geq1}$, or their relevant transformations $\{g(z_t)\}_{t\geq1}$ in practice. Since practitioners do not observe the true probability densities, they typically substitute them with appropriate nonparametric estimates in analysis, leading to inevitable estimation errors. As noted by \cite{seong2021functional}, neglecting these estimation errors without proper treatment results in inconsistency of standard estimators used in functional linear models. In Section \ref{Sec_empirics}, we consider a specific empirical example involving density-valued functional observations, providing a more detailed discussion based on the existing literature.

Specifically, we assume that $\tilde{x}_t$ is a measurement of $x_t$ with an additive error $e_t$, as follows:
	\begin{equation*}
		\tilde x_t  = x_t + e_t,
	\end{equation*}
	where	$e_t$ may generally be correlated with the variables $u_t^N$ and $u_t^S$ appearing \eqref{eqreg1} and \eqref{eqreg1}, and it may also be serially correlated. A key assumption which we employ for our asymptotic analysis is that $e_{t-\kappa}$ (and also $e_{t+\kappa}$) for some finite $\kappa>0$ has asymptotically negligible sample (cross-)covariances with $u_t^N$, $u_t^S$, and $e_t$. Given that $e_{t-\kappa}$ or $e_{t+\kappa}$ is (primarily) smoothing error or random disturbance associated with a variable observed at a different time, this assumption is practically reasonable and likely satisfied even for small positive $\kappa$. Similarly, $y_t$ can suffer from  error contamination, but as in conventional multivariate regression models, its measurement error is absorbed into $u_t$. This only changes the interpretation of $u_t$ in the subsequent analysis. In the presence of measurement errors $e_t$, we may rewrite \eqref{eqreg1} as follows:
	\begin{equation}
	y_{t} = f(\tilde{x}_t) + \tilde{u}_t, \quad \tilde{u}_t = u_t - f (e_t).  \label{eqreg1add} 
	\end{equation}
	We introduce assumptions on the data-generating mechanism.  Below, for notational convenience, we let  $\widetilde{\mathcal H}=  \mathcal H_y \times \mathcal H$ be the (Cartesian) product Hilbert space equipped with the inner product $\langle (h_1,h_2), (\ell_1,\ell_2) \rangle_{\widetilde{\mathcal H}} = \langle h_1,\ell_1 \rangle +   \langle h_2,\ell_2 \rangle$ (note that we let $\langle \cdot,\cdot \rangle$ to denote the inner product on either $\mathcal H_y$ or $\mathcal H$ to simplify notation, so the former is the inner product on $\mathcal H_y$). 
	Observing that $\mathcal H$ can be orthogonally decomposed by $\mathcal H^N$ and $\mathcal H^S$, we write $\mathcal H = \mathcal H^N \times \mathcal H^S$ and also write any $h\in \mathcal H$ as $(\PP^N h, \PP^S h)$; of course, in this case, for any $(h_1,h_2) \in  \mathcal H$ 
	and $(\ell_1,\ell_2) \in  \mathcal H$, 
	the inner product on this product space can simply be represented by $\langle h_1,\ell_1 \rangle + \langle h_2,\ell_2 \rangle$ with the inner product associated with $\mathcal H$. 	We employ the following assumptions throughout: in the assumption below, we consider $\mathcal H$-valued process $\mathcal E_t^x = (u_t^N, u_t^S)$ and  $\widetilde{\mathcal H}$-valued process  $\mathcal E_t = (u_t, \mathcal E_t^x)$.
	
	\begin{assumptions}\label{assum1}  $\mathcal H_y = \mathbb{R}$ or $\mathcal H$, and the following are satisfied:
		\begin{enumerate}[label=(\alph*)] 
			\item \label{assum1a} $\{x_t\}_{t\geq 1}$ satisfies \eqref{psdecom}, $d_N = \dim(\mathcal H^N) < \infty$ and $d_N$ is known. 
			\item\label{assum1b}  $\{\mathcal E_t^x\}_{\geq 1}$ is stationary and geometrically strongly mixing.
			\item \label{assum1b2}  $T^{-1}\sum_{t=1}^T \mathcal E_{t}^x \otimes \mathcal E_{t+\ell}^x = \mathbb{E}[\mathcal E_{t}^x \otimes \mathcal E_{t+\ell}^x]  + O_p(T^{-1/2})$ for any fixed integer $\ell$.
			\item \label{assum1d}	For any $k \geq 1$ and  $v_1,\ldots,v_k \in \widetilde{\mathcal H}$, 	$T^{-1}  \sum_{t=1}^T \left(\sum_{s=1}^t \mathcal E_{k,s}\right) \mathcal E_{k,t}'$ converges in distribution to $\int_{0}^1 W_k(s) d W_k(s)' + \sum_{j=0}^\infty \mathbb{E}[\mathcal E_{k,t-j}\mathcal E_{k,t}']$, where $\mathcal E_{k,t} = (\langle \mathcal E_{t}, v_1 \rangle_{\widetilde{\mathcal H}}, \langle \mathcal E_{t}, v_2 \rangle_{\widetilde{\mathcal H}}, \ldots, \langle \mathcal E_{t}, v_k \rangle_{\widetilde{\mathcal H}})'$, $W_k$ is the $k$-dimensional Brownian motion whose covariance operator is given by $ 
			\sum_{j=-\infty }^\infty \mathbb{E}[\mathcal E_{k,t-j}\mathcal E_{k,t}']$. Moreover, $\sup_{t}\mathbb{E}[\|u_t\|^{2+\delta}] <\infty$ for some $\delta>0$ and $T^{-1/2} \sum_{t=1}^T u_t$ converges weakly to a Brownian motion $W_u$ in $\mathcal H^y$. 
\end{enumerate}
\end{assumptions}
We also require assumptions on measurement errors. As detailed in Section~\ref{Sec_econometrics2}, our estimator relies on the lag-$\kappa$ autocovariance operator, with $\kappa \geq 1$ for the error-contaminated case, and also $\kappa = 0$ allowed in the error-free case. Accordingly, we impose the following assumption:
 \begin{assumpE}\label{assum1c}  One of the following holds:
            \begin{enumerate}[label=(\alph*)]
  \item \label{assum1ci} (Error-contaminated case) $\kappa \geq 1$ and  $\mathbb{E}[e_{t}\otimes z_{t+\ell}]=0$ for any $|\ell|\geq \kappa$, where $z_t = e_t$, $u_t^N$ and $u_t^S$; moreover, $\{e_t\}_{t\geq 1}$ is stationary and geometrically strongly mixing, $\mathbb{E}[\|e_t\|^4] < \infty$, 	$T^{-1} \sum_{t=1}^T e_t = O_p(T^{-1/2})$, and $T^{-1} \sum_{t=1}^T e_{t}\otimes z_{t+\ell} = \mathbb{E}[e_{t}\otimes z_{t+\ell}] + O_p(T^{-1/2})$ for any $|\ell|\geq \kappa$.
  \item (Error-free case) \label{assum1cii} $\kappa = 0$ and $e_t = 0$ for all $t$ almost surely.
\end{enumerate}
	\end{assumpE}


	Comments on Assumptions \ref{assum1} and \ref{assum1c} are in order. We assume that $\mathcal H_y = \mathbb{R}$ (resp.\ $\mathcal H_y = \mathcal H$) if $y_t$ is scalar-valued (resp.\ function-valued). In the function-valued case, $y_t$ may be defined not on $[a_1,a_2]$, as $x_t$ is, but on a different interval, say $[b_1,b_2]$; extending the subsequent theoretical results to this case is straightforward, and hence assuming $\mathcal H_y = \mathcal H$ entails no loss of generality.
    In Assumption~\ref{assum1}\ref{assum1a}, we assume that \(d_N\) is finite. This condition has been widely employed in the literature on nonstationary functional time series and seems empirically relevant (\citealp{Chang2016152,NSS,seoshang22}). A wide class of functional time series satisfies this condition (see Remark~\ref{remrepresen}).
	For convenience, we assume that \(d_N\) is known, even though it is not the case in most empirical applications. However, replacing \(d_N\) with a consistent estimator does not affect the asymptotic results to be developed. Moreover, we show in Section \ref{Sec_VRtest} of the Supplementary Material that the variance-ratio testing procedure of \cite{NSS} can be used in our setting, allowing for measurement errors.
	Assumption~\ref{assum1}\ref{assum1b} is employed to facilitate our theoretical analysis based on useful limit theorems in the existing literature (see, e.g., \citealp{Bosq2000}). Given Assumption~\ref{assum1}\ref{assum1b}, Assumption~\ref{assum1}\ref{assum1b2} does not appear restrictive, and some primitive sufficient conditions can be found in \citet[Chapter 2]{Bosq2000}.    
	Assumption~\ref{assum1}\ref{assum1d} is a technical condition required for our asymptotic analysis. Similar assumptions were employed by \cite{seo2020functional} in the study of the functional principal component analysis (FPCA) for cointegrated functional time series. Sufficient, but not restrictive, conditions for the weak convergence results stated in Assumption~\ref{assum1}\ref{assum1d} can be found in, e.g., \cite{berkes2013weak} and \cite{seo2020functional}.

	
Assumption~\ref{assum1c} states requirements on the measurement errors.
    Although our primary focus is on the error-contaminated case (Assumption~\ref{assum1c}\ref{assum1ci}), we will also show how the theoretical results simplify in the error-free case (Assumption~\ref{assum1c}\ref{assum1cii}) when applying the standard covariance-based approach (see Remark \ref{rema1}). If \(e_t\) is serially independent and also independent of \(u_s^S\) and \(u_s^N\) for every \(s\) and \(t\), then, noting that (i) \(\mathbb{E}[e_t \otimes u_s^N] = \mathbb{E}[e_t \otimes u_s^S] = 0\) for all \(s\) and \(t\) in the considered scenario and (ii) \(e_t \otimes z_t\) is a Hilbert-valued random variable (see Theorems 2.7 and 2.16 of \citealp{Bosq2000}), the conditions in Assumption~\ref{assum1c}\ref{assum1ci} are satisfied under mild conditions. However, Assumption~\ref{assum1c}\ref{assum1ci} is not restricted to such a case and allows more general cases;  specifically, \(e_t\) is assumed to be uncorrelated with \(z_{t+\ell}\) if \(|\ell|\) is sufficiently large, while no restriction is imposed on \(\mathbb{E}[e_t \otimes z_{t+\ell}]\) if \(|\ell|\) is small. That is, for our theoretical investigation of the proposed method, we only require each of \(e_t\), \(u_t^N\), and \(u_t^S\) to be uncorrelated with non-adjacent past or future measurement errors \(e_s\). This not only appears to be a mild assumption but is also reasonable for most empirical applications.

	\begin{remarks}\label{rema1}
Practitioners may find it useful to examine the theoretical results for the standard FPCA-based estimator (corresponding to the case with $\kappa=0$, as will be shown) in the absence of measurement errors, since functional data may sometimes be observed accurately. To the authors' knowledge, even in this simplified setting, no statistical theory has been established for nonstationary $y_t$ and $x_t$ (although  \citealp{Hu2016} studied the nonstationary functional AR(1) model). Accordingly, the subsequent results for $\kappa=0$ and the error-free case are also novel, motivating our explicit consideration of this scenario.
	\end{remarks}
	
	\begin{remarks}\label{remrepresen}
		Suppose that $X_t$ satisfies a functional ARMA($p,q$) law of motion (\citealp{klepsch2017prediction}): for some iid sequence $\{\varepsilon_t\}_{t \in \mathbb{Z}}$, 	$\Phi(L)X_t = \Theta(L)\varepsilon_t$,  
		where $\Phi(L) = I-\Phi_1L-\cdots-\Phi_p L^p$, $\Theta(L) = I-\Theta_1L-\cdots-\Theta_q L^q$ ($L$ denotes the lag operator),  $\Phi_1,\ldots,\Phi_p$ and $\Theta_1,\ldots,\Theta_q$ are all bounded linear operators. If we further assume that $\Phi_1, \ldots, \Phi_p$ are compact (a common assumption in the literature) and that there exists a unit root in the AR polynomial (i.e., $\Phi(1)$ is not invertible but $\Phi(z)$ is invertible for all other $z$ with $|z|<1+\eta$ for some $\eta>0$), then, according to a functional version of the Granger–Johansen representation theorem (see, e.g., \citealp{BS2018, Franchi2017b, seo_2022, seo_2023_fred}), it follows that $\mathcal H^N$ associated with the functional ARMA law of motion must possess a finite-dimensional nonstationary component; that is, $d_N = \dim(\mathcal H^N) < \infty$.
	\end{remarks}
	
We introduce additional notation. When these quantities are well-defined, we let $\llambda\edex{j}{A}$ be the $j$-th largest eigenvalue of a compact operator $A$, $\varv\edex{j}{A} $ be the corresponding eigenvector, and $\PPi\edex{j}{A} $ be the orthogonal projection onto  $\spn\{\varv\edex{j}{A}\}$; that is, 
	\begin{equation*}
		\llambda\edex{j}{A}  \varv\edex{j}{A}  = A \varv\edex{j}{A}   \quad \text{and} \quad \PPi\edex{j}{A}   = \varv\edex{j}{A}  \otimes  \varv\edex{j}{A}.
	\end{equation*}
	We also let $\Omega$ be defined by
	\begin{equation*}
		\Omega = 	\sum_{j=-\infty }^\infty \mathbb{E}[\mathcal E^x_{t-j} \otimes \mathcal E^x_{t}].
	\end{equation*}
	Under Assumption \ref{assum1}\ref{assum1c}, the above is a well-defined bounded linear operator acting on $\mathcal H$ (see Section 2.3. of \citealp{BSS2017}). 
	We hereafter let $\mathfrak F_t$ be the filtration given by 
	\begin{equation} \label{eqfiltration}
		\mathfrak F_{t}=\sigma( \{u_s\}_{s\leq t},  \{u_s^N\}_{s\leq t+1}, \{u_s^S\}_{s\leq t+1}).
	\end{equation}
	\section{Estimation and inference}\label{Sec_econometrics2}
	\subsection{Autocovariance-based FPCA} \label{sec_proposed_estimator}
	We first define  the following operators for any nonnegative integer $\kappa\geq 0$:
	\begin{equation*}
		\widehat{C}_\kappa = \frac{1}{T} \sum_{t=1}^T \tilde{x}_{t-\kappa} \otimes \tilde{x}_t, \quad \widehat{D}_{\kappa} = \widehat{C}_{\kappa}^\ast \widehat{C}_{\kappa}.
	\end{equation*}
	Here, $\widehat{C}_{\kappa}$ is the so-called lag-$\kappa$ sample autocovariance operator and $\widehat{D}_{\kappa}$, by construction, is a nonnegative self-adjoint compact operator. As such, it allows the following spectral representation:  
	\begin{equation}\label{eqspectralD}
		\widehat{D}_{\kappa} = \sum_{j=1}^\infty \llambda\edex{j}{\widehat{D}_{\kappa}}\PPi\edex{j}{\widehat{D}_{\kappa}}, \quad \llambda\edex{j}{\widehat{D}_{\kappa}} \geq 0.
	\end{equation} We then define its inverse on the restricted domain $\ran (\sum_{j=1}^{\KK} \PPi\edex{j}{\widehat{D}_{\kappa}})$ for $\KK > 0$ 
	as follows:
	\begin{equation*}
		(\widehat{D}_\kappa)^{-1}_{\KK} =  \sum_{j=1}^\KK {\llambda^{-1}\edex{j}{\widehat{D}_{\kappa}}} \PPi\edex{j}{\widehat{D}_{\kappa}}. 
	\end{equation*}
	Our proposed estimator is constructed based on the following sample operator: for some random element $z_t$, 
	\begin{equation} \label{eqestimator}
		\left(\frac{1}{T}\sum_{t=1}^T  \tilde{x}_{t-\kappa}  \otimes z_t \right)\widehat{C}_\kappa (\widehat{D}_\kappa)^{-1}_\KK, \quad \kappa \geq 0.
	\end{equation}
	In the case where $\kappa=0$ (and thus $\widehat{C}_\kappa (\widehat{D}_\kappa)^{-1}_\KK= \sum_{j=1}^\KK 
	\llambda^{-1}\edex{j}{\widehat{C}_{0}}\PPi\edex{j}{\widehat{C}_{0}}$) and $z_t=y_t$, \eqref{eqestimator} becomes identical to the standard FPCA-based estimator considered in the literature concerning stationary functional time series (see e.g., \citealp{Park2012397}). However, as the results of \cite{seong2021functional} suggest, this estimator is affected by measurement errors and may not be a consistent estimator of $f$. In our context, $f$ on the subspace $\mathcal H^S =\ran \PP^S$ is not generally consistently estimated, which results from the fact that the sample covariance $\PP^S\widehat{C}_0\PP^S$ suffers from non-negligible contamination by measurement errors (note that $\PP^S\widehat{C}_0\PP^S$ contains the component $T^{-1}\sum_{t=1}^T \PP^S e_t\otimes \PP^S e_t$, which is non-negligible) and thus is an inconsistent estimator of its true counterpart (i.e., $\mathbb{E}[\PP^Sx_t\otimes \PP^Sx_t]$). On the other hand, under Assumption \ref{assum1c}\ref{assum1ci}, 
	we may deduce that $\PP^S\widehat{C}_{\kappa}\PP^S$ for $\kappa \geq 1$ does not suffer from such serious contamination. This is why we will mainly consider the case $\kappa \geq 1$ and construct our proposed estimator using the sample autocovariance operator. Computing $(\widehat{D}_\kappa)^{-1}_\KK$ requires determining the number of retained eigenvectors, $\KK$. Subsequently, we will require $\KK$ to grow without bound depending on certain sample eigenvalues, but for now we assume only the following, required for the first few main results: 
	
	\begin{assumptions}\label{assum2}  $\KK \geq d_N$.
	\end{assumptions}
In our asymptotic analysis, we decompose $f$ as follows:
	\begin{equation}\notag 
		f = f^N + f^S, \quad \text{where} \quad f^N = f\PP^N \quad \text{and} \quad f^S = f\PP^S.
	\end{equation}
	We then consistently estimate each summand. For our purposes, it is important to obtain consistent estimators of $\PP^N$ and $\PP^S$.   
	We first show that such estimators can be obtained from the eigenvectors of $\widehat{D}_{\kappa}$. In the theorem below and hereafter, we let 
	\begin{align} \label{eqprojections}
		\widehat{\PP}_\kappa^{N} = \sum_{j=1}^{d_N} \PPi\edex{j}{\widehat{D}_{\kappa}} \quad \text{and} \quad \widehat{\PP}_\kappa^{S} = I- \widehat{\PP}_\kappa^{N}.
	\end{align}
	
	\begin{theorems}\label{thm0}
Suppose that Assumption~\ref{assum1} holds, and that either Assumption~\ref{assum1c}\ref{assum1ci} (with $\kappa\geq 1$)  or Assumption~\ref{assum1c}\ref{assum1cii} (with $\kappa = 0$)  is satisfied. Then,
		\begin{align}\label{eqthm1}
			T (\widehat{\PP}_\kappa^{N}   - \PP^N) - \Upsilon_T   &\to_{\pp}  \mathcal A_\kappa^\ast + \mathcal A_\kappa,  \\
			T (\widehat{\PP}_\kappa^{S} -\PP^S)  + \Upsilon_T &\to_{\pp}  - (\mathcal A_\kappa^\ast + \mathcal A_\kappa),\label{eqthm1a}
		\end{align}
		where $\Upsilon_T = O_p(1)$ (see Remark \ref{remadd1} for a detailed expression of $\Upsilon_T$), \begin{equation}\notag 
			\mathcal A_\kappa =_d  \left(\int W^N \otimes W^N\right)^\dag \left(\int dW^S \otimes W^N + \sum_{j \geq -\kappa}\mathbb{E}[u^S_t \otimes u^N_{t-j}] \right),
		\end{equation}	
		and $W^N$ (resp.\ $W^S$) is Brownian motion in $\mathcal H$ whose covariance operator is $\PP^N\Omega \PP^N$ (resp.\ $\PP^S \Omega \PP^S$). If there is no measurement error (i.e., $e_t=0$), then $\Upsilon_T = 0$.
	\end{theorems}
	
	\begin{remarks}\label{remadd1}
		In Theorem \ref{thm0}, \eqref{eqthm1a} follows directly from  \eqref{eqthm1} and the fact that $	T (\widehat{\PP}_\kappa^{N}   - \PP^N) = -T (\widehat{\PP}_\kappa^{S} -\PP^S)$. Moreover, from our proof of Theorem \ref{thm0}, we obtain $	\Upsilon_T = \YYYY_T + \YYYY_T^\ast$, where
		\begin{equation}\label{equpsilon}
		{\YYYY}_T = \left(T^{-2}\PP^N  \widehat{D}_\kappa \PP^N\right)^\dag (T^{-1} \PP^N  \widehat{C}_\kappa^\ast \PP^N)\left(T^{-1}\sum_{t=\kappa+1}^T \PP^S e_{t-\kappa} \otimes \PP^N {x}_{t}\right)
		\end{equation}
and  $(T^{-2}\PP^N  \widehat{D}_\kappa \PP^N)^\dag$ denotes the Moore-Penrose inverse of $T^{-2}\PP^N  \widehat{D}_\kappa \PP^N$, which is well defined (see the proof of Theorem 3.1 of \citealp{seo2020functional}); it is also shown that $	{\YYYY}_T $ is asymptotically non-negligible. The expression of $\Upsilon_T$ tells us that if we consider a special case where measurement errors are concentrated on $\mathcal H^N$ (i.e., $\PP^S e_t = 0$ for all $t$), then $\Upsilon_T = 0$.
	\end{remarks}
	
	In the case where there is no measurement error and 
    $\kappa = 0$, we have $\varv\edex{j}{\widehat{D}_{0}} = \varv\edex{j}{\widehat{C}_{0}}$ and also $\Upsilon_T = 0$. This special case corresponds to Theorem 3.1 of \cite{seo2020functional}, which concerns the FPCA of cointegrated functional time series, and Theorem \ref{thm0} can therefore be regarded as a suitable generalization of that result toward an autocovariance-based FPCA method that is robust to measurement errors. Theorem \ref{thm0} shows that the estimator $	\widehat{\PP}_{\kappa}^{N}$ is super-consistent, and, as shown in our proof of Theorem \ref{thm0}, the asymptotic bias remains and is of order $T^{-1}$; a similar result holds for $\widehat{\PP}_{\kappa}^{S}$.

	The projection estimators $	\widehat{\PP}_{\kappa}^{N}$ and $	\widehat{\PP}_{\kappa}^{S}$ give us a natural decomposition of $\widehat{D}_{\kappa}$. In the subsequent sections, we consider the decomposition of $\widehat{D}_{\kappa}$ in \eqref{eqspectralD} into the sum of $\widehat{D}_{\kappa}^{N}$ and $\widehat{D}_{\kappa}^{S}$ given \eqref{eqDdecom} below; this equation not only defines $\widehat{D}_{\kappa}^{N}$ and $\widehat{D}_{\kappa}^{S}$, but also highlights some of their key properties:
	\begin{equation}\label{eqDdecom}
		\widehat{D}_{\kappa}^{N}=	\widehat{D}_{\kappa} \widehat{\PP}_{\kappa}^{N} = \widehat{\PP}_{\kappa}^{N} \widehat{D}_{\kappa} = \sum_{j=1}^{d_N} \llambda\edex{j}{\widehat{D}_{\kappa}}\PPi\edex{j}{\widehat{D}_{\kappa}} \,\text{ and } \, 	\widehat{D}_{\kappa}^{S}=	\widehat{D}_{\kappa} \widehat{\PP}_{\kappa}^{S} = \widehat{\PP}_{\kappa}^{S} \widehat{D}_{\kappa}  = \sum_{j=d_N+1}^{\infty} \llambda\edex{j}{\widehat{D}_{\kappa}}\PPi\edex{j}{\widehat{D}_{\kappa}}.
	\end{equation}
The properties above, along with the asymptotic properties of $\widehat{\PP}_\kappa^{N}$ and $\widehat{\PP}_\kappa^{S}$ in Theorem \ref{thm0}, play a crucial role in the asymptotic analysis of our proposed estimator to be discussed.
	
	%
	\subsection{Proposed estimator} \label{sec_proposed_estimator2}
	Note that $f = f^N + f^S$, where $f^N$ captures how the persistent (nonstationary) component in $x_t$ affects $y_t$, while $f^S$ reflects the effect of the transitory (stationary) component. We propose an estimator for each of these two components, with the projections defined in \eqref{eqprojections} playing a key role. Specifically, we propose estimators of $f$, $f^N$, and $f^S$, as follows:
	\begin{align}
		&\widehat{f}_{\kappa} = \widehat{f^N_{\kappa}} + \widehat{f^S_{\kappa}}, \notag\\
		&\widehat{f^N_{\kappa}} = \left(\frac{1}{T}\sum_{t=1}^T  \tilde{x}_{t-\kappa}  \otimes y_t \right)\widehat{C}_\kappa (\widehat{D}_\kappa)^{-1}_{\KK}\widehat{\PP}_\kappa^N,  \label{eqpropdecom1} \\
		&\widehat{f^S_{\kappa}}  = \left(\frac{1}{T}\sum_{t=1}^T  \tilde{x}_{t-\kappa}  \otimes (y_t-\widehat{f^N_{\kappa}} \tilde{x}_t)\right)\widehat{C}_\kappa(\widehat{D}_\kappa)^{-1}_{\KK}\widehat{\PP}_\kappa^S, \label{eqpropdecom2}
	\end{align}
	where we note that 
	\begin{equation} \label{eqaddnotation}
		(\widehat{D}_\kappa)^{-1}_{\KK}\widehat{\PP}_\kappa^N   = \sum_{j=1}^{d_N} {\llambda^{-1}\edex{j}{\widehat{D}_{\kappa}}} \PPi\edex{j}{\widehat{D}_{\kappa}} \quad \text{and} \quad (\widehat{D}_\kappa)^{-1}_{\KK}\widehat{\PP}_\kappa^S  = \sum_{j=d_N+1}^\KK  {\llambda^{-1}\edex{j}{\widehat{D}_{\kappa}}} \PPi\edex{j}{\widehat{D}_{\kappa}},
	\end{equation}
	and these may be viewed as the inverses of $\widehat{D}_{\kappa}^{N}$ and $\widehat{D}_{\kappa}^{S}$ (see \eqref{eqDdecom}) in a restricted domain. 
	
	
The following theorem establishes the consistency of $\widehat{f^N_{\kappa}}$ as an estimator of $f^N (= f\PP^N)$ and details its limiting behavior: in the theorem below, $W^N$, $W^S$, and $W^u$ are defined as in Theorem \ref{thm0} and Assumption \ref{assum1}, and recall that $\to_{\pp}$ denotes convergence in probability with respect to the usual operator norm for operator-valued sequences (see Section \ref{Sec_prelim2}).

	\begin{theorems}\label{thm1}
		Suppose that Assumptions \ref{assum1} and \ref{assum2} hold, along with either \ref{assum1c}\ref{assum1ci} (for $\kappa \geq 1$) or \ref{assum1c}\ref{assum1cii} (for $\kappa = 0$). Further, assume that $\{u_t\}_{t \geq 1}$ is a martingale difference sequence with respect to $\mathfrak F_{t}$ defined in \eqref{eqfiltration}.  Then as $T\to \infty$, 
		$\widehat{f^N_{\kappa}} 
		\to_{\pp} f^N$ and 
		\begin{align}
			T(\widehat{f^N_{\kappa}} - f^N) + \YY_T \to_{\pp} f(\mathcal A_{\kappa} + \mathcal A_{\kappa}^\ast ) + V_{2} V_1^\dag, \label{thm1eq1}
		\end{align}
		where $\YY_T = O_p(1)$  (see Remark \ref{remadd1a} for a detailed expression of $\YY_T$), $V_1 =_d \int W^N \otimes W^N$ and $V_{2} =_d \int W^N \otimes d W^u$. If there is no measurement error (i.e., $e_t=0$), then $\YY_T = 0$.
	\end{theorems}
	Theorem \ref{thm1} demonstrates that the proposed estimator $\widehat{f^N_{\kappa}}$ is a consistent estimator of $f^N$, and the asymptotic bias is of order $T^{-1}$; this result parallels that of the standard least squares-type estimator for the cointegrating relationship in the finite-dimensional case.
	We present remarks containing complementary results to Theorem \ref{thm1}.
	\begin{remarks}
		It may be deduced from our proofs of Theorems \ref{thm0} and \ref{thm1} that the explicit expression of $\YY_T$ in Theorem \ref{thm1} is given as follows: 
		\begin{equation}\label{eqyy}
			\YY_T =  \left(T^{-1}\sum_{t=1}^T   {\PP}^N  x_{t-\kappa} \otimes  f(e_t)\right){\widehat{\QQ}^N_{\kappa}  \widehat{C}_\kappa \widehat{\PP}^N_{\kappa} } ({\widehat{\PP}^N_{\kappa} \widehat{D}_\kappa \widehat{\PP}^N_{\kappa}})_{\KK}^{-1} - f(\Upsilon_T), 
		\end{equation}
		where $\Upsilon_T$ is given in Theorem \ref{thm0} and Remark \ref{remadd1}, and $\YY_T=O_p(1)$ is easily deduced from our proof.  From \eqref{equpsilon} and \eqref{eqyy}, we know that this $O_p(1)$ term results from (i) $T^{-1}\sum_{t=\kappa+1}^T e_{t-\kappa} \otimes \PP^N {x}_{t}$ and (ii) $T^{-1}\sum_{t=1}^T   {\PP}^N  x_{t-\kappa} \otimes  f(e_t)$ appearing in our asymptotic analysis. If these two are asymptotically negligible, $\YY_T$ in \eqref{thm1eq1} disappears; however, in the presence of measurement errors, (i) and (ii) are not generally negligible. 
	\end{remarks}

    
	\begin{remarks}\label{remadd1a}
		In Theorem \ref{thm1}, we assume that $u_t$ is a martingale difference with respect to $\mathfrak{F}_{t}$. A more general result can be obtained, without requiring the martingale difference condition. This only requires replacing $V_2$ in Theorem \ref{assum2} with $$V_{2} =_d \int W^N \otimes d W^u - \sum_{j \geq \kappa}\mathbb{E}[u_{t-j}^N \otimes u_{t}].$$ In fact, our proof of Theorem \ref{thm1} given in Section \ref{app: sec: d} of the Supplementary Material accommodates this more general case.
	\end{remarks}
	We next study the asymptotic properties of $\widehat{f^S_{\kappa}}$ as an estimator of $f^S(=f\PP^S)$.  As with  standard FPCA-based estimators, our proposed estimator given in \eqref{eqpropdecom2} is defined on a finite-dimensional eigenspace of $\widehat{D}_\kappa^S$. Using the result that $\widehat{\PP}^S_{\kappa} - \PP^S = O_p(T^{-1})$ (see Theorem \ref{thm0}), we may deduce that $\widehat{D}_\kappa^S$ is a consistent estimator of $D_{\kappa}^S$, defined below:
	\begin{equation*}
		D_{\kappa}^S = (C^S_{\kappa})^\ast C^{S}_{\kappa}, \quad \text{with} \quad	C^S_{\kappa} = \mathbb{E}[\PP^S x_{t-\kappa} \otimes \PP^S x_{t}].
	\end{equation*}
Note that our estimator $\widehat{f^S_{\kappa}}$ is defined on a $(\KK-d_N)$-dimensional eigenspace of $\widehat{D}_{\kappa}$. For this estimator to be a consistent estimator of $f^S$ defined on the entire $\mathcal H^S$, we need some conditions on $D_{\kappa}^S$ and $f^S$. Moreover, it is also necessary to let $\KK$ grow without bound. The required conditions are summarized below: 
	\begin{assumptions}\label{assum3} $D_{\kappa}^S$, $f^S$ and $\KK_S$ (defined as $\KK_S:=\KK-d_N$) satisfy the following: \begin{enumerate}[(a)] \item \label{assum3a}$D_{\kappa}^S$ is injective on $\mathcal H^S$ (i.e., $\ker D_{\kappa}^S \cap \mathcal H^S = \{0\}$), and $\sum_{j=1}^\infty \| f^S(g_j) \|^2 <\infty$  for any orthonormal basis $\{g_j\}_{j\geq 1}$ (meaning that $f^S$ is a Hilbert-Schmidt operator if $\mathcal H_y = \mathcal H$). \item\label{assum3b}   $\KK_S = \# \{j : \llambda\edex{j}{\widehat{D}^S_{\kappa}} > \alpha\}$ and $\alpha = a_1T^{-a_2}$ for some $a_1>0$ and $a_2\in (0,1/2)$.
    \end{enumerate}
	\end{assumptions}
	\noindent Assumption \ref{assum3}\ref{assum3a} contains requirements similar to those employed by \cite{seong2021functional} for functional linear models. The decision rule for $\KK_S$ in Assumption \ref{assum3}\ref{assum3b} adapts a commonly used approach, considered reasonable in practice for FPCA-based estimators (see, e.g., Section 3.1 and Remark 2 of the aforementioned paper). From the properties in \eqref{eqDdecom}, we have 
\begin{equation*}
\llambda\edex{j}{\widehat{D}^S_{\kappa}} = \llambda\edex{j+d_N}{\widehat{D}_{\kappa}},
\end{equation*}
    and hence computing $\KK_{S}$ according to Assumption \ref{assum3}\ref{assum3b} does not require additional calculation of eigenvalues associated with $\widehat{D}^S_{\kappa}$.    
	We next give the asymptotic properties of $\widehat{f^S_{\kappa}}$ as an estimator of $f^S$. 
	In the theorem below and hereafter, we let $\widetilde{C}_0^S = \mathbb{E}[\PP^{S} \tilde x_{t} \otimes \PP^{S} \tilde x_{t}]$, $\widetilde{C}_{u} = \mathbb{E}[\tilde{u}_t\otimes \tilde{u}_t]$, $\ttau\edex{j}{{D}_{\kappa}^S} = \max\{(\llambda\edex{j-1}{{D}_{\kappa}^S}-\llambda\edex{j}{{D}_{\kappa}^S})^{-1},(\llambda\edex{j}{{D}_{\kappa}^S}-\llambda\edex{j+1}{{D}_{\kappa}^S})^{-1}\}$, 
	\begin{equation}
	\widehat{\PP}^{\KK_S}_{\kappa} = \widehat{\PP}^{\KK}_{\kappa}\widehat{\PP}^{S}_{\kappa}  = \sum_{j=d_N+1}^{\KK} \PPi\edex{j}{\widehat{D}_{\kappa}} \quad\text{and}\quad 		\theta_{\KK_S}(\zeta) = \langle \zeta,  (D_\kappa^S)_{\KK_S}^{-1}   (C_{\kappa}^S)^\ast \widetilde{C}_{0}^S C_{\kappa}^S(D_\kappa^S)_{\KK_S}^{-1} (\zeta) \rangle,\label{eqtheta}
	\end{equation}
	where
	\begin{equation} \label{eqdkappa}
		(D_\kappa^S)_{\KK_S}^{-1} = \sum_{j=1}^{\KK_S} \llambda^{-1}\edex{j}{{D}_{\kappa}^S}\PPi\edex{j}{{D}_{\kappa}^S}.
	\end{equation}
	Our next result studies the asymptotic properties of $\widehat{f^S_{\kappa}}$: in the theorem below, $N(0,A)$ denotes zero-mean Gaussian random element taking values in $\mathcal H_y$ with (co)variance $A$.
	\begin{theorems}\label{thm2}
		Suppose that  Assumptions \ref{assum1}-\ref{assum3} hold, along with either \ref{assum1c}\ref{assum1ci} (for $\kappa \geq 1$) or \ref{assum1c}\ref{assum1cii} (for $\kappa = 0$). Further assume that $u_t$ is a martingale difference with respect to $\mathfrak F_{t}$  in \eqref{eqfiltration}, and
		\begin{equation} \label{eqcondition_a}
			\llambda\edex{1}{D_{\kappa}^S}>\llambda\edex{2}{D_{\kappa}^S}>\cdots > 0 \,\, \text{ and } \,\,  T^{-1/2}\alpha^{-1/2}\sum_{j=1}^{\KK_S}\ttau\edex{j}{D_{\kappa}^S} \to_p 0.
		\end{equation}
		Then,  $\widehat{f^S_{\kappa}} \to_p f^S$.  Moreover, for any $\zeta \in \mathcal H$, the following holds: 
		\begin{equation} \label{eqthm2}
			\sqrt{T/\theta_{\KK_S}(\zeta)}(\hat{f}_{\kappa}(\zeta)  -f\widehat{\PP}_\kappa^{\KK}(\zeta)) = 		\sqrt{T/\theta_{\KK_S}(\zeta)}(\widehat{f^S_{\kappa}}(\zeta)  -f^{S}\widehat{\PP}_\kappa^{\KK_S}(\zeta))  + o_p(1)\to_d N(0,\widetilde{C}_u).
		\end{equation}
	\end{theorems}
	Even if the condition given by \eqref{eqcondition_a} in Theorem \ref{thm2} requires that the eigenvalues of $D_{\kappa}^S$ are distinct, it does not place any other essential restrictions on the eigenstructure of $D_{\kappa}^S$. Given that $\sum_{j=1}^{\KK_S}\ttau\edex{j}{D_{\kappa}^S}$ increases in $\KK_S$ (and thus $\alpha^{-1})$, this condition merely requires $\alpha$ to decay to zero at a sufficiently slower rate. In fact, assumptions similar to \eqref{eqcondition_a} are standard and widely used in the literature on functional linear models (see e.g., \citealp{Park2012397,seong2021functional}). Moreover, it is possible to relax the assumption of distinct eigenvalues in Theorem \ref{thm2} under a different set of assumptions, which is detailed in Remark \ref{remadd}. 
	
	From Theorems \ref{thm1} and \ref{thm2}, we know that the proposed estimator $\hat{f}_{\kappa}$ is consistent under the employed assumptions, i.e., $\hat{f}_{\kappa} \to_p f$. 
	Moreover, as described by \eqref{eqthm2}, we find that our estimator $\widehat{f}_{\kappa}$ is asymptotically normal in a certain sense. However, unlike in a finite-dimensional setting, there are some limitations associated with the asymptotic normality given by \eqref{eqthm2}. First, $\widehat{f}_{\kappa}$ is centered at a random biased operator $f\widehat{\PP}^{\KK}_\kappa$, but not $f$, and (ii) the convergence is established in a pointwise manner at each point $\zeta \in \mathcal H$ but not uniformly over the entire space $\mathcal H$. As noted by \citet[Section 3.2]{seong2021functional}, these limitations are in fact common in the literature concerning FPCA-based estimation of the functional linear model; see also Theorem 3.10 of \cite{Hu2016}.  
	
	Obviously, \eqref{eqthm2} may be used for inference on  $f\widehat{\PP}^{\KK}_\kappa (\zeta)$, where $\widehat{\PP}^{\KK}_\kappa(\zeta)$ is naturally understood as the optimal approximation of $\zeta$ using the eigenvectors of $\widehat{D}_{\kappa}$. For example, when $\mathcal H_y = \mathcal H$ (and hence $y_t$ is function-valued), we may construct the 95\% confidence interval of $\langle f\widehat{\PP}^{\KK}_\kappa (\zeta), \varphi \rangle$ for some $\varphi \in \mathcal H_y$ using the asymptotic normality result \eqref{eqthm2}, as follows::
	\begin{equation}\label{localasymp01}
		\langle \hat{f}_{\kappa} \widehat{\PP}^{\KK}_{\kappa}(\zeta), \varphi \rangle \pm 1.96 \sqrt{{\theta}_{\KK_S} \langle \widetilde{C}_{u}\varphi,\varphi \rangle/T},
	\end{equation}
	where the unknown quantities ${\theta}_{\KK_S}$ and $\widetilde{C}_{u}$ can be replaced by reasonable estimators that can be easily computed from our proposed estimator $\hat{f}_{\kappa}$ without affecting asymptotic validity (see Corollary \ref{cor1} of the Supplementary Material). 
	However, practitioners may want to avoid being interfered by a random projection $\widehat{\PP}^{\KK}_\kappa$ and implement a direct statistic inference on $\langle f (\zeta), \varphi\rangle$ rather than on  $\langle f\widehat{\PP}^{\KK}_\kappa (\zeta), \varphi\rangle$.  We will show that, under additional assumptions (requring ``smoothness'' of $f$ and $\zeta$), the asymptotic normality result \eqref{eqthm2} still holds even when $f\widehat{\PP}^{\KK}_\kappa$ is replaced by $f$. This allows statistical inference without the influence of the random projection $\widehat{\PP}^{\KK}_\kappa$, as discussed in more detail in the next section.  

	
	\begin{remarks} \label{remadd}
		In Theorem \ref{thm2}, we require that the eigenvalues of $D_{\kappa}^S$ are distinct. Even if similar assumptions have been widely adopted in the literature on functional linear models, practitioners may want to relax this restriction. In fact, it can be shown that Theorem \ref{thm2} holds when \eqref{eqcondition_a} is replaced by the following conditions:
		(i) $\alpha$ and $\KK_S$ are chosen so that $\llambda\edex{{\KK}_S}{D_{\kappa}^S} \neq \llambda\edex{{\KK}_S+1}{D_{\kappa}^S}$ and $T^{1/2}(\llambda\edex{{\KK}_S}{D_{\kappa}^S} -\llambda\edex{{\KK}_S+1}{D_{\kappa}^S}) \to_p \infty$ and  (ii) $\sqrt{\frac{{\KK}_S}{T}}\llambda^{-1}\edex{{\KK}_S}{D_{\kappa}^S}(\llambda\edex{{\KK}_S}{D_{\kappa}^S} -\llambda\edex{{\KK}_S+1}{D_{\kappa}^S})^{-1} \to_p 0$. Our proof of Theorem \ref{thm2} provides more details on how these conditions can replace \eqref{eqcondition_a}. 
		Note that in the two conditions, we only require the last eigenvalue appearing in \eqref{eqdkappa} to be distinct from the next one, thus allowing arbitrary repetition of other eigenvalues. 
	\end{remarks}
	
	\begin{remarks}\label{remadd2}
		As is well known (see \citealp{seong2021functional}), $\theta_{\KK}(\zeta)$ in Theorem \ref{thm2} may converge or diverge depending on $\zeta$, so it is impossible to find a sequence $c_T$ such that $c_T(\widehat{f}_{\kappa}(\zeta)-f\widehat{\PP}_\kappa^{\KK}(\zeta))$ converges  uniformly in $\zeta$. As also noted by \citet[Theorem 3.1]{Mas2007}, it is generally impossible to find a sequence $c_T$ such that $c_T(\widehat{f}_{\kappa}(\zeta)-f(\zeta))$ converges uniformly in $\zeta$. 
	\end{remarks}

    \begin{remarks}\label{remadd3}
One may consider using the standard FPCA-based estimator (corresponding to $\kappa=0$) even in the presence of measurement errors. With only a slight modification of our proof of Theorem \ref{thm1}, it can be shown that \(\hat{f}_0^N\) consistently estimates \(f^N\). Therefore, a simple modification of the standard FPCA-based estimator yields a consistent estimator of \(f^N\). However, as can be deduced from our proof of Theorem~\ref{thm2} and from the existing results of \cite{Chen_et_al_2020} and \cite{seong2021functional}, $\hat{f}_{0}^S$ is inconsistent for $f^S$ in this case, and hence $\hat{f}_0$ is inconsistent as an estimator of $f$. 
     \end{remarks}
	
	\subsection{Statistical inference: local confidence bands of a partial effect} \label{sec_inference}
We consider the following assumptions on $\PP^Sx_t$, $f$ and $\zeta$, which are similar to the conditions employed by \cite{seong2021functional}: below, for $j, \ell\geq 1$,  $\varpi_{t}(j,\ell)=\langle \PP_Sx_t, \varv\edex{j}{D_{\kappa}^S} \rangle\langle \PP_Sx_{t-\kappa},\varv\edex{\ell}{E_{\kappa}^S} \rangle-\mathbb{E}[\langle \PP_Sx_t,\varv\edex{j}{D_{\kappa}^S} \rangle\langle \PP_Sx_{t-\kappa},\varv\edex{\ell}{E_{\kappa}^S}\rangle]$ and  $E_{\kappa}^S = C^{S}_{\kappa}(C^S_{\kappa})^\ast$. 
	\begin{assumptions}\label{assum2add}  There exist $c>0$, $\rho>2$, $\varsigma>1/2$, $\gamma > 1/2$ and  $\delta_{\zeta} > 1/2$ satisfying the following:
		\begin{enumerate}[label=(\alph*)]  
			\item \label{assum2a} $\llambda\edex{j}{D_{\kappa}^S}\leq cj^{-\rho}$, $\llambda\edex{j}{D_{\kappa}^S}-\llambda\edex{j+1}{D_{\kappa}^S}\geq c j^{-\rho-1}$, $\langle f (\varv\edex{j}{D_{\kappa}^S}),\varv\edex{\ell}{E_{\kappa}^S} \rangle \leq c j^{-\varsigma} \ell^{-\gamma}$, 	 $\mathbb{E}[\varpi_{t}(j,\ell)\varpi_{t-s}(j,\ell)]\leq c s^{-m}\mathbb{E}[\varpi_{t}^2(j,\ell)]$ for $m>1$,  $\mathbb{E}[\langle \PP^Sx_t,\varv\edex{j}{D_{\kappa}^S} \rangle^4] \leq c \llambda\edex{j}{D_{\kappa}^S}$, $\mathbb{E}[\langle \PP^Sx_{t},\varv\edex{j}{E_{\kappa}^S} \rangle^4] \leq c \llambda\edex{j}{E_{\kappa}^S}$, and $\langle \varv\edex{j}{D_{\kappa}^S},\zeta \rangle \leq c j^{-\delta_{\zeta}}$.  
			\item \label{assum2c} $\varsigma +\delta_{\zeta} > \rho/2 + 2$ and $T\alpha^{2\varsigma+2\delta_{\zeta}-1}=O(1)$. 	 		
		\end{enumerate} 
	\end{assumptions}
	Assumption \ref{assum2add}\ref{assum2a} summarizes the technical conditions needed to establish the desired results. Similar requirements have been employed in the literature on functional linear models (see e.g., \citealp{Hall2007,imaizumi2018,seong2021functional}). Noting that, under this condition, $\ttau\edex{j}{D_{\kappa}^S}\leq  c j^{\rho+1}$ and $\sum_{j=1}^{M} j^{\rho+1} = O(M^{\rho+2})$ for positive integer $M$, one may observe that,  under Assumption \ref{assum2add}\ref{assum2a}, the conditions given by \eqref{eqcondition_a} may be replaced with the following sufficient condition: $T^{-1/2}\alpha^{-1/2} \KK_S^{\rho+2} \to_p 0$.
	Given that $\alpha = a_1 T^{-a_2}$ for some $a_1 > 0$ and $a_2 \in (0, 1/2)$, Assumption \ref{assum2add}\ref{assum2c} requires that $\|f(\varv\edex{j}{D_{\kappa}^S})\|$ and $\langle \zeta, \varv\edex{j}{D_{\kappa}^S} \rangle$ decay to zero at a sufficiently fast rate as $j$ increases, implying that $f$ and $\zeta$ are sufficiently smooth with respect to the eigenvectors $\varv\edex{j}{D_{\kappa}^S}$. 
	\begin{theorems}\label{prop1} 
		Suppose that the assumptions in Theorem \ref{thm2} hold along with Assumption \ref{assum2add} and $\theta_{\KK_S}(\zeta) \to_p \infty$. Then,
		\begin{equation} \label{eqthm2a}
			\sqrt{T/\theta_{\KK_S}(\zeta)}(\hat{f}_{\kappa}(\zeta)-f(\zeta))  \to_d N(0,\widetilde{C}_{u}).
		\end{equation}
	\end{theorems} 
	\begin{remarks}
		The above theorem requires that $\theta_{\KK_S}(\zeta)\to\infty$. This is likely to be true for many possible choices of $\zeta$. As discussed in the literature (see, e.g., Remark 4 of \citealp{seong2021functional}, and references therein), $\theta_{\KK_S}(\zeta)$ is convergent only on a strict subspace of $\mathcal H$; thus, unless $\zeta$ lies entirely within this subspace, we have $\mathbb{P}\{\theta_{\KK_S}(\zeta) < c < \infty\} \to 0$ as $\KK \to \infty$.
	\end{remarks}
	
	
	Even if all the assumptions required for Theorem \ref{prop1} hold, $\hat{f}_{\kappa}-f$ converges to a Gaussian random element at a rate depending on $\zeta$ and thus it is not generally possible to construct a uniform confidence band of $f$ from Theorem \ref{prop1} (see Remark \ref{remadd2}). However, it may be possible to construct a local (or locally approximate) confidence band, which is naturally interpreted. 
	We first note that $f(\zeta)$ may be understood as a partial effect on $y_t$ of a perturbation $\zeta$ in $x_t$, which is often of interest in practice. If $\mathcal H_y = \mathbb{R}$ and hence $y_t$ is real-valued, $f(\zeta)$ is a real-valued effect on $y_t$ of a perturbation $\zeta$, 
and in this case we may directly use \eqref{eqthm2a} for statistical inference by replacing $\widetilde{C}_{u}$ and $\theta_{\KK_S}$ with their sample counterparts (see Corollary \ref{cor1} of the Supplementary Material). Now suppose that $\mathcal H_y = \mathcal H$. In this case, $f(\zeta)$ is a function defined on $[a_1,a_2]$, and we may construct a sequence of confidence intervals for local averages of $f(\zeta)$. Specifically, let $\mathcal I_j = (b_{j+1}- b_j)^{-1} 1\{u \in [b_j,b_{j+1}]\}$ for some $b_j$ and $b_{j+1}$ with $a_1 \leq b_j < b_{j+1} \leq a_2$. Then
	$\langle f(\zeta), \mathcal I_j \rangle = (b_{j+1}- b_j)^{-1} \int_{b_j}^{b_{j+1}} f(\zeta)(s) ds$ computes the local average of $f(\zeta)$ on the interval $[b_j,b_{j+1}]$. Using the results given in Theorem \ref{prop1}, we know that
	\begin{equation} \label{eqappconf} 
		\sqrt{T/\theta_{\KK_S}(\zeta)}(\langle (\hat{f}_{\kappa}(\zeta)-f(\zeta)), \mathcal I_j \rangle \to_d N(0,\langle\mathcal  I_j, \widetilde{C}_{u} \mathcal I_j \rangle).  
	\end{equation}
	Note that $\langle\mathcal I_j, \widetilde{C}_{u} \mathcal I_j \rangle$ and $\theta_{\KK_S}$ can also be replaced by their sample counterparts (Corollary \ref{cor1} of the Supplementary Material), enabling construction of an asymptotically valid confidence band for the local average via \eqref{eqappconf}. This can be applied to overlapping or non-overlapping sequences of intervals $\{\mathcal I_j\}_{j=1}^{M}$ with $\cup_{j=1}^M \mathcal I_j = [a_1,a_2]$.  These confidence intervals are readily interpretable and can be useful in practice.
	
	\subsection{The model with an intercept} \label{sec_det}
	In the previous sections, we developed statistical inferential methods for the case where $\mathbb{E}[y_t] = \mathbb{E}[x_t] = 0$ for simplicity. However, in practice, a nonstationary time series may include a nonzero intercept, and hence our observations $x_t$ and $y_t$ may satisfy $\mathbb{E}[x_t] = \mu_x$ and $\mathbb{E}[y_t] = \mu_y$ for some unknown $\mu_x$ and $\mu_y$. To accommodate this scenario, one may consider the model with a deterministic term as follows: 
	\begin{equation}
		y_t = \mu + f(\tilde x_t) + \tilde{u}_t, \label{eqreg1deter}
	\end{equation}
	where $\mu = f(\mu_x) - \mu_y \in \mathcal{H}$.
	With a straightforward modification,   we can still achieve consistent estimation of $f$ and extend the statistical inference on $f(\zeta)$ given in Section \ref{sec_inference}.  More specifically, inference for this case can be implemented using the centered (demeaned) variables ${y}_{c,t}=y_t-\bar{y}_T$ and $\tilde{x}_{c,t}=x_t +e_t -\bar{x}_T-\bar{e}_T$, where $\bar{y}_T = T^{-1}\sum_{t=1}^T y_t$, and $\bar{x}_T$ and $\bar{e}_T$ are similarly defined. The proposed estimator is given as follows: $\widehat{f}_{c,\kappa} = \widehat{f}_{c,\kappa}^N + \widehat{f}_{c,\kappa}^S,$ where
	\begin{align}
		&\widehat{f}_{c,\kappa}^N = \left(\frac{1}{T}\sum_{t=1}^T  \tilde{x}_{c,t-\kappa}  \otimes y_{c,t} \right)\widehat{C}_{c,\kappa} (\widehat{D}_{c,\kappa})^{-1}_{\KK}\widehat{\PP}_{c,\kappa}^N, \notag \\ 
		&\widehat{f}_{c,\kappa}^S  = \left(\frac{1}{T}\sum_{t=1}^T  \tilde{x}_{c,t-\kappa}  \otimes (y_{c,t}-\widehat{f^N_{\kappa}} \tilde{x}_{c,t})\right)\widehat{C}_{c,\kappa}(\widehat{D}_{c,\kappa})^{-1}_{\KK}\widehat{\PP}_{c,\kappa}^S,  \notag  
	\end{align}
	where $\widehat{C}_{c,\kappa}$, $\widehat{D}_{c,\kappa}$, $\widehat{\PP}_{c,\kappa}^N$, and $\widehat{\PP}_{c,\kappa}^S$ are similarly computed as $\widehat{C}_{\kappa}$, $\widehat{D}_{\kappa}$, $\widehat{\PP}_{\kappa}^N$, $\widehat{\PP}_{\kappa}^S$, respectively, but with the centered variables. The consistency of the estimator can be established with only slight modifications, and the pointwise asymptotic normality can also be achieved as follows:
	\begin{align} \label{eqthm2aapp}
		\sqrt{T/\theta_{c,\KK_S}(\zeta)}(\hat{f}_{c,\kappa}(\zeta)-f(\zeta))  \to_d N(0,\widetilde{C}_{u}),
	\end{align}
	where $\theta_{c,\KK_S}$ is defined similarly to $\theta_{\KK_S}$ in \eqref{eqtheta}, but with $C_{\kappa}^S$ and the other operators ($\widetilde{C}_{0}^S$, $D_\kappa^S$, $(D_\kappa^S)_{\KK_S}^{-1}$) depending on $C_{\kappa}^S$ being computed from the centered variables $\PP^S\tilde{x}_{t} - \mathbb{E}[\PP^S\tilde{x}_{t}]$ ($t=1,\ldots,T$)  rather than $\PP^S\tilde x_{t}$. As in the previous case,  $\theta_{c,\KK_{S}}$ and $\widetilde{C}_{u}$ can be replaced by their sample counterparts without affecting the asymptotic result in \eqref{eqthm2aapp}, enabling statistical inference on $f(\zeta)$ in practice. A more detailed discussion including theoretical justification of these results is given in Section \ref{ap_sec_supp2} of the Supplementary Material.
	%
	
	

\section{Numerical studies}\label{Sec_empirics}
\subsection{Monte Carlo simulation}
We conducted simulation experiments to compare the autocovariance-based estimator with the standard covariance-based one.  We compare estimator accuracy and coverage probabilities of confidence intervals for $\langle f(\zeta),\varphi \rangle$ for $\zeta \in \mathcal H$ and $\varphi \in \mathcal H_y$ across estimators. The simulation results support using the autocovariance-based approach with error-contaminated data. However, we postpone these details to Section \ref{Sec_monte_carlo} of the Supplementary Material to focus on real-data analysis of the economic impact of climate change, which more effectively illustrates our methods' empirical relevance and usefulness.
\subsection{Empirical Applications: Economic Impact of Climate Change}

We present an empirical study examining the economic impact of climate change using our proposed method. A large body of research in climate economics has shown that rising temperatures negatively affect economic growth (e.g., \citealp{dell2012temperature}; \citealp{burke2015global}; \citealp{newell2021gdp}; \citealp{CruzHansberg2023}), and estimating these effects is of central importance because they inform the economic costs of climate change and policy responses. Motivated by this literature, we apply our proposed method to global climate and economic data to demonstrate its empirical usefulness in providing statistical evidence for this relationship along with broader estimates of the economic impact of climate change.

We consider appropriate transformations of the probability densities of gross regional product (GRP) growth rates ($y_t$), as a measure of regional economic activity, and land temperature anomalies ($x_t$), commonly used as indicators of climate change. Because these time series are expected to be nonstationary, contaminated by measurement errors, and to exhibit nonzero unconditional means, we employ model \eqref{eqreg1deter}, with $u_t$ absorbing measurement errors in the dependent variable.

\subsubsection{Raw data and functional data in analysis}\label{Sec_empirics1}
\indent We use non-infilled gridded land temperature anomaly data from a collaborative product of the Climatic Research Unit at the University of East Anglia, the Met Office Hadley Centre, and the National Centre for Atmospheric Science (CRUTEM.5.0.2.0, \citealp{osborn2021land}). We estimate spatially distributed temperature anomaly densities for 1951–2019 using a Gaussian kernel with Silverman’s bandwidth. To avoid COVID-19-related distortions in $y_t$, data from 2020 onward are excluded. At each time $t$, the distribution’s support is restricted to the range containing 99\% of the total probability mass, $[-5.80, 6.68]$, thereby excluding outliers as in \cite{chang2020evaluating}.

\indent For the GRP growth rates, we employ both the real GRP data of \cite{leonie_wenz_2023_7573249} for 1960–2019 and the real GDP in millions of 2021 international dollars, converted using Purchasing Power Parities, from the Conference Board Total Economy Database (TED) for 1950–2019.\footnote{The Conference Board Total Economy Database™ (April 2022) - Output, Labor and Labor Productivity, 1950-2022, downloaded from https://www.conference-board.org/data/economydatabase/total-economy-database-productivity on April 13, 2023.} \cite{leonie_wenz_2023_7573249} provide subnational economic output data for over 1,661 regions across 83 countries, enabling panel and cross-sectional regression analyses that reduce coverage bias and increase the number of observations. Based on this approach, we spatially disaggregate TED's country-level real GDP into the regional product levels from 1950 to 2019. Using these data, we estimate panel fixed-effects models to remove persistent regional heterogeneity and long-term structural changes, and then relate the residual component of regional growth to climate variables. Detailed procedures for spatial disaggregation, density generation (on support $[-0.105, 0.092]$, excluding  extreme observations), and panel specifications are provided in Section~\ref{AP_WGRP} of the Supplementary Material.

\indent Figure \ref{Fig:GTemp_dist} shows the densities of land temperature anomalies and the temperature-related components of GRP growth rates, along with their first two central moments from 1951 to 2019. The mean of land temperature anomalies shows a persistent upward trend, with a rising standard deviation indicating greater variability. In contrast, the mean of temperature-related regional growth turned negative after the mid-1980s, while its standard deviation remained largely unchanged, suggesting stable dispersion in regional growth responses. 

\begin{figure}[t]
	\begin{center}
		\includegraphics[height=0.25\textwidth, width=0.45\textwidth]{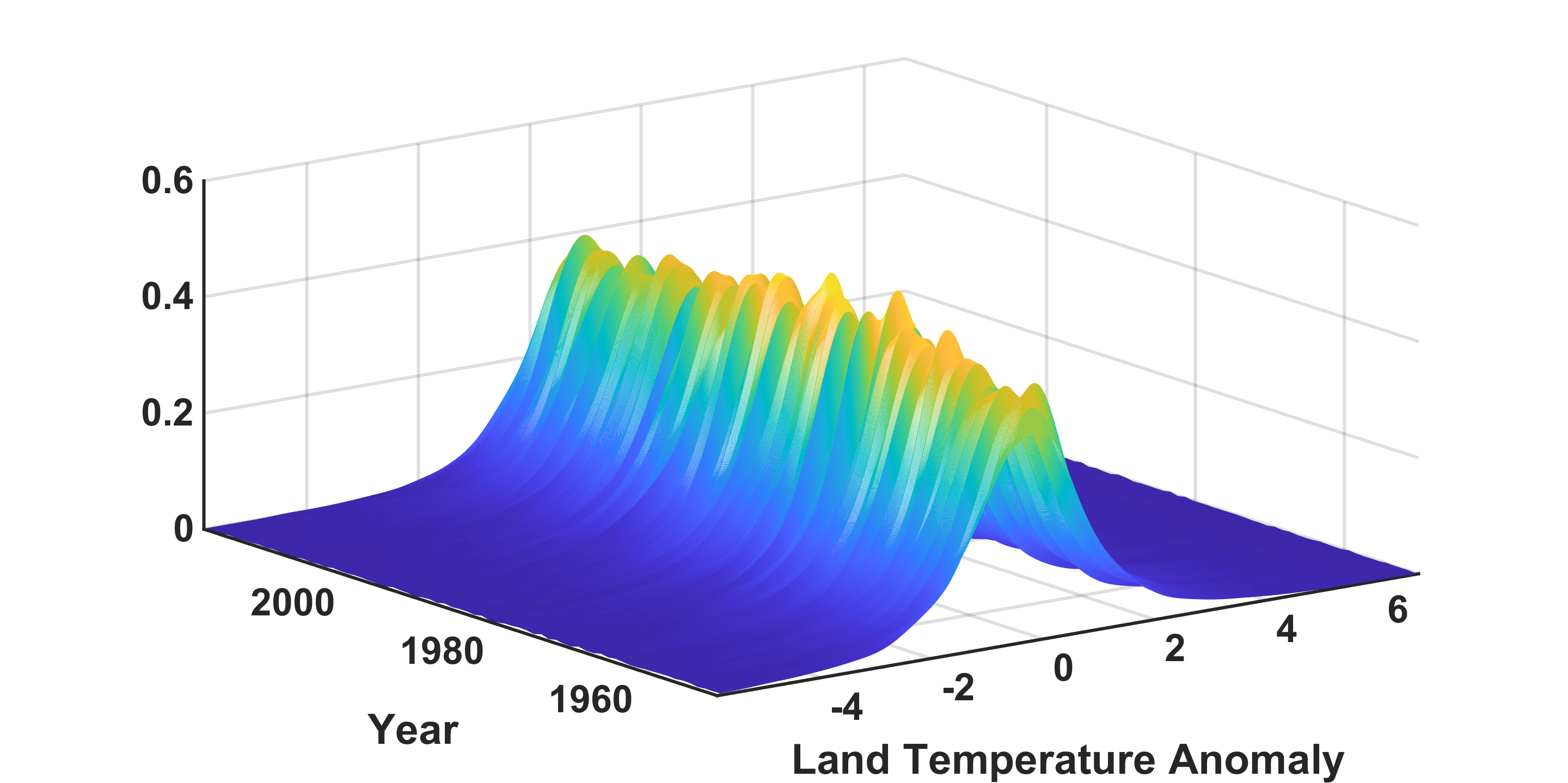}
	    \includegraphics[height=0.25\textwidth, width=0.45\textwidth]{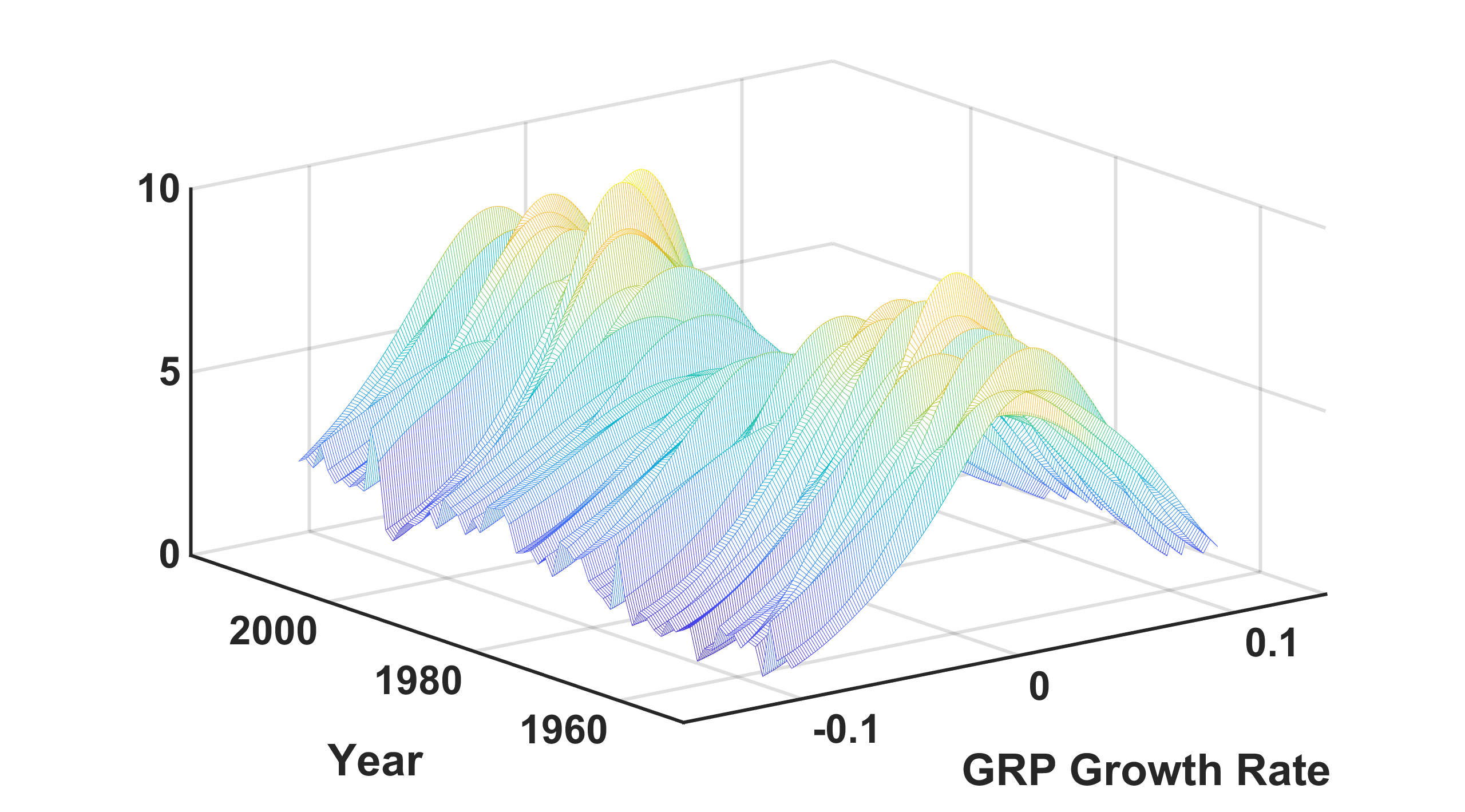}
		\includegraphics[height=0.2\textwidth, width=0.45\textwidth]{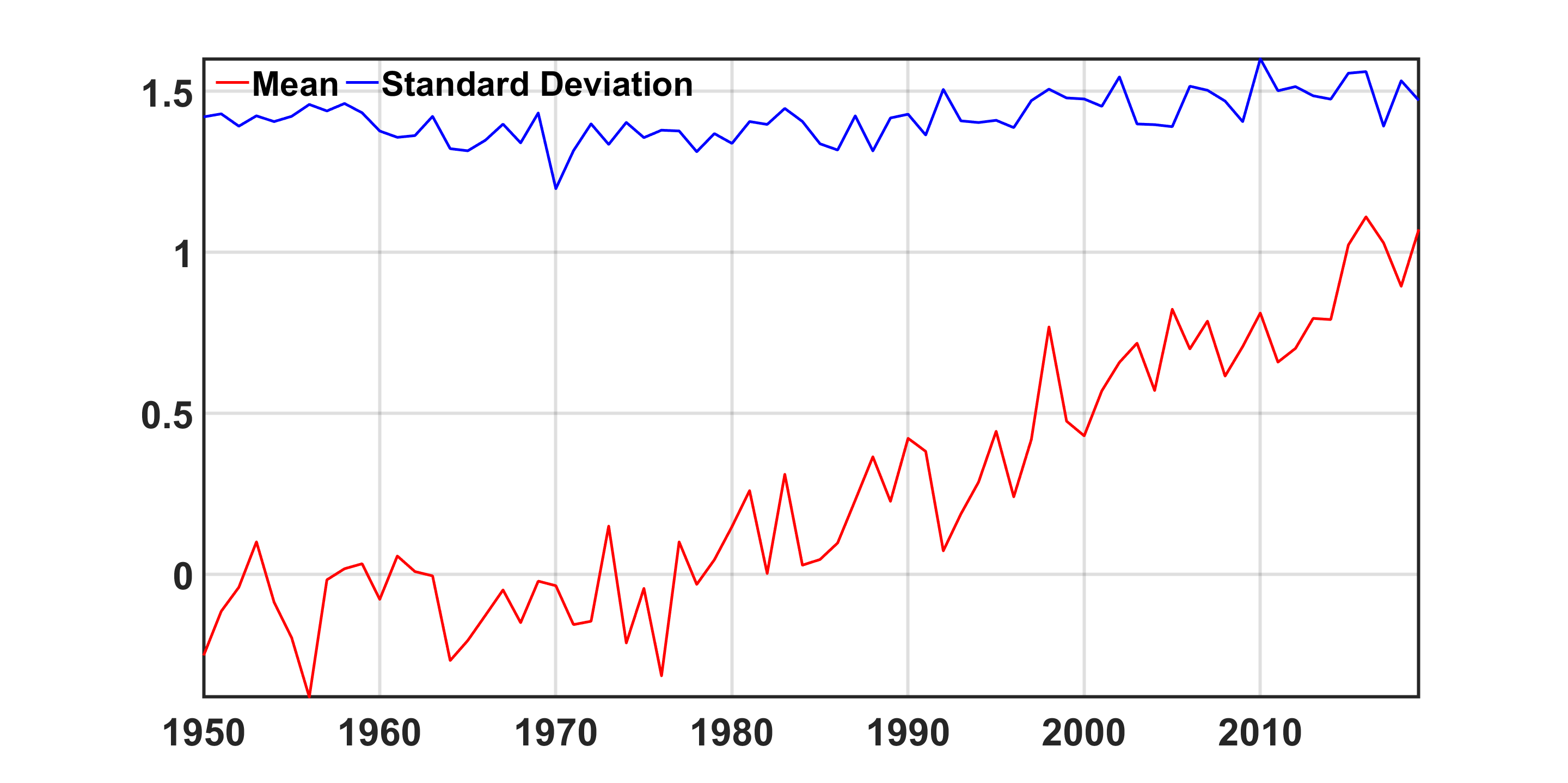}
            \includegraphics[height=0.2\textwidth, width=0.45\textwidth]{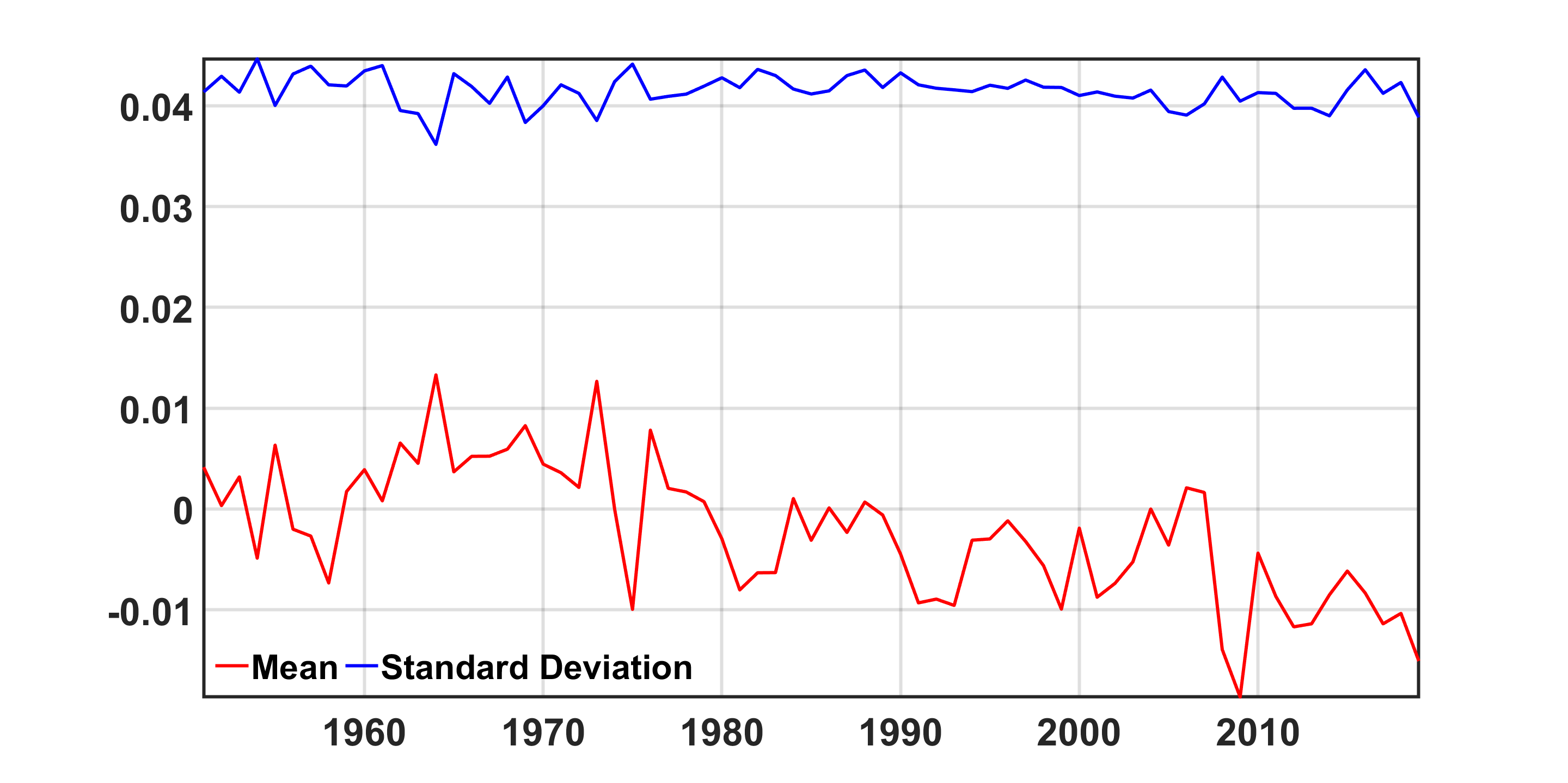}
		\caption{Probability density functions of land temperature anomalies (top left) and temperature-related regional growth rates (top right) with the corresponding sample mean and standard deviation processes from 1951 to 2019 (bottom).}
		\label{Fig:GTemp_dist}
	\end{center}
\end{figure}
\indent The literature notes that treating probability densities with support $[a_1,a_2]$ (without transformation) as standard Hilbert-valued elements is inadvisable (e.g., \citealp{petersen2016}), since densities do not form a linear space; this issue is particularly pronounced for nonstationary density-valued time series (\citealp{seo2019cointegrated}). To implement our framework, we apply the centered log-ratio (CLR) transformation of each density $g$, given by $g\mapsto \log  g(u)-(a_2-a_1)^{-1} \int_{a_1}^{a_2} \log g(s)ds$ which, under regularity conditions, maps (in a bijective manner) the density $g$ into the subspace of  $L^2[a_1,a_2]$ orthogonal to constant functions (\citealp{egozcue2006hilbert}), enabling direct application of our methods.\footnote{Since the CLR transformation involves $\log g(s)$, problems arise when $g(s)=0$. As is common in practice (e.g., \citealp{seoshang22}), this is avoided by adding a small constant to $g(s)$. In our study, the densities are constructed on a restricted domain excluding a few extreme values, so this issue does not occur.} 
The resulting CLR-transformed densities of GRP growth rates (resp.\ land temperature anomalies), generated from the raw data, are interpreted as measures of their true counterparts with measurement errors, and they are treated as functional data $y_t$ (resp. $\tilde{x}_t$) in \eqref{eqreg1deter}.\footnote{We assume that both time series include intercepts but no deterministic time trends, as in \cite{chang2020evaluating}.}
\subsubsection{Nonstationarity and testing procedure for $d_N$}\label{Sec_empirics2}
We examine the nonstationarity of the CLR-transformed time series computed in the previous section and estimate the nonstationarity dimension $d_N$ of $\{x_t\}_{t\geq 1}$, an input to our inferential methods. We apply the variance-ratio testing procedure of \cite{NSS}, which is shown to be robust to measurement errors in Section \ref{Sec_VRtest} of the Supplementary Material. The procedure determines $d_N$ by testing $H_0: d_N = d_0$ against $H_1: d_N < d_0$ sequentially for $d_0 = d_{\max}, \ldots, 1$ until the null is not rejected for the first time, where $d_{\max}$ is a prespecified upper bound for $d_N$. The results with $d_{\max}=5$, presented in Table \ref{tab:Nonstat_test0}, identify $d_N = 2$  
at the standard 10\% or 5\% significance level. This testing procedure can also be useful for a different purpose in our empirical analysis. Our model requires that the stochastic trends of $y_t$ are explained by those of $x_t$ ($f^N$ captures this). A straightforward extension of this testing procedure can be used as a diagnostic check to see whether this is the case in practice, and the testing results are supportive of our empirical analysis (see Remark \ref{remvr2} of the Supplementary Material for detailed discussion and results).
\begin{table}[t]
	\centering
	\caption{Testing results on $d_N$ of the CLR transformed densities of land temperature anomalies}
	\begin{tabular*}{1\linewidth}{@{\extracolsep{\fill}}ccrcccc}
		\hline
		\multicolumn{2}{c}{$d_0$} & {$5$} & {$4$} & {$3$} & {$2$} & {$1$} \\
		\hline    
		\multicolumn{2}{c}{Test Statistics}  & 7247.24 & 3216.9 & 1214.36 & 177.39 & 11.73 \\
        \multicolumn{2}{c}{$p$-values (\%)}  & $<$0.1 &  $<$0.1 & 0.3 & 26.1 & 87.3 \\
		\hline    
	\end{tabular*}\\ %
	\label{tab:Nonstat_test0}%
	\begin{minipage}{1\linewidth}
				\footnotesize{Notes: 
				$H_0:d_N=d_0$ is tested sequentially against $H_1:d_N<d_0$ for $d_0 = 5,\ldots,1$ using the procedure in Section \ref{Sec_VRtest}. $p$-values are computed from the quantiles of 100,000 Monte Carlo draws from the asymptotic null distribution.}
	\end{minipage}
\end{table}%

\subsubsection{Estimation results: economic impact of climate change}\label{Sec_empirics3}
We present estimation results on the economic impact of climate change. Our main interest is in the slope parameter $f$ in the model with an intercept \eqref{eqreg1deter}. The estimation uses the CLR-transformed time series, with the threshold $\alpha$ set as in our simulation experiments (Section \ref{Sec_monte_carlo2} of the Supplementary Material), yielding $\KK=4$ in this study. We use $d_N=2$, as estimated in Section \ref{Sec_empirics2}. Since $f^N$ can be viewed as capturing a stable long-run relationship between persistent stochastic trends, it is called the long-run response function, while $f^S$ is called the short-run response function, as in the literature (e.g., \citealp{meierrieks2023temperature}); accordingly, $f$ is referred to as the total-run response function.

We compute the estimators for $\kappa = 1$ and $\kappa = 0$ for comparison. The estimator $\hat{f}_{\kappa}$ is an operator that maps a function to another. Although it can be visualized (since $f$ is Hilbert–Schmidt in the present setup, the estimated Hilbert–Schmidt kernel can be plotted in three dimensions), such a plot is unlikely to yield meaningful insights for practitioners concerned with the economic implications of climate-related scenarios or major events. Instead, we consider a functional change $\zeta$ in $x_t$, interpreted as a global warming shock to the world economy, and estimate its partial effect $f(\zeta)$ to quantify the economic damages resulting from the shock. The hypothetical global warming shock $\zeta$ can produce permanent effects, transitory effects, or both on regional economic growth. Permanent and transitory impacts are measured based on long-run (climate change) and short-run (interannual weather) variations of the functional changes, respectively. The estimated total-run response function thus illustrates how global warming collectively impacts the spatial distribution of regional growth rates. Note that while measurement errors do not affect the consistency of the long-run response function estimator $\hat{f}_{\kappa}^N$ for either $\kappa=0$ or $\kappa=1$, they do affect the consistency of the short-run response function estimator $\hat{f}_{\kappa}^S$ (see Remark \ref{remadd3}). Thus, setting a positive $\kappa$ is necessary for robust statistical inference on the total-run response function.

\begin{figure}[t]
	\begin{center}
		\includegraphics[height=0.25\textwidth, width=0.45\textwidth]{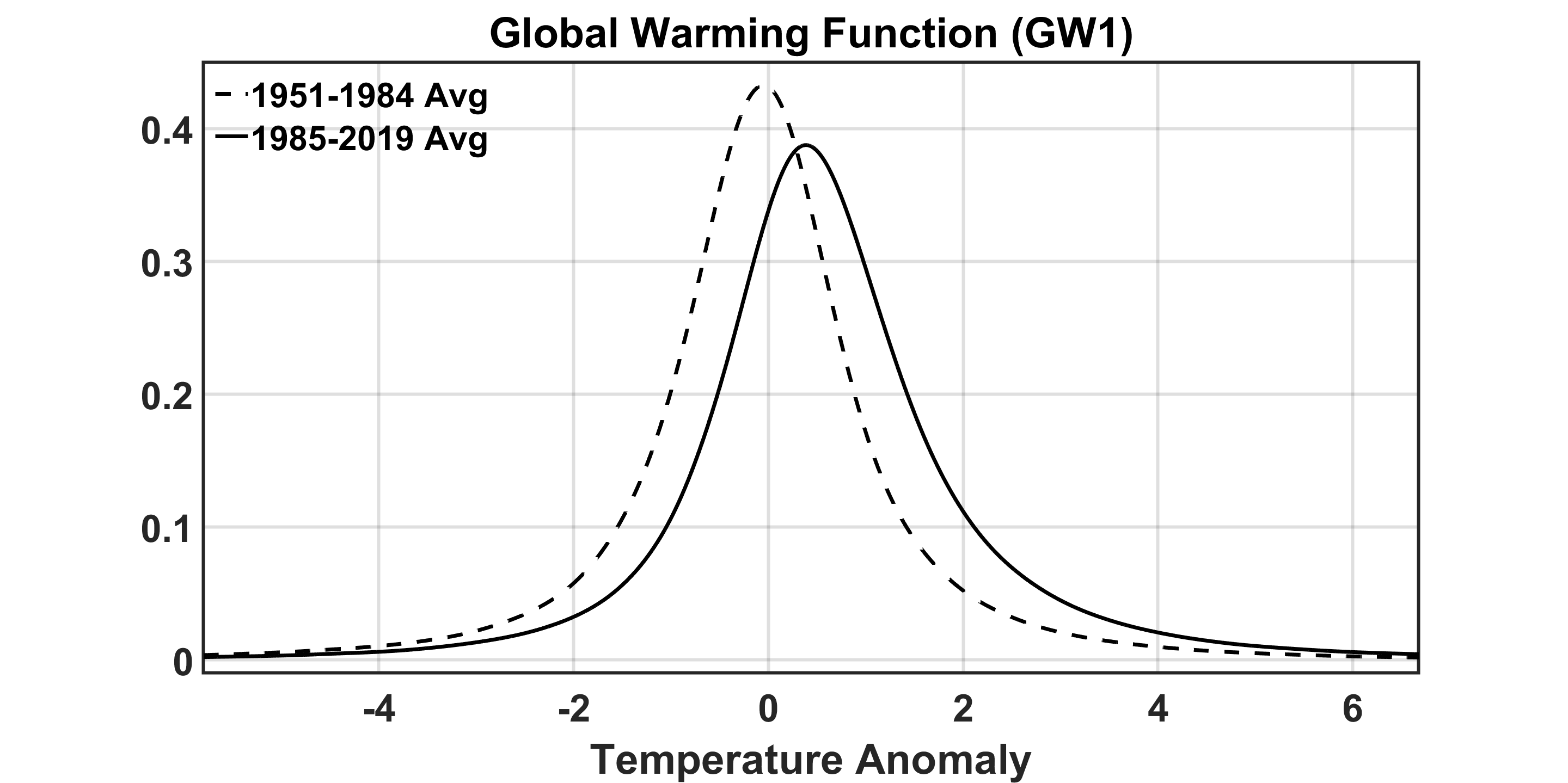}
           \includegraphics[height=0.25\textwidth, width=0.45\textwidth]{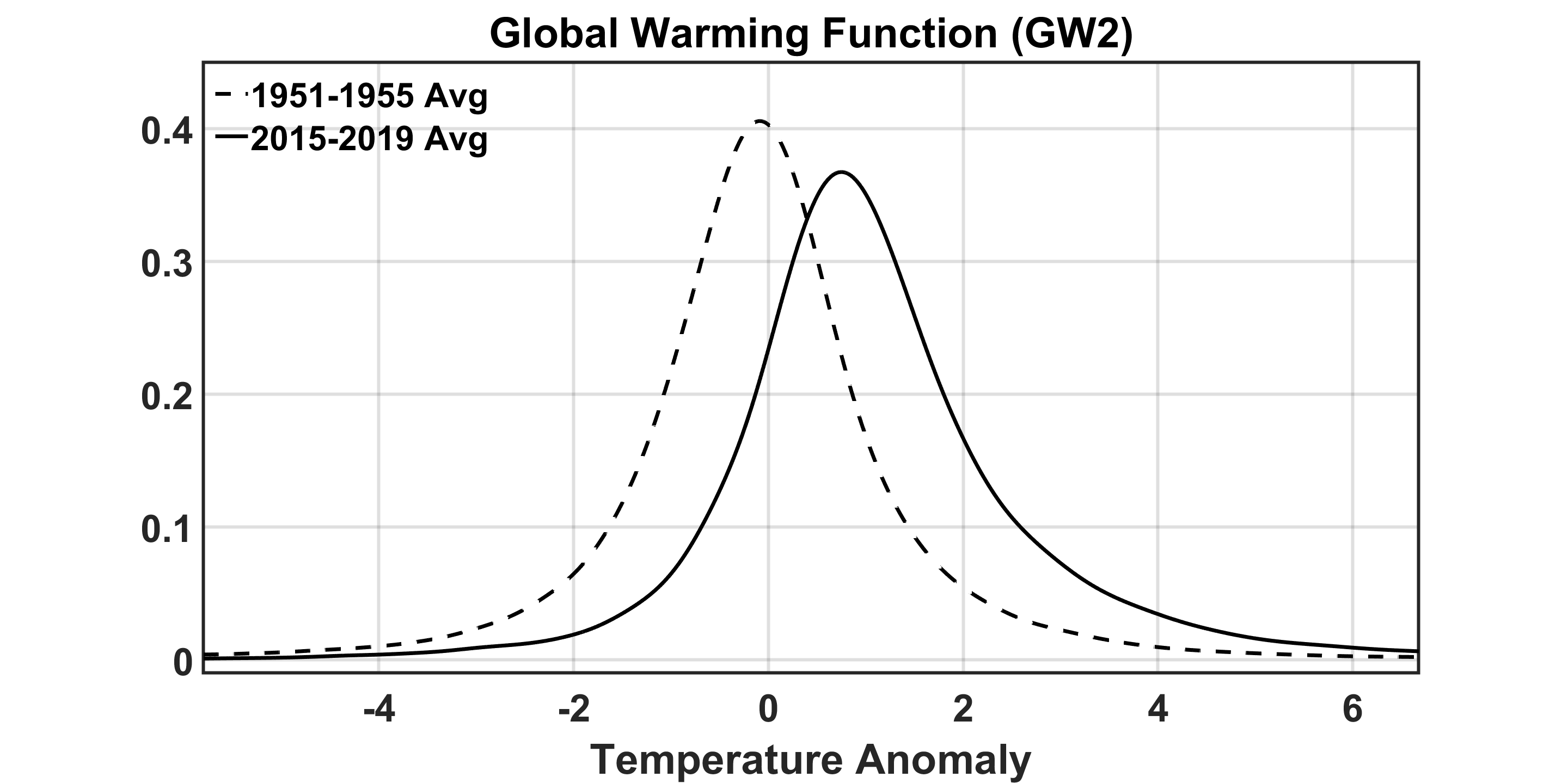}
		\caption{Averaged probability density functions: first half vs. second half (left) and first 5 years vs. last 5 years (right) of the sample period.}
		\label{Fig:GW_Shocks}
	\end{center}
\end{figure}

To construct a representative global warming function, at first, we compute the mean difference between the first and second halves of the density estimates for the land temperature anomaly (hereafter, GW1). As shown in the left panel of Figure \ref{Fig:GW_Shocks}, global warming can be conceptualized in statistical terms as a probabilistic shift from negative to positive anomalies, capturing the long-run distributional change in the Earth’s land surface temperature over the past 70 years. Previous studies estimate the break date for the northern hemisphere temperature anomaly at 1985 in the NASA dataset and 1984 in the HadCRUT3 dataset (\citealp{estrada2013statistically}; \citealp{Estrada2019}). Given the close similarity in statistical properties between the land temperature anomaly and the northern hemisphere series (\citealp{chang2020evaluating}), we adopt 1985 as a credible break date marking the onset of global warming in GW1. Of course, practitioners consider an alternative conceptualization of global warming. For example, one can define global warming as the distributional shift over the first and last five years of the sample period (hereafter, GW2); see  the right panel of Figure \ref{Fig:GW_Shocks}. In this setting, GW2 serves as a complementary measure, offering greater robustness to interannual variability and to uncertainties in the precise timing of the structural break. While Figure \ref{Fig:GW_Shocks} shows the probability densities, we use the model with CLR-transformed densities due to mathematical issues noted in the literature (e.g., \citealp{petersen2016,seo2019cointegrated}). Since the CLR map is bijective, we consider the CLR transformations of the densities in each panel of Figure \ref{Fig:GW_Shocks} and define $\zeta$ as the difference between the CLR-transformed densities. This is treated as a global warming shock in the model. 


\begin{figure}[t]
	\begin{center}
		\includegraphics[height=0.32\textwidth, width=0.496\textwidth]{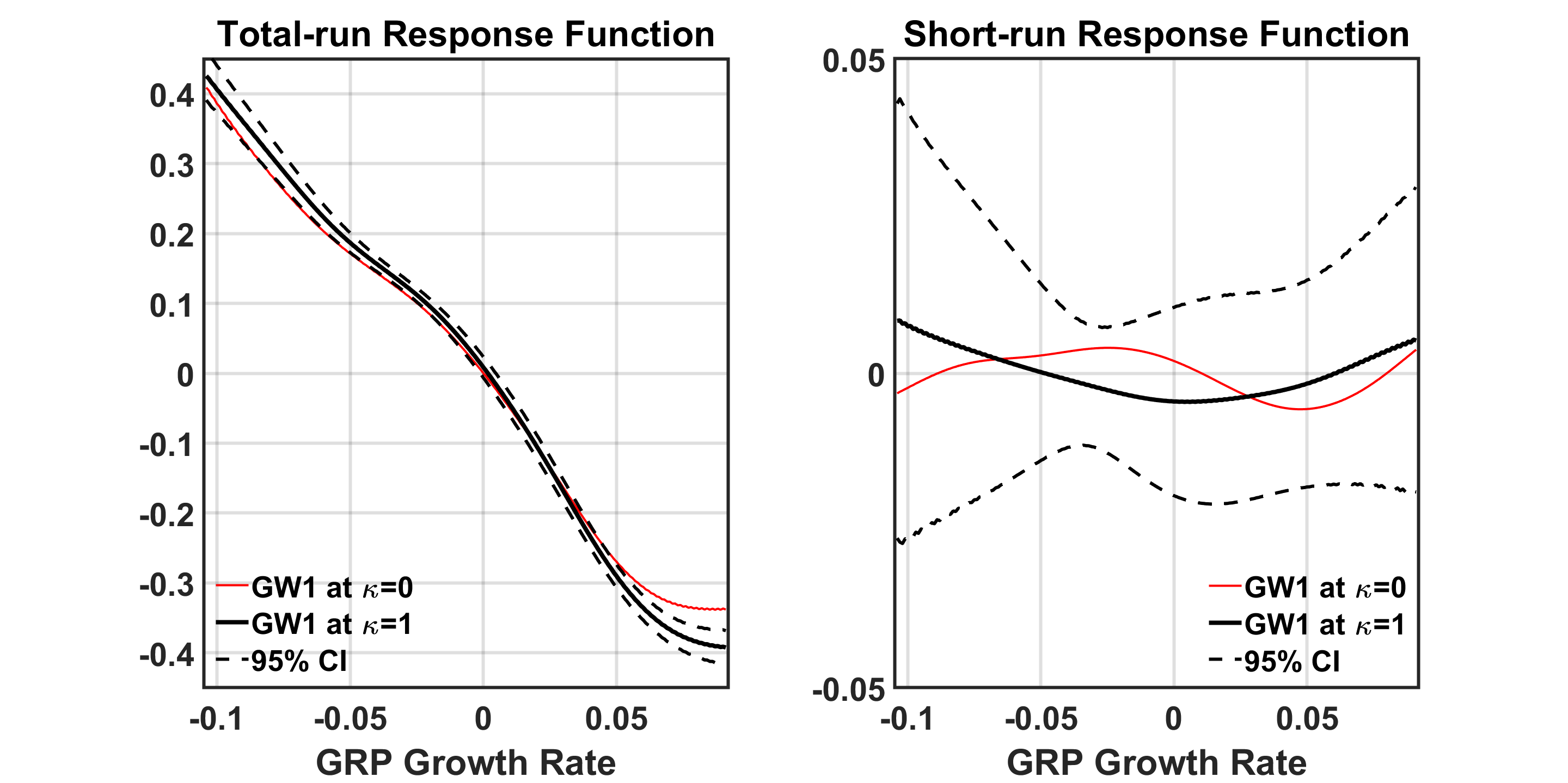}
            \includegraphics[height=0.32\textwidth, width=0.496\textwidth]{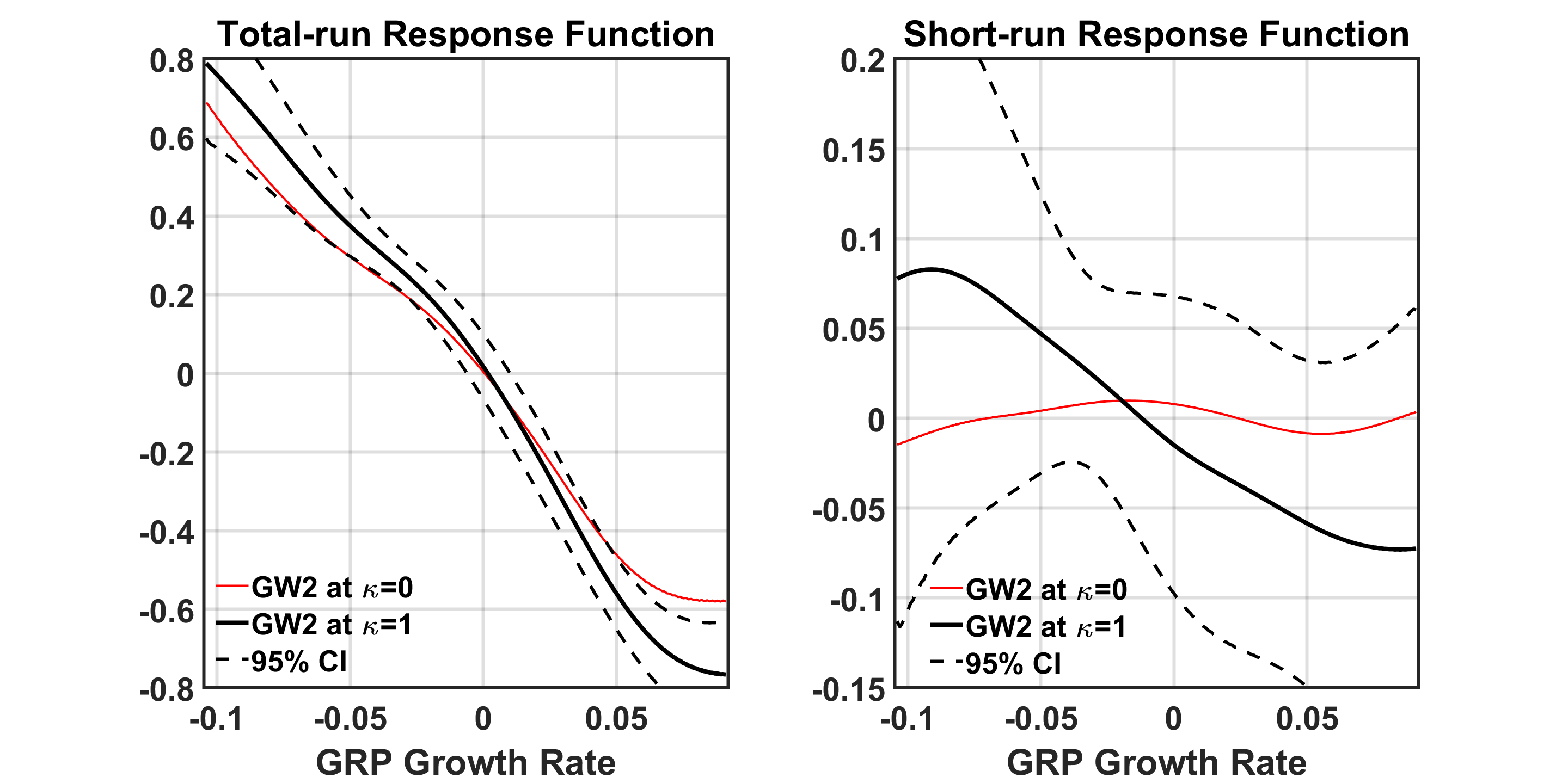}
		\caption{Total-run response function for GW1 (first) and GW2 (third), and short-run response function for GW1 (second) and GW2 (fourth). Dashed lines indicate 95\% confidence bands for the locally averaged response function at $\kappa = 1$.}
		\label{Fig:LR_Resp}
	\end{center}
\end{figure}

\indent The estimate $\hat{f}_{\kappa}(\zeta)$ captures the effect of the generated global warming shock on the CLR-transformed density of regional growth rates. Figure~\ref{Fig:LR_Resp} presents the estimated total-run $(\hat{f}_{\kappa} (\zeta))$ and short-run $(\hat{f}_{\kappa}^S (\zeta))$ responses to the considered global warming shock. Since the regressor is likely contaminated by measurement errors, statistical inference is conducted for the estimates with $\kappa = 1$, using the theoretical results in Theorem~\ref{prop1} (and Corollary~\ref{cor1} in the Supplementary Material). Specifically, the local confidence interval is obtained by estimating the pointwise standard error from the residual covariance within a one-grid bandwidth neighborhood and scaling it by the normal critical value at each point (see Section~\ref{sec_inference} and~\eqref{eqappconf}). The 95\% confidence intervals for the locally averaged response functions indicate that, while the short-run effects are statistically insignificant, global warming has a significant total-run impact on regional economic growth (potentially due to the limited sample size and the slower convergence rate of $\hat{f}_{\kappa}^S$ compared to $\hat{f}_{\kappa}^N$). 

\indent The downward slope of $\hat{f}_{\kappa}(\zeta)$ indicates that global warming reduces the share of regions with high temperature-related economic growth while increasing the share with lower growth. In other words, as land temperatures rise, the distribution of regional growth shifts toward weaker outcomes. The slope is generally steeper under GW2 than under GW1, indicating that the magnitude of the climate-induced shift in regional growth outcomes is more pronounced when global warming is defined by the first-versus-last 5-year contrast. When measurement errors in the functional data are accounted for (with $\kappa = 1$), the slope of $\hat{f}_{\kappa}(\zeta)$ becomes steeper at the right tail compared to the error-free case (with $\kappa = 0$). This discrepancy likely reflects bias from measurement errors, implying that the magnitude of climate-related economic impacts is underestimated when such errors are ignored.

\indent 
From a practical perspective, it is more informative to visualize the distributional effect implied by $f(\zeta)$ in terms of changes in the probability density of GRP growth rates (noting that $f(\zeta)$ represents an effect on the CLR-transformed density). This is achieved by (i) fixing a reference density and its CLR transformation $y_{\text{ref}}$, and (ii) inverting the CLR-valued quantity $y_{\text{ref}} + \hat{f}_{\kappa}(\zeta)$ back into the corresponding probability density, then comparing it with the reference density. For the inversion, the inverse CLR transformation $g(s) \mapsto \exp(g(s))/\int_{a_1}^{a_2} \exp(g(u))du$ is applied \citep{egozcue2006hilbert}. Moreover, scaled global warming shocks $\zeta_{q} = q \zeta$ for $q \geq 0$ and their distributional effects are considered to examine how the reference density changes as the global warming shock intensifies or diminishes.

\begin{figure}[t]
	\begin{center}
	\hspace{-1.8em}	\includegraphics[height=0.34\textwidth, width=0.36\textwidth]{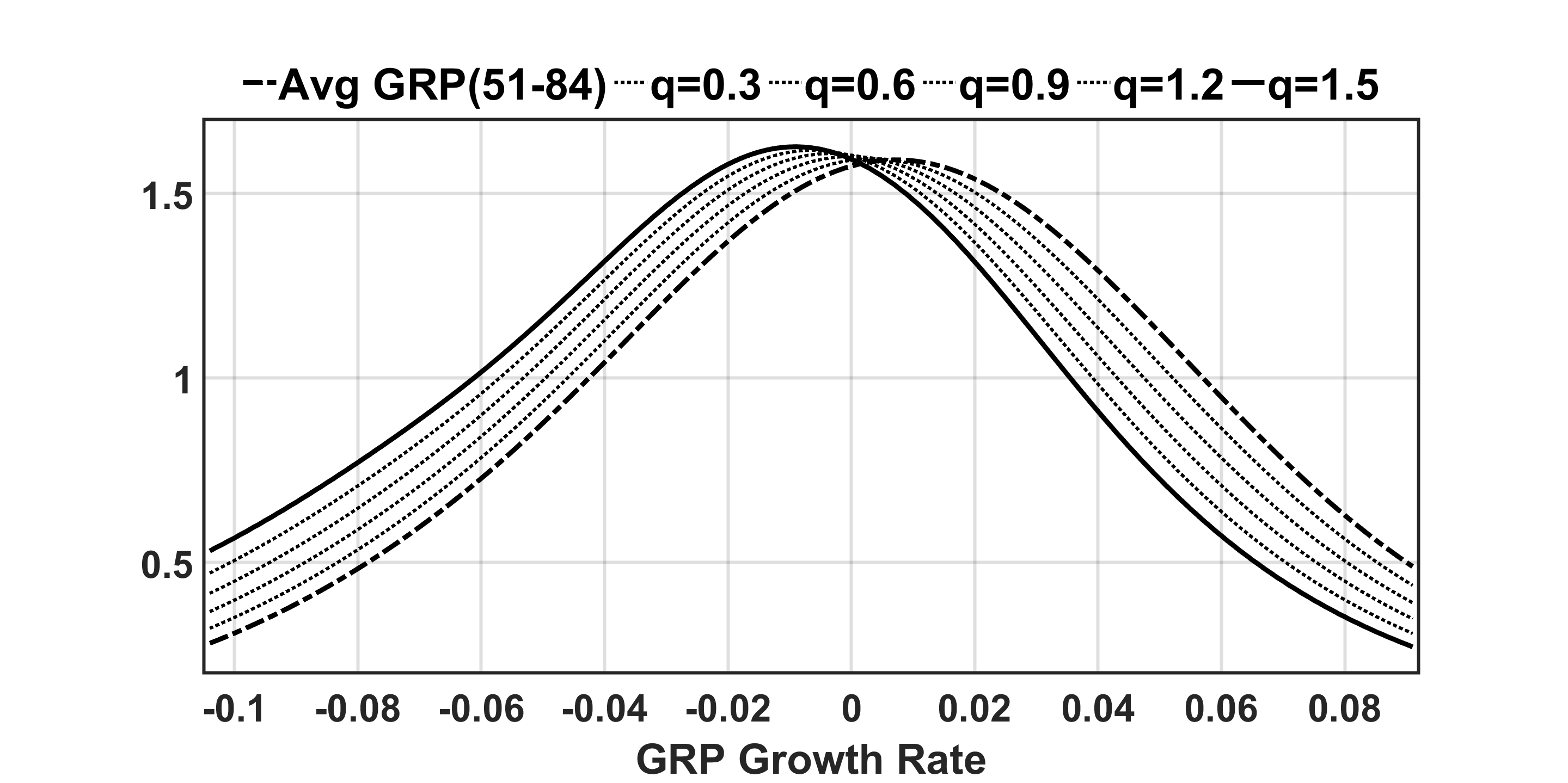} \hspace{-1.8em}
         \includegraphics[height=0.34\textwidth, width=0.36\textwidth]{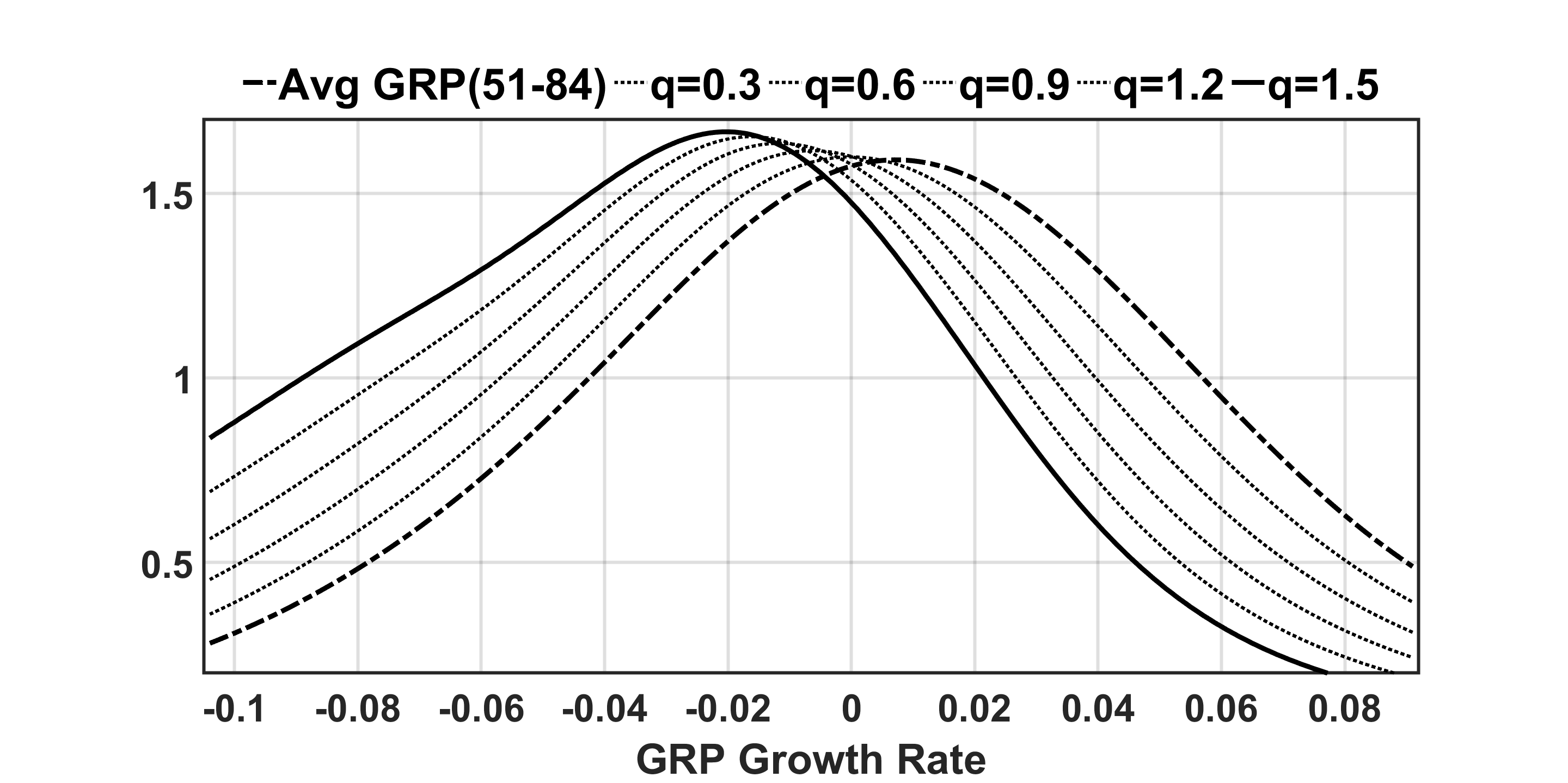} 
\hspace{-1.8em}
		\includegraphics[height=0.34\textwidth, width=0.37\textwidth]{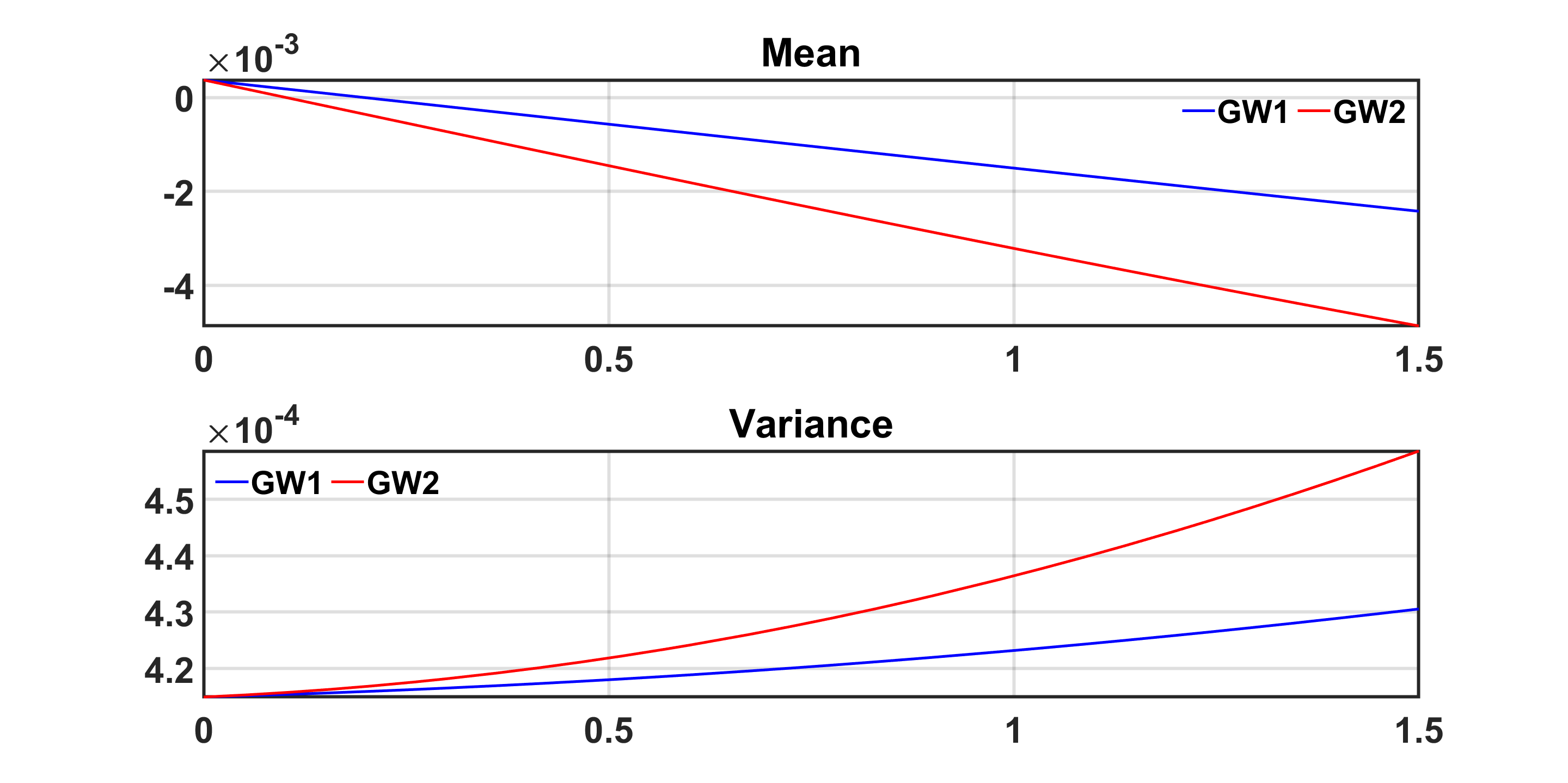}
		\caption{Shifts in the probability density of regional growth rate under $q$-scaled GW1 (left) and GW2 (middle) shocks; mean and variance over time (right).}
		\label{Fig:GRP_Resp2}
	\end{center}
\end{figure}

\indent The left and middle panels of Figure~\ref{Fig:GRP_Resp2} show the result when the reference density is set to the average density of $y_t$ over the period 1951–1984, $q$ increases from $0$ to $1.5$, and $\zeta$ is constructed from GW1 or GW2. As $q$ increases, both shocks shift the mass of the distribution leftward and modestly widen it, reflecting lower average growth rates and greater dispersion across regions. The right panel summarizes these changes in terms of the first two moments. The mean declines approximately linearly with $q$, while the variance increases at an accelerating rate. Across all scales, GW2 produces more pronounced changes than GW1 in both the mean and variance, indicating a stronger impact on the central tendency and dispersion of regional growth rates. Taken together, these results suggest that stronger global-warming shocks are associated with slower average growth and increased dispersion, demonstrating the usefulness of our approach as a practical tool for policymakers to evaluate the adverse economic impacts of climate change. 

\section{Concluding Remarks}\label{Sec_conclude}
This paper develops regression models for nonstationary and potentially error-contaminated functional time series and introduces a novel autocovariance-based inferential method. The methodology is broadly applicable to problems involving nonstationary functional data. Not only to illustrate our approach, but also for its intrinsic importance, we apply our methodology to assess the economic impact of climate change. Our analysis provides empirical evidence that global warming has a negative effect on regional economic growth.

\appendix
\section*{Appendix}
\normalsize{This supplementary material contains mathematical preliminaries (Section \ref{Sec_mathprelim}), simulation results (Section \ref{Sec_monte_carlo}), theoretical results that complement those in the main article (Section \ref{ap_sec_sup}), proofs (Section \ref{app: sec: d}), and details on the generated probability densities used in Section \ref{Sec_empirics} of the main article (Section \ref{AP_WGRP}).}

\onehalfspacing 

\renewcommand{\theequation}{\thesection.\arabic{equation}}	
\appendix
\renewcommand{\thesection}{\Alph{section}}
\renewcommand{\thetheorems}{\thesection.\arabic{theorems}}
\renewcommand{\thecorollarys}{\thesection.\arabic{corollarys}}
\renewcommand{\thelemmas}{\thesection.\arabic{lemmas}}
\renewcommand{\thepropositions}{\thesection.\arabic{propositions}}
\renewcommand{\theassumptions}{\thesection.\arabic{assumptions}}

\setcounter{propositions}{0}
\setcounter{corollarys}{0}
\setcounter{theorems}{0}
\setcounter{assumptions}{0}
\setcounter{lemmas}{0} 

\section{Mathematical preliminaries}\label{Sec_mathprelim}
\subsection{Bounded linear operators on Hilbert spaces}\label{Sec_prelim2}
For any Hilbert spaces $\mathcal H_1$ (equipped with inner product $\langle \cdot,\cdot \rangle_1$ and norm $\|\cdot\|_1$) and $\mathcal H_2$ (equipped with inner product $\langle \cdot,\cdot \rangle_2$ and norm $\|\cdot\|_2$), let $\mathcal L_{\mathcal H_1,\mathcal H_2}$ denote the normed space of continuous linear operators from $\mathcal H_1$ to $\mathcal H_2$, equipped with the uniform operator norm $\|A\|_{\opnorm}= \sup_{\|x\|_1\leq 1} \|A(x)\|_2$ for $A \in \mathcal L_{\mathcal H_1,\mathcal H_2}$. Let $\otimes$ denote the operation of tensor product associated with $\mathcal H_1$, $\mathcal H_2$, or both, i.e., for any $\zeta_k \in \mathcal H_k$ and $\zeta_\ell \in \mathcal H_\ell$, 
\begin{equation} \label{eqtensor}
	\zeta_k\otimes \zeta_\ell (\cdot) = \langle 	\zeta_k,\cdot \rangle_k \zeta_\ell, 
\end{equation}
which is a map from $\mathcal H_k$ to $\mathcal H_\ell$ for $k \in \{1,2\}$ and $\ell \in \{1,2\}$.
For any $A \in \mathcal L_{\mathcal H_1,\mathcal H_2}$, the range and kernel are denoted by $\ran A$ and $\ker A$ respectively; that is, $\ran A = \{A\zeta : \zeta \in \mathcal H_1\}$ and $\ker A = \{\zeta \in \mathcal H_1 : A\zeta = 0\}$. The adjoint $A^\ast$ of $A$ is the unique element of $\mathcal L_{\mathcal H_1,\mathcal H_2}$ satisfying that $\langle A \zeta_1 ,\zeta_2 \rangle_2=\langle \zeta_1 ,  A^\ast \zeta_2 \rangle_1$ for all $\zeta_1 \in \mathcal H_1$ and $\zeta_2 \in \mathcal H_2$. 

If there is no risk of confusion, we let $\mathcal L_{\mathcal H_1}$ denote $\mathcal L_{\mathcal H_1,\mathcal H_1}$. 
If $A=A^\ast$, $A$ is said to be self-adjoint. We say $A \in \mathcal L_{\mathcal H_1}$ is nonnegative (resp.\ positive) if $\langle A \zeta,\zeta \rangle_1 \geq 0$ (resp.\ $\langle A \zeta,\zeta \rangle_1 > 0$) for all $\zeta\in \mathcal H_1$.   An element $A \in \mathcal L_{\mathcal H_1}$ is called compact if $A = \sum_{j=1}^\infty a_j\zeta_{1j} \otimes \zeta_{2j} $ for some orthonormal bases $\{\zeta_{1j}\}_{j \geq 1}$ and $\{\zeta_{2j}\}_{j \geq 1}$ and some sequence of real numbers $\{a_j\}_{j \geq 1}$ tending to zero. If $A$ is compact and its Hilbert-Schmidt norm, defined by $\|A\|_{\HS} = (\sum_{j=1} ^\infty \Vert A \zeta_j \Vert_1 ^2)^{1/2}$ for any orthonormal basis $\{\zeta_j\}_{j\geq 1}$, is finite, then it is called a Hilbert-Schmidt operator.

\subsection{Random elements of Hilbert spaces}\label{Sec_prelim1}
Let $(\mathbb S,\mathbb F, \mathbb P)$ be the probability space, and let $\mathcal H_1$ and $\mathcal H_2$ be the Hilbert spaces considered in Section \ref{Sec_prelim2}; each of $\mathcal H_1$ and $\mathcal H_2$ is assumed to be equipped with the usual Borel $\sigma$-field. We call $X$ an $\mathcal H_1$-valued random variable if it is a measurable map from $\mathbb S$ to $\mathcal H_1$. $X$ is square-integrable if $\mathbb{E} [\|X\|_1^2] < \infty$. For such a random element $X$, the unique element $\mathbb{E}[X] \in \mathcal H_1$ satisfying $\mathbb{E}[\langle X,\zeta\rangle_1] = \langle \mathbb{E}[X],\zeta\rangle_1$ for every $\zeta \in \mathcal H_1$ is called the expectation of $X$, and the operator defined by $C_X= \mathbb{E}[(X-\mathbb{E}[X]) \otimes (X-\mathbb{E}[X])]$ is called the covariance operator of $X$. 
Let $Y$ be another square-integrable $\mathcal H_2$-valued random variable. If $\mathbb{E}[\|X\|_1\|Y\|_2] < \infty$, the cross-covariance operator $C_{XY}  =   \mathbb{E}[(X-\mathbb{E}[X]) \otimes (Y-\mathbb{E}[Y])]$ is well defined.

\section{Simulation study}\label{Sec_monte_carlo}
\subsection{Simulation data generating process}  \label{Sec_monte_carlo1}   
We investigate the finite sample performance of the proposed estimator in simulation experiments. We assume that, if there were no measurement errors whatsoever, the observations would be given by $x_t= \mu_x + x_t^0$ and $y_t = \mu_y + y_t^0$, and the relationship $y_t^0 = f(x_t^0) + u_t$ holds, where $y_t^0$, $x_t^0$, and $u_t$ are generated processes whose unconditional means are zero. This leads to the model with an intercept $y_t= \mu +  f(x_t) + u_t$ with  $\mu = \mu_y - f(\mu_x)$  (see \eqref{eqreg1deter}). Noting that $x_t^0$ can be written as \begin{equation} \label{eqsim1}
	x_t^0  = \sum_{j=1}^{\infty} \langle x_t^0, v_j\rangle v_j
\end{equation} for an orthonormal basis $\{v_j\}_{j=1}^\infty$ (to be specified later) of $\mathcal H$ and assuming that $\mathcal H^N = \spn\{v_1,\ldots,v_{d_N}\}$, we simulate realizations of $x_t^0$ by generating $\langle x_t^0, v_j\rangle$ as a real-valued nonstationary (resp.\ stationary) process for each $j \leq d_N$ (resp.\ $j\geq d_N+1$). 
Specifically, we generate $\langle x_t^0,v_j \rangle$ using the following AR(1) law of motion: for some $\beta_j \in (-1,1)$ and $\sigma_{\varepsilon,j}>0$,
\begin{eqnarray}
	&\Delta \langle x_t^0,v_j \rangle = \beta_j^N 	\Delta \langle x_{t-1}^0,v_j \rangle + \sigma_{\varepsilon,j}\varepsilon_{j,t}, \quad &j=1,\ldots,d_N,  \label{simdgp1}\\
	&\langle x_t^0,v_j \rangle  =  \beta_{j}^S  \langle x_{t-1}^0,v_j \rangle + \sigma_{\varepsilon,j}\varepsilon_{j,t}, \quad &j\geq d_N+1, \label{simdgp2}
\end{eqnarray}
where $\varepsilon_{j,t}$ is iid $N(0,1)$ across $j$ and $t$, and also independent of any other variables. As will be detailed, $\sigma_{\varepsilon,j}$ is set to decay to zero as $j$ increases, and thus the time series $\langle x_t^0,v_j \rangle$ in \eqref{simdgp2} becomes more important in determining the properties of the stationary components of $x_t^0$ when $j$ is smaller. We let $\beta_j^N$ be randomly determined in each simulation run, specifically as $\beta_j^N = s_j U_j^N$, where $U_j^N$ is a uniform random variable supported on $[-0.5, 0.5]$ (i.e., $U[-0.5, 0.5]$), and $s_j$ is a Rademacher random variable independent of $U_j^N$; both sequences are independent across $j$. Moreover, given that (i) $\beta_j^S$ governs the correlation between $\langle \PP^S x_t^0, v_j \rangle$ and $\langle \PP^S x_{t-\kappa}^0, v_j \rangle$ and (ii) stationary time series tend to exhibit positive autocorrelation in many applications, we let $\beta_j^S$ be drawn independently from $U[0.5, 0.9]$ for $j \leq M$, and from $U[-0.9, 0.9]$ for $j \geq M+1$, for some $M > 0$ to be specified, in each repetition of the simulation experiment; combined with the decay of $\sigma_{\varepsilon,j}$, this ensures that the dominant part of the stationary components generally exhibits positive autocorrelation.
The parameter $\sigma_{\varepsilon,j}$ determines the scale of $\langle x_t^0, v_j \rangle$, which must decay to zero sufficiently fast for $C_{\kappa}^S$ to be a compact operator and hence well defined. We consider two simulation designs for this sequence, motivated by the setups in \cite{seong2021functional}.
In the first design, referred to as the exponential design, we set $\sigma_{\varepsilon,j} = 1$ for $j \leq d_N + m$ and  $\sigma_{\varepsilon,j} = (0.8)^{j- d_N-m}$ for $j =d_N+m+1, \ldots, d_N+M$, where $m$ (resp.\ $M$) is a moderately (resp.\ sufficiently) large integer. Given the required decay rate of the eigenvalues of $C_{\kappa}^S$ for our theoretical development, it is natural to consider the case where $\sigma_{\varepsilon,j}$ decreases geometrically for $j \geq M$; accordingly, we set $\sigma_{\varepsilon,j} = \sigma_{\varepsilon,M} (j - M)^{-2}$ for $j \geq M + 1$. We use $m=7$ and $M = 20$ throughout the simulation experiments.
In the second design, referred to as the sparse design, we set $\sigma_{\varepsilon,j} = 1$ for $j \leq d_N + m$, and $\sigma_{\varepsilon,j} = (0.1)^{j - d_N - m}$ for $j = d_N + m+1, \ldots, d_N + M$. As in the exponential design, we set $\sigma_{\varepsilon,j} = \sigma_{\varepsilon,M}  (j - M)^{-2}$ for $j \geq M + 1$.
To generate $x_t^0$ as a function using \eqref{eqsim1} under these two simulation designs, we use the Fourier basis functions, excluding the constant function, as in our empirical study in Section \ref{Sec_empirics}. We then construct  $x_t= \mu_x + x_t^0$ by setting $\mu_x = \sum_{j=1}^{10} c_{x,j}v_j$, where $c_{x,j}$ is iid $N(0,1)$ across $j$  in each repetition of the simulation experiment. 

Similarly, we write $y_t^0  =\sum_{j=1}^{\infty} \langle y_t^0, w_j \rangle w_j$ for an orthonormal basis $\{w_j\}_{j=1}^\infty$ of $\mathcal{H}_y$. We simulate $y_t$ by generating the coefficients $\langle y_t^0, w_j \rangle$ and using the Fourier basis functions, excluding the constant function, for $\{w_j\}_{j=1}^\infty$. 
Throughout this simulation study, we assume that the linear map $f$ is defined by the following property: $f(v_j) = \gamma_j w_j$  for $\gamma_j \neq 0$  for each $j$, with $\gamma_j$ tending to zero as $j$ increases. We set $\gamma_j = a_j U_j^\gamma$, where $U_j^\gamma$ is generated independently from $U(-1,1)$, $a_j = 1$ for $j \leq d_N+m$ and $a_j = (0.8)^{j-d_N-m}$ for $j \geq d_N+m+1$; the decay of $a_j$ is introduced to ensure the summability condition in Assumption \ref{assum3}. Since $y_t^0=f(x_t^0) + u_t$ holds, it may be deduced that,  we in this case have $\langle y_t^0,w_j \rangle = \gamma_j \langle x_t^0,v_j \rangle + \langle u_t,w_j \rangle$ for each $j$. We generate $y_t^0$ using this relationship and $u_t$ generated as follows:   \begin{align*}
u_t &= \sum_{j=1}^\infty \langle u_t,w_j \rangle w_j, \quad \langle u_t,w_j \rangle \sim_{iid} N(0,\sigma_{u,j}^2), 
\end{align*}
where $\sigma_{u,j}$ is generated by the same mechanism as that of $\sigma_{\varepsilon,j}$.   We then construct $y_t^0$ similarly as $x_t^0$, and then generate $y_t=\mu_y + y_t^0$ by setting $\mu_y = \sum_{j=1}^{10} c_{y,j}w_j$, where $c_{y,j}$ is iid $N(0,1)$ across $j$ in each simulation run. 


Based on the above simulation DGP, constructed without any measurement errors, we consider the presence of measurement errors in $x_t$ and $y_t$, and hence assume that we can only observe  $\tilde{x}_t = x_t  + e_t$ and $\tilde{y}_t = y_t  + e_{y,t}$ with additive measurement errors $e_{t}$ and $e_{y,t}$. In this case, from the equation $y_t=\mu+f(x_t)+u_t$, we find that the following holds in the presence of measurement errors:
\begin{equation}\label{simdgp01}
\tilde{y}_t = \mu + f(\tilde x_t) + \tilde{u}_t, \quad \tilde{u}_t = u_t - f(e_t) + e_{y,t}.
\end{equation}
The coefficient $f$ in the above equation is estimated using the generated observations $\{\tilde{x}_t\}_{t=1}^T$ and $\{\tilde{y}_t\}_{t=1}^T$ in each simulation run. 
We set  $e_t = \sigma_{e} \eta_t$, where $\eta_t = \sum_{j=1}^{d_N+1} c_{\eta,j} v_j$ and  $c_{\eta,j}$ is iid $N(0,1)$ across $j$. The scalar $\sigma_e$ serves as a scale factor that controls the magnitude of the measurement error, and we let it depend on (the magnitude of) $\mathcal E_t^x$. Specifically, we choose $\sigma_e$ so that the nuclear norm of the covariance operator of $e_t$ matches $0\%$ (i.e., no measurement errors in $x_t$), $50\%$  and $100\%$ of that of $\mathcal E_t^x = (\Delta \PP^N x_t, \PP^S x_t)$, which can be generated from the simulation DGP. In each simulation run, the nuclear norm of the covariance operator of $\mathcal{E}_t^x$ (i.e., the sum of its eigenvalues) is approximated by the average of the corresponding sample estimates, computed from the simulated sequence of $\mathcal{E}_t^x$ based on \eqref{simdgp1} and \eqref{simdgp2}; we use 800 repetitions to calculate the average.
We similarly set $e_{y,t} = \sigma_{e,y} \xi_{t}$, where $\xi_t = \sum_{j=1}^{d_N+1} c_{\xi,j} w_j$ and  $c_{\xi,j}$ is iid $N(0,1)$ across $j$. However, since $e_t$ and $u_t$ are independent in the simulation setup, a larger measurement error in $x_t$ unintentionally tends to be accompanied by a larger magnitude (variance) of $\tilde{u}_t = u_t-f(e_t)+e_{y,t}$, which in turn may affect the finite sample performance of the considered estimator and statistical inference. We thus choose $\sigma_{e,y}$ so that the nuclear norm of the covariance operator of $e_{y,t}$ matches $(100-\alpha)\%$ of that of $\mathcal E_t^x = (\Delta \PP^N x_t, \PP^S x_t)$ when $\sigma_e$ matches $\alpha\%$ of it for $e_t$. This approach allows us to control significant changes in the magnitude of the error $\tilde{u}_t$ in \eqref{simdgp01}. Of course, unlike $e_{t}$, $e_{y,t}$ alone does not cause problematic issues, such as inconsistency, in the standard covariance-based estimator $\hat{f}_0$. Thus, the case with $\sigma_{e}=0$ corresponds to the error-free scenario, which we consider in Assumption \ref{assum1c}.

\subsection{Simulation results}  \label{Sec_monte_carlo2}   
We examine the finite sample performance of our proposed estimators using the simulation DGP introduced in Section \ref{Sec_monte_carlo1}. As in our empirical application, we consider the case where \( d_N = 2 \), and compute our estimators with $\kappa=0$ and $\kappa =1$ to compare them in a few different scenarios regarding the scales of measurement errors. The tuning parameter $\KK$ follows a pre-specified choice rule for the entire simulation experiments; specifically, given that \( \KK_S = \KK - d_N  > 0\) is required (note that this is a minimal requirement for nonzero $\hat{f}_{\kappa}^S$ to be defined), we set \( \KK \) as $\KK = d_N + \max_j\{ \tilde{\lambda}_{j} > 0.4\, T^{-0.2}\}$, where \( \tilde{\lambda}_j \) is a scale-adjusted eigenvalue defined by
$\tilde{\lambda}_j = {\llambda\edex{j}{\widehat{D}_{\kappa}^S}}/{\sum_{j=1}^\infty \llambda\edex{j}{\widehat{D}_{\kappa}^S}}.$\footnote{Noting that any \( \KK \) satisfying Assumptions \ref{assum2} and \ref{assum3b} necessarily depends on the scale of the functional observation \( x_t \), the choice of \( \KK \) based on scale-adjusted eigenvalues was previously considered by \cite{seong2021functional} as a scale-invariant selection in functional regression with stationary regressors.} As a measure of the inaccuracy of the estimator $\hat{f}_{\kappa}$ for $f$, we compute the Hilbert–Schmidt norm of $\hat{f}_{\kappa} - f$, which can be calculated as $\sqrt{\sum_{j=1}^\infty \|\hat{f}_{\kappa}(v_j) - f(v_j)\|^2}$ for any arbitrary orthonormal basis $\{v_j\}_{j=1}^\infty$ of $\mathcal{H}$. The simulation results are reported in Table \ref{tabsim1}. As may easily be expected from our theoretical results, the proposed estimator $\widehat{f}_0$ performs better than $\hat{f}_1$ when there are no measurement errors. However, in the presence of measurement errors, $\hat{f}_0$ performs worse than $\hat{f}_1$, and the performance gap increases with the scale of measurement errors. The performance of $\hat{f}_1$ appears to be robust in the considered simulation setup, regardless of the scale of measurement errors. Overall, the simulation results support our theoretical findings in Section \ref{Sec_econometrics2}.

In addition to evaluating the overall accuracy of the estimators, we also examine the coverage probability of the 95\% confidence interval for $\langle f(\zeta), \varphi \rangle$, based on the pointwise asymptotic normality result \eqref{localasymp01}, which follows from Theorems \ref{thm1} and \ref{prop1} (and their extension discussed in Section \ref{sec_det}). We let $\zeta = \sum_{j=1}^\infty c_{\zeta,j}v_j$, where $c_{\zeta,j} = 1$ for $j = 1, \ldots, 9$, and $c_{\zeta,j} = (j - 8)^{-2}$ thereafter, and let $\varphi$ be $w_1$. The results are reported in Table \ref{tabsim1coverage}. As expected, $\hat{f}_0$ yields poor coverage probabilities in the presence of measurement errors in $x_t$, while $\hat{f}_1$ provides improved results, and its performance is robust across different scales of measurement errors in $x_t$. 

We also examined the finite sample performance under a different set of parameters and obtained qualitatively similar results. As an example, we report the simulation results for the case where $d_N = 3$ in Tables \ref{tabsim2} and \ref{tabsim1coverage2}. 

\begin{table}[H]
	\renewcommand{\arraystretch}{0.85}
	\caption{Finite sample performance when $d_N=2$, the average Hilbert Schmidt norm of $\hat{f}_{\kappa}-f$}
	\label{tabsim1} 
	\vskip -8pt
	\small
	\begin{tabular*}{\linewidth}{@{\extracolsep{\fill}}lrrrrrrrrrr}
		\toprule
		& &	\multicolumn{4}{c}{Exponential design} && \multicolumn{4}{c}{Sparse design}\\\midrule
		&\multicolumn{10}{c}{Scale of measurement errors: $0\%$}\\\midrule
		Estimators & &$T=100$& $200$ & $400$ & $800$&  &$T=100$& $200$ & $400$  & $800$ \\
		\midrule
		$\kappa=0$ & &0.957& 0.918& 0.879& 0.830&& 0.941& 0.904& 0.863& 0.807
		\\
		$\kappa=1$ & &  1.019& 0.966& 0.932& 0.901 &&1.021& 0.971 &0.937& 0.910
		\\ 
		\midrule
		&\multicolumn{10}{c}{Scale of measurement errors: $50\%$}\\\midrule
		Estimators & &$T=100$& $200$ & $400$ & $800$&  &$T=100$& $200$ & $400$ & $800$ \\
		\midrule
		$\kappa=0$ & & 1.072& 1.051& 1.029& 1.005&& 1.031& 1.001& 0.972& 0.940
		\\
		$\kappa=1$ & &  1.010& 0.957& 0.926& 0.897&& 1.010& 0.964& 0.937& 0.907
		\\ 
		\midrule
		&\multicolumn{10}{c}{Scale of measurement errors: $100\%$}\\\midrule
		Estimators & &$T=100$& $200$ & $400$ & $800$&  &$T=100$& $200$ & $400$& $800$  \\
		\midrule
		$\kappa=0$ & &   1.184& 1.179& 1.171& 1.154&& 1.135& 1.120& 1.100& 1.076
		\\
		$\kappa=1$ & &  1.001& 0.948& 0.920& 0.894&& 0.990& 0.956& 0.933& 0.905
		\\ 
		\midrule
	\end{tabular*} 
	\vskip 4pt
	{\footnotesize Notes: The average Hilbert Schmidt norm of $\hat{f}-f$ is computed from 3000 Monte Carlo replications. The scale of measurement errors is given by $\alpha \%$ when $\sigma_e$ is chosen so that the nuclear norm of the covariance operator of $e_t$ matches $\alpha \%$ of that of $\mathcal E_t^x = (\Delta \PP^N x_t, \PP^S x_t)$.}
\end{table}

\begin{table}[H]
	\renewcommand{\arraystretch}{0.85}
	\caption{The coverage probability of the 95\% confidence interval for $\langle f(\zeta),\varphi \rangle$ when $d_N=2$.}
	\label{tabsim1coverage} 
	\vskip -8pt
	\small
	\begin{tabular*}{\linewidth}{@{\extracolsep{\fill}}lrrrrrrrrrr}
		\toprule
		& &	\multicolumn{4}{c}{Exponential design} && \multicolumn{4}{c}{Sparse design}\\\midrule
		&\multicolumn{10}{c}{Scale of measurement errors: $0\%$}\\\midrule
		Estimators & &$T=100$& $200$ & $400$ & $800$&  &$T=100$& $200$ & $400$  & $800$ \\
		\midrule
		$\kappa=0$ & &0.768& 0.854& 0.917& 0.934&& 0.798& 0.868& 0.918& 0.940
		\\
		$\kappa=1$ & &0.774& 0.854 &0.917& 0.930&& 0.805& 0.867& 0.922& 0.935
		\\ 
		\midrule
		&\multicolumn{10}{c}{Scale of measurement errors: $50\%$}\\\midrule
		Estimators & &$T=100$& $200$ & $400$ & $800$&  &$T=100$& $200$ & $400$ & $800$ \\
		\midrule
		$\kappa=0$ & & 0.615 &0.689& 0.763& 0.838&& 0.677& 0.765 &0.831 &0.883		
		\\
		$\kappa=1$ & & 0.738& 0.843 &0.914& 0.928&& 0.779& 0.864 &0.927 &0.933
		\\ 
		\midrule
		&\multicolumn{10}{c}{Scale of measurement errors: $100\%$}\\\midrule
		Estimators & &$T=100$& $200$ & $400$ & $800$&  &$T=100$& $200$ & $400$& $800$  \\
		\midrule
		$\kappa=0$ & & 0.158& 0.222& 0.307& 0.404 &&0.238& 0.335& 0.440& 0.576
		\\
		$\kappa=1$ & & 0.582& 0.781& 0.892& 0.928 &&0.668& 0.818& 0.892 &0.937
		\\ 
		\midrule
	\end{tabular*} 
	\vskip 4pt
	{\footnotesize Notes: The coverage probability is computed from 3000 Monte Carlo replications. The scale of measurement errors is determined as described in Table \ref{tabsim1}. }
\end{table}

\begin{table}[H]
	\renewcommand{\arraystretch}{0.85}
	\caption{Finite sample performance when $d_N=3$, the average Hilbert Schmidt norm of $\hat{f}_{\kappa}-f$}
	\label{tabsim2} 
	\vskip -8pt
	\small
	\begin{tabular*}{\linewidth}{@{\extracolsep{\fill}}lrrrrrrrrrr}
		\toprule
		& &	\multicolumn{4}{c}{Exponential design} && \multicolumn{4}{c}{Sparse design}\\\midrule
		&\multicolumn{10}{c}{Scale of measurement errors: $0\%$}\\\midrule
		Estimators & &$T=100$& $200$ & $400$ & $800$&  &$T=100$& $200$ & $400$  & $800$ \\
		\midrule
		$\kappa=0$ & & 0.936& 0.896& 0.863& 0.819&& 0.916 &0.878& 0.848& 0.797
		\\
		$\kappa=1$ & & 1.021& 0.951& 0.919& 0.894&& 1.026& 0.954 &0.929& 0.902
		\\ 
		\midrule
		&\multicolumn{10}{c}{Scale of measurement errors: $50\%$}\\\midrule
		Estimators & &$T=100$& $200$ & $400$ & $800$&  &$T=100$& $200$ & $400$ & $800$ \\
		\midrule
		$\kappa=0$ & &  1.035& 1.009& 0.987& 0.957&& 0.995& 0.969& 0.938& 0.903		
		\\
		$\kappa=1$ & & 1.005& 0.943& 0.915& 0.891&& 1.007& 0.946& 0.925& 0.900
		\\ 
		\midrule
		&\multicolumn{10}{c}{Scale of measurement errors: $100\%$}\\\midrule
		Estimators & &$T=100$& $200$ & $400$ & $800$&  &$T=100$& $200$ & $400$& $800$  \\
		\midrule
		$\kappa=0$ & & 1.143&1.135& 1.122& 1.100&& 1.096& 1.080 &1.058 &1.030
		\\
		$\kappa=1$ & &  0.989 &0.931 &0.909& 0.887&& 0.981&0.936& 0.920& 0.899
		\\ 
		\midrule
	\end{tabular*} 
	\vskip 4pt
	{\footnotesize Notes: The average Hilbert Schmidt norm of $\hat{f}_{\kappa}-f$ is computed from 3000 Monte Carlo replication. The scale of measurement errors is determined as described in Table \ref{tabsim1}. }
\end{table}

\begin{table}[H]
	\renewcommand{\arraystretch}{0.85}
	\caption{The coverage probability of the 95\% confidence interval for $\langle f(\zeta),\varphi \rangle$ when $d_N=3$.}
	\label{tabsim1coverage2} 
	\vskip -8pt
	\small
	\begin{tabular*}{\linewidth}{@{\extracolsep{\fill}}lrrrrrrrrrr}
		\toprule
		& &	\multicolumn{4}{c}{Exponential design} && \multicolumn{4}{c}{Sparse design}\\\midrule
		&\multicolumn{10}{c}{Scale of measurement errors: $0\%$}\\\midrule
		Estimators & &$T=100$& $200$ & $400$ & $800$&  &$T=100$& $200$ & $400$  & $800$ \\
		\midrule
		$\kappa=0$ & & 0.733& 0.808& 0.884& 0.926&& 0.767& 0.838& 0.906& 0.924
		\\
		$\kappa=1$ & &  0.735& 0.811 &0.887& 0.928&& 0.772& 0.840& 0.907 &0.931
		\\ 
		\midrule
		&\multicolumn{10}{c}{Scale of measurement errors: $50\%$}\\\midrule
		Estimators & &$T=100$& $200$ & $400$ & $800$&  &$T=100$& $200$ & $400$ & $800$ \\
		\midrule
		$\kappa=0$ & & 0.577 &0.657& 0.740& 0.806&& 0.641 &0.716& 0.794& 0.866		
		\\
		$\kappa=1$ & & 0.700& 0.805 &0.878& 0.918&& 0.751& 0.830& 0.896& 0.928
		\\ 
		\midrule
		&\multicolumn{10}{c}{Scale of measurement errors: $100\%$}\\\midrule
		Estimators & &$T=100$& $200$ & $400$ & $800$&  &$T=100$& $200$ & $400$& $800$  \\
		\midrule
		$\kappa=0$ & & 0.162&  0.224&  0.305&  0.435&& 0.227& 0.320& 0.429& 0.572
		\\
		$\kappa=1$ & &  0.517&  0.712&  0.842&  0.914&& 0.632& 0.762& 0.872& 0.928
		\\ 
		\midrule
	\end{tabular*} 
	\vskip 4pt
	{\footnotesize Notes:  The coverage probability is computed from 3000 Monte Carlo replications. The scale of measurement errors is determined as described in Table \ref{tabsim1}.}
\end{table}

\newpage 
\section{Supplementary theoretical results}\label{ap_sec_sup}
We provide some theoretical results, which complement to the main results developed in Section \ref{Sec_econometrics2}. The proofs of the results presented in this section will be given in Section \ref{ap_sec_sup_proof}.
\subsection{Supplement to the pointwise asymptotic normality results} \label{ap_sec_supp}
\subsubsection{Sample counterparts of $\theta_{\KK_S}$ and $\widetilde{C}_{u}$  for feasible inference}\label{ap_sec_supp1}
Note that $\theta_{\KK_S}$ and $\widetilde{C}_u$, given in Theorems \ref{thm2} and \ref{prop1}, are unknown, and thus the asymptotic results stated therein cannot be directly used for inference in practice. However, these unknown quantities can be replaced by reasonable estimators, which makes the results more useful in practice. To obtain the desired results, we introduce some additional notation. Let $$\widehat{\theta}_{\KK_S}(\zeta) = \langle \zeta,  (\widehat{D}_\kappa^S)_{\KK_S}^{-1}   (\widehat C_{\kappa}^S)^\ast \widehat{C}_{0}^S \widehat C_{\kappa}^S(\widehat D_\kappa^S)_{\KK_S}^{-1} (\zeta) \rangle \quad \text{and} \quad \widehat{C}_u = T^{-1}\sum_{t=1}^T \hat{u}_t \otimes \hat{u}_t,$$ where $\hat{u}_t = y_t - \hat{f}_{\kappa}(\tilde{x}_{t})$ is the residual from the model, $\widehat{C}_\kappa^S = \widehat{C}_\kappa \widehat{\PP}^S_{\kappa}$, $\widehat{C}_0^S = T^{-1}\sum_{t=1}^T  \widehat{\PP}^S_{\kappa} \tilde x_{t} \otimes \widehat{\PP}^S_{\kappa} \tilde x_t$,  $\widehat{\PP}^S_{\kappa}$ is defined in \eqref{eqprojections}, and  $(\widehat D_\kappa^S)_{\KK_S}^{-1}$ is defined as
\begin{equation} \label{eqapppf01}
	(\widehat D_\kappa^S)_{\KK_S}^{-1} = \sum_{j=d_N+1}^{\KK} \llambda^{-1}\edex{j}{\widehat D_\kappa}  \PPi\edex{j}{\widehat D_\kappa} = 
    \sum_{j=1}^{\KK_S} \llambda^{-1}\edex{j}{\widehat D_\kappa^S}   \PPi\edex{j}{\widehat D_\kappa^S}.  
\end{equation} 
Note that $\widehat{\theta}_{\KK_S}(\zeta)$ and $\widehat{C}_u$ can be computed from the given data and the residuals obtained using the proposed estimator $\hat{f}_{\kappa}$.  We provide the desired results below:
\begin{corollarys} \label{cor1} Let the assumptions in Theorem \ref{thm2} be satisfied. Then the following hold:
	\begin{enumerate}[(i)]
		\item \label{cor1a}\eqref{eqthm2} and \eqref{eqthm2a} hold if ${\theta}_{\KK_S}(\zeta)$ is replaced with $\widehat{\theta}_{\KK_S}(\zeta)$.
		\item \label{cor1b}
		$\widehat{C}_u\to_p \widetilde{C}_u$.
	\end{enumerate}
\end{corollarys}


In Corollary \ref{cor1}, we use the assumptions employed for Theorem \ref{thm2}, including the assumption of distinct eigenvalues, \eqref{eqcondition_a}. However, as discussed in Remark \ref{remadd}, this condition can be replaced by the two conditions given in Remark \ref{remadd}, allowing for the repetition of an eigenvalue; see the proof of Corollary \ref{cor1} given in Section \ref{ap_sec_sup_proof}.

\subsubsection{Pointwise asymptotic normality in the model with an intercept}\label{ap_sec_supp2}

In this section, we consider the model and estimator briefly discussed in Section \ref{sec_det} and extend the  inferential methods developed in Section \ref{sec_inference} to this case. We first introduce a set of assumptions adapted from those in the previous sections. To this end, let
$D_{c,\kappa}^S=(C^S_{c,\kappa})^\ast C^S_{c,\kappa}$ and $E_{c,\kappa}^S= C^S_{c,\kappa}(C^S_{c,\kappa})^\ast$, where $C^S_{c,\kappa}=\mathbb{E}[(\PP^S x_{t-\kappa} - \mu_{x,S}) \otimes (\PP^Sx_{t} - \mu_{x,S})]$ and $\mu_{x,S} = \PP^S\mu_x=\mathbb{E}[\PP^S x_{t}] = \mathbb{E}[u_t^S]$, which is nonzero in general, while, in the considered model, $\mathbb{E}[u_t^N]=\mathbb{E}[\PP^N\Delta x_t] = 0$. 
We also define $
\theta_{c,\KK_S}(\zeta) = \langle \zeta,  (D_\kappa^S)_{c,\KK_S}^{-1}   (C_{c,\kappa}^S)^\ast \widetilde{C}_{c,0}^S C_{c,\kappa}^S(D_{c,\kappa}^S)_{c,\KK_S}^{-1} (\zeta) \rangle$, where $\widetilde{C}_{c,0}^S = \mathbb{E}[(\PP^{S} \tilde x_{t} - \mu_{x,S}) \otimes (\PP^{S} \tilde x_{t}- \mu_{x,S})]$), and  $\varpi_{c,t}(j,\ell)=\langle \PP^Sx_t-\mu_{x,S}, \varv\edex{j}{D_{c,\kappa}^S} \rangle\langle \PP^Sx_{t-\kappa}-\mu_{x,S},\varv\edex{\ell}{E_{c,\kappa}^S} \rangle-\mathbb{E}[\langle \PP^Sx_t-\mu_{x,S},\varv\edex{j}{D_{c,\kappa}^S} \rangle\langle \PP^Sx_{t-\kappa}-\mu_{x,S},\varv\edex{\ell}{E_{c,\kappa}^S}\rangle]$.

\begin{assumptions}\label{assumapp}	The following hold:\begin{enumerate}[(a)]
		\item\label{assumapp1} The model \eqref{eqreg1deter} in Section \ref{sec_det} holds with Assumptions \ref{assum1}, \ref{assum2}, and \ref{assum1c}, with $x_t$ and $u_t^S$ being replaced by $x_t-\mu_x$ and $u_t^S - \mu_{x,S}$, respectively.
		\item \label{assumapp2}  Assumption \ref{assum3} holds with $D_{\kappa}^S$ (resp.\ $\widehat{D}_{\kappa}^S$) replaced by $D_{c,\kappa}^S$ (resp.\ $\widehat{D}_{c,\kappa}^S$).  
		\item \label{assumapp3}  Assumption \ref{assum2add} holds with $D_{\kappa}^S$, $E_{\kappa}^S$ and $\varpi_t(j,\ell)$ replaced by $D_{c,\kappa}^S$, $E_{c,\kappa}^S$ and  $\varpi_{c,t}(j,\ell)$.  
	\end{enumerate}
\end{assumptions}
\noindent 
Consistency and the pointwise asymptotic normality of the considered estimator are established as follows:
\begin{corollarys}\label{thm2app}
	Suppose that  Assumptions \ref{assumapp}\ref{assumapp1}-\ref{assumapp2} hold, $u_t$ is a martingale difference with respect to $\mathfrak F_{t}$ given in \eqref{eqfiltration}, and the following holds:
	\begin{equation*} 
		\llambda\edex{1}{D_{c,\kappa}^S}>\llambda\edex{2}{D_{c,\kappa}^S}>\cdots > 0 \quad \text{and} \quad T^{-1/2}\alpha^{-1/2}\sum_{j=1}^{\KK_S}\ttau\edex{j}{D_{c,\kappa}^S} \to_p 0.
	\end{equation*}
	Then, $\hat{f}_{c,\kappa}$ is consistent  (i.e., $\hat{f}_{c,\kappa}\to_p f  $).  Moreover, if  Assumptions \ref{assumapp}\ref{assumapp3} additionally holds and $\theta_{c,\KK_S}(\zeta)\to_p \infty$, then 
	\begin{equation*} 
		\sqrt{T/\theta_{c,\KK_S}(\zeta)}(\hat{f}_{c,\kappa}(\zeta)-f(\zeta))  \to_d N(0,\widetilde{C}_{u}).
	\end{equation*}
\end{corollarys}

As discussed in Section \ref{ap_sec_supp1}, for feasible statistical inference,
${\theta}_{c,\KK_S}(\zeta)$ and $\widetilde{C}_{u}$ can be replaced by their sample counterparts, given by 
$$\widehat{\theta}_{c,\KK_S}(\zeta) = \langle \zeta,  (\widehat{D}_{c,\kappa}^S)_{\KK_S}^{-1}   (\widehat C_{c,\kappa}^S)^\ast \widehat{C}_{c,0}^S \widehat C_{c,\kappa}^S(\widehat D_{c,\kappa}^S)_{\KK_S}^{-1} (\zeta) \rangle \quad \text{and} \quad \widehat{C}_{c,u} = T^{-1}\sum_{t=1}^T \hat{u}_{c,t} \otimes \hat{u}_{c,t},$$
where $\hat{u}_{c,t} = y_{c,t} -\hat{f}_{c,\kappa}(\tilde{x}_{c,t})$, $\widehat{C}_{c,\kappa}^S = \widehat{C}_{c,\kappa} \widehat{\PP}^S_{c,\kappa}$,  $\widehat{C}_{c,0}^S = T^{-1}\sum_{t=1}^T  \widehat{\PP}^S_{c,\kappa} \tilde{x}_{c,t} \otimes \widehat{\PP}^S_{\kappa} \tilde{x}_{c,t},$ and $(\widehat{D}_{c,\kappa}^S)_{\KK_S}^{-1}$ is defined as $ (\widehat{D}_{\kappa}^S)_{\KK_S}^{-1}$ in \eqref{eqapppf01} but with $\widehat{D}_{\kappa}$ being replaced by $\widehat{D}_{c,\kappa}$ (or $\widehat{D}_{\kappa}^S$ being replaced by $\widehat{D}_{c,\kappa}^S$). The theoretical justification of this replacement is parallel to that in Section \ref{ap_sec_supp1}, and will therefore be omitted.

\subsection{Robustness of the variance-ratio testing procedure for $d_N$}\label{Sec_VRtest}
We keep the notation introduced in Section \ref{Sec_econometrics}. Consider testing the hypotheses
\begin{align}
	H_0: d_N = d_0 \quad \text{against} \quad H_1 : d_N \leq d_0 - 1, \label{eqhypo0}
\end{align}
for some $d_0>0$. Hereafter, let $$\widehat{K}_0 = T^{-1} \sum_{t=1}^T (\sum_{s=1}^t \tilde{x}_t \otimes \sum_{s=1}^t \tilde{x}_t).$$ We consider the variance-ratio (VR) test statistic, proposed by \cite{NSS} and further generalized by \cite{NSS2}, for examining \eqref{eqhypo0}. The test statistic is computed from the following eigenproblem:
\[
\gamma_j \widehat{\PP}_{\ell} \widehat{K}_0 \widehat{\PP}_{\ell} \phi_j = \widehat{\PP}_{\ell} \widehat{C}_0 \widehat{\PP}_{\ell} \phi_j,
\]
where $\{\gamma\}_{j \geq1}$ are the eigenvalues ordered from smallest to largest, $\{\phi_j\}_{j\geq 1}$ are the corresponding eigenvectors,  $\widehat{\PP}_{\ell} = \sum_{j=1}^\ell \PPi\edex{j}{\widehat{C}_0}$ and $\ell \geq d_0$. 
The VR test statistic for examining \eqref{eqhypo0} is then given by
\begin{equation}\label{NielsenVR_stat}
	\widehat{\mathcal T}_{d_0} = T^2 \sum_{j=1}^{d_0} \gamma_j.
\end{equation}
We will show that the presence of measurement errors $e_t$ does not affect consistency of the VR test of \cite{NSS}. We here only consider the case when there is no deterministic component and thus $\mathbb{E}[x_t] = 0$. Extension to the case with a nonzero intercept and/or a linear trend requires only a slight modification, as shown by \cite{NSS}. 
\begin{lemmas} \label{lemapp1}Suppose that Assumption \ref{assum1} holds. Then $T^{-1} \widehat{C}_0 = T^{-2}\sum_{t=1}^T x_t\otimes x_t + o_p(1)$ and $T^{-3} \widehat{K}_0 = T^{-4}\sum_{t=1}^T (\sum_{s=1}^t {x}_t \otimes \sum_{s=1}^t {x}_t) + o_p(1)$.
\end{lemmas}

The robustness of the VR testing procedure to the presence of measurement errors is established by the following proposition:
\begin{propositions} \label{propapp1}Let the assumptions in Lemma \ref{lemapp1} hold and $\widetilde{C}_0^S =\mathbb{E}[\PP^S \tilde{x}_t \otimes \PP^S \tilde{x}_t ]$ allows $\ell$ nonzero eigenvalues. 
	Then, $\widehat{\mathcal T}_{d_0}$ given in \eqref{NielsenVR_stat} satisfies the following:
	\begin{eqnarray*}
		&\widehat{\mathcal T}_{d_0} &\to_d \mathrm{tr} \left( \left(\int V_{d_0}V_{d_0}'\right)^{-1}\left(\int W_{d_0}W_{d_0}'\right)\right)   \quad \text{under $H_0$ of \eqref{eqhypo0}}, \\
		&\widehat{\mathcal T}_{d_0} &\to_p \infty \quad \text{under $H_1$ of \eqref{eqhypo0}}, 
	\end{eqnarray*}
	where $W_{d_0}$ is $d_0$-dimensional standard Brownian motion, $V_{d_0}(r)=\int_{0}^{r} W_{d_0}(s) ds$, and $\mathrm{tr}(A)$ denotes the trace of a square matrix $A$.   
\end{propositions}

The asymptotic null distribution of $\widehat{\mathcal T}_{d_0}$ depends only on $d_0$ and thus its quantiles can be tabulated with standard simulation methods. For some reasonable upper bound ${d}_{\max}$ of $d_N$, we may repeat the proposed test for $d_0={d}_{\max},{d}_{\max}-1,\ldots,1$, and let $\hat{d}_N$ be the value of $d_0$ when $H_0$ is not rejected for the first time (if $H_0$ is rejected for all $d_0={d}_{\max},{d}_{\max}-1,\ldots,1$, then $\hat{d}_N=0$). From Theorem 2 of \cite{NSS}, it is immediate to show the following: for any fixed significance level $\eta\in (0,1)$ used in the testing procedure, 
\begin{align}
	\mathbb{P}\{\hat{d}_N =d_N\} \to_p 1-\eta \quad \text{and}\quad \mathbb{P}\{\hat{d}_N > d_N\} \to_p 0.
\end{align}
Moreover, if $\eta$ is chosen such that $\eta\to 0$ as $T\to \infty$, $\mathbb{P}\{\hat{d}_N =d_N\} \to 1$. The proposed testing procedure extends the VR testing procedure proposed by \cite{NSS} by allowing for measurement errors and by adopting a slightly weaker assumption on $\widetilde{C}_0^S$, which in their paper is assumed to be positive definite on $\mathcal H_S$. The proofs are given in Section \ref{ap_sec_sup_proof}; however, as shown there, the results follow from moderate modifications of the proofs in \cite{NSS}.
\begin{remarks}\label{remvr}
  The VR test can be adapted to models with an intercept by constructing the test statistics from the centered variables $\tilde{x}_{c,t}$ defined in Section \ref{sec_det}. In this case, the limiting behavior described in Proposition \ref{propapp1} still holds, with $W_{d_0}$ interpreted as a $d_0$-dimensional centered Brownian motion, as detailed in \cite{NSS}.
\end{remarks}

As discussed, for the consistency of the VR testing procedure, we need a conjectured upper bound $d_{\max}$ of $d_N$ and also $\ell \geq d_N$ (see \citealp{NSS}, Section 3.5). In the empirical study in Section \ref{Sec_empirics}, where this testing procedure is applied, we set $d_{\max} = \ell = 5$. 

\begin{remarks} \label{remvr2}
In our proposed model, the stochastic trends of $y_t$ are explained by those of $x_t$ via $f^N$. As an extension, the VR testing procedure may be used as a diagnostic check to see if this holds in the given dataset. For example, if this is true, then we have $\hat{u}_t^N \coloneqq y_t - \hat{f}_{\kappa}^N(\tilde{x}_t) = u_t^N + O_p(T^{-1})$ (where ${u}_t^{N} = y_t - {f}^N_{\kappa}(\tilde{x}_t) = f^S (\tilde{x}_t) + u_t$) due to Theorem \ref{thm1}, and hence the nonstationarity dimension associated with $u_t^N$ must equal zero. By a straightforward modification of Lemma \ref{lemapp1} and Proposition \ref{propapp1}, it is immediately deduced that, if the stochastic trends of $y_t$ are explained by those of $x_t$, then the VR testing procedure applied to $\hat{u}_t^N$ needs to conclude that its nonstationarity dimension equals zero. If the model with an intercept discussed in Section \ref{sec_det} is considered as in our empirical analysis in Section \ref{Sec_empirics1}, this diagnostic check is applied to $\hat{u}_t^N \coloneqq {y}_{c,t} - \hat{f}_{\kappa}^N(\tilde{x}_{c,t})$ using the VR testing procedure, adjusted for an intercept, discussed in Remark \ref{remvr}. 
 We applied this diagnostic check with $d_{\max} = \ell = 3$. These values are smaller than those employed in Table \ref{tab:Nonstat_test0} but this is natural since the number of stochastic trends is expected to be smaller in this case; when the same VR testing procedure applied to $\tilde{x}_t$ is applied to $y_t$, we obtain $p$-values for $d_0 = 5, \ldots, 1$ of 0.3\%, 1.3\%, 6.3\%, 62.6\%, and 75.4\%, strongly supporting the presence of two (or possibly three) stochastic trends, with the number of stochastic trends in $u_t^N$ not exceeding this.
We obtained test statistics of  $3458.5,820.95$ and $101.09$ for $d_0=3,2,$ and $1$, respectively, and the corresponding $p$-values are less than $0.1\%$ for $d_0=3$ and $2$, and $4.7\%$ for $d_0=1$, which strongly supports that $u_t^N$ does not have a stochastic trend.  
\end{remarks}

\section{Proofs}\label{app: sec: d}
It will be convenient to introduce some notation in addition to that in Section \ref{sec_proposed_estimator}. 
Note first that, similar to $\widehat{D}_{\kappa}$, $\widehat{E}_{\kappa}$ (defined in Assumption \ref{assum2add})  allows the following spectral decomposition: 
\begin{equation*}
	 \widehat{E}_{\kappa} =  \sum_{j=1}^\infty \llambda\edex{j}{\widehat{E}_{\kappa}}  \PPi\edex{j}{\widehat{E}_{\kappa}}, \quad \llambda\edex{j}{\widehat{E}_{\kappa}}  = \llambda\edex{j}{\widehat{D}_{\kappa}}.
\end{equation*}
Combining this with the spectral representation of $\widehat{D}_{\kappa}$, we know  $\widehat{C}_{\kappa}$ allows the following representation:
\begin{equation*}
	 \widehat{C}_{\kappa} =  \sum_{j=1}^\infty \sqrt{\llambda\edex{j}{\widehat{D}_{\kappa}}} \varv\edex{j}{\widehat{D}_{\kappa}}\otimes \varv\edex{j}{\widehat{E}_{\kappa}};
\end{equation*}
see \citep[pp.\ 117-118]{Bosq2000}.
We define 
\begin{equation*}
	\widehat{\QQ}_{\kappa}^N = \sum_{j=1}^{d_N} \PPi\edex{j}{\widehat{E}_{\kappa}}, \quad \widehat{\QQ}_{\kappa}^S= I-	\widehat{\QQ}_{\kappa}^N.
\end{equation*}
Moreover, we let 
\begin{equation*}
	\widehat{\QQ}_{\kappa}^{\KK} = \sum_{j=1}^{\KK} \PPi\edex{j}{\widehat{E}_{\kappa}}, \quad \widehat{\QQ}_{\kappa}^{\KK_S} =  \sum_{j=d_N+1}^{\KK}  \PPi\edex{j}{\widehat{E}_{\kappa}}.
\end{equation*}

\subsection{Proof of the results in Section \ref{sec_proposed_estimator} on autocovariance-based FPCA}
\begin{proofs}[Proof of Theorem \ref{thm0}]
		Since $\PP^N+\PP^S = I$, we note the identity 
	\begin{equation} \label{eq001}
		\widehat{\PP}^N_\kappa  - \PP^N = 	{\PP}^S \widehat{\PP}^N_\kappa  + \PP^N \widehat{\PP}^N_\kappa - \PP^N = \PP^S\widehat{\PP}^N_\kappa- \PP^N\widehat{\PP}^S_\kappa.
	\end{equation}
	Since $\widehat{\PP}^N_\kappa$ is the projection onto the span of the first $d_N$ leading eigenvectors of $\widehat{D}_{\kappa}$,
	\begin{equation} \label{eq001a}
		{\PP}^N\widehat{D}_{\kappa}\widehat{\PP}^S_\kappa= {\PP}^N\widehat{D}_\kappa{\PP}^N\widehat{\PP}^S_\kappa +  {\PP}^N\widehat{D}_\kappa {\PP}^S\widehat{\PP}^S_\kappa ={\PP}^N \widehat{\Lambda},
	\end{equation}
	where $\widehat{\Lambda} = \sum_{j=d_N+1}^{\infty}  \llambda\edex{j}{\widehat{D}_{\kappa}} \PPi\edex{j}{\widehat{D}_{\kappa}}$.  
	From \eqref{eq001a} and the fact that $\widehat{\Lambda}=\widehat{\PP}^S_\kappa\widehat{\Lambda}$, we obtain
	\begin{equation}
		T \PP^N\widehat{\PP}^S_\kappa  = -\left(T^{-2}\PP^N  \widehat{D}_\kappa \PP^N\right)^\dag T^{-1} \PP^N  \widehat{D}_\kappa \PP^S + \left(T^{-2}\PP^N  \widehat{D}_\kappa \PP^N\right)^\dag  T^{-1} \PP^N\widehat{\PP}^S_\kappa  \widehat{\Lambda}, \label{eq002}
	\end{equation}
	where $(T^{-2}\PP^N  \widehat{D}_\kappa \PP^N)^\dag$ denotes the Moore-Penrose inverse of $T^{-2}\PP^N  \widehat{D}_\kappa \PP^N$, which is well defined since $T^{-2}\PP^N  \widehat{D}_\kappa \PP^N$ is a finite rank operator (see proof of Theorem 3.1 of \citealp{seo2020functional}); more generally, we hereafter let $A^\dag$ denote the Moore-Penrose inverse of $A$ if it is well-defined. Since $I=\PP^N+\PP^S$, we note that $\PP^N\widehat{D}_\kappa \PP^N = \PP^N\widehat{C}_\kappa^\ast \PP^N\widehat{C}_\kappa \PP^N+\PP^N\widehat{C}_\kappa^\ast \PP^S\widehat{C}_\kappa \PP^N$, and hence  
	\begin{align}\label{eqadd03}
		T^{-2}\PP^N  \widehat{D}_\kappa \PP^N &= \left(T^{-1} \PP^N\widehat{C}_\kappa^\ast \PP^N \right)\left(T^{-1} \PP^N\widehat{C}_\kappa \PP^N\right) + \left(T^{-1}\PP^N\widehat{C}_\kappa^\ast \PP^S \right)\left(T^{-1}\PP^S\widehat{C}_\kappa \PP^N\right).
	\end{align}
	Since $\sup_{1\leq t\leq T}\|\PP^N\tilde{x}_t\| = O_p(T^{-1/2})$ (see the proof of Lemma 1 of \citealp{NSS}), we find that $\|T^{-1}\PP^N\widehat{C}_\kappa \PP^S\|_{\opnorm}= \|T^{-2}\sum_{t=\kappa+1}^T \PP^S {x}_{t-\kappa}\otimes \PP^N{x}_t\|_{\opnorm} +  o_p(1) = o_p(1)$. 
	Furthermore, $\|T^{-1}\PP^N \widehat{C}_\kappa \PP^N - T^{-2}\sum_{t=1}^T \PP^N x_t \otimes   \PP^N x_t\|_{\opnorm} = o_p(1)$ since $\kappa$ is finite, and we know from nearly identical arguments used in the proof of Theorem 3.1 of \cite{seo2020functional} that $T^{-2}\sum_{t=1}^T \PP^N x_t \otimes   \PP^N x_t \to_{\pp}  V_1 =_d \int W^N \otimes W^N$. Combining these results, the following is established (due to the finiteness of $\kappa$): $T^{-1}\sum_{t=1}^T \PP^N x_t \otimes   \PP^N x_{t-\kappa} \to_{\pp}  V_1$. This result, combined with the definition of $\widehat{D}_{\kappa}$ and \eqref{eqadd03}, implies that 
	\begin{equation*} 
		T^{-2}\PP^N  \widehat{D}_\kappa \PP^N \to_{\pp} V_1^\ast V_1 \, (= V_1V_1). 
	\end{equation*}   
	Furthermore, from the same arguments used to derive (S6.7) in \cite{seo2020functional}, we can also find that 
	\begin{equation}\label{eqadd02}
		(T^{-2}\PP^N  \widehat{D}_\kappa \PP^N)^{\dagger}  \to_{\pp} (V_1^\ast V_1)^\dag.
	\end{equation}
	We next observe that
	$$\PP^N\widehat{D}_\kappa \PP^S = \PP^N\widehat{C}_\kappa^\ast \PP^N\widehat{C}_\kappa \PP^S+\PP^N\widehat{C}_\kappa^\ast \PP^S\widehat{C}_\kappa \PP^S.$$
	As shown above, $T^{-1}\PP^N\widehat{C}_\kappa^\ast \PP^S $ is $o_p(1)$, and from this result, we note that $\|T^{-1}\PP^N\widehat{C}_\kappa^\ast \PP^S\widehat{C}_\kappa \PP^S\|_{\opnorm}=\|(T^{-1}\PP^N\widehat{C}_\kappa^\ast \PP^S )(\PP^S\widehat{C}_\kappa \PP^S)\|_{\opnorm}=o_p(1)$. Thus, we have
	\begin{equation}\label{eqaddpf01}
		T^{-1}\PP^N  \widehat{D}_\kappa \PP^S = \left(T^{-1}\PP^N\widehat{C}_\kappa^\ast \PP^N \right)\left(\PP^N\widehat{C}_\kappa \PP^S\right)+ o_p(1).
	\end{equation}
	We now obtain the limiting behavior of $\PP^N\widehat{C}_\kappa \PP^S$. Since $T^{-1}\sum_{t=1}^T \PP^S  e_{t-\kappa} \otimes \PP^N e_t + T^{-1}\sum_{t=1}^T \PP^S  x_{t-\kappa} \otimes \PP^N e_t = o_p(1)$ under Assumption \ref{assum1c}, we find that $\PP^N\widehat{C}_\kappa \PP^S = T^{-1}\sum_{t=1}^T \PP^S \tilde{x}_{t-\kappa} \otimes \PP^N \tilde{x}_{t}= T^{-1}\sum_{t=1}^T \PP^S {x}_{t-\kappa} \otimes \PP^N {x}_{t}+  T^{-1}\sum_{t=1}^T \PP^S  e_{t-\kappa} \otimes \PP^N {x}_{t} + o_p(1)$. We also observe that  
	\begin{equation} \label{eqpf000}
		\frac{1}{T}\sum_{t=1}^T \PP^S {x}_{t-\kappa} \otimes \PP^N {x}_{t} =\frac{1}{T}\sum_{t=1}^T \PP^S x_t \otimes \PP^N x_t - \frac{1}{T}\sum_{t=\kappa+1}^T  (\Delta \PP^S {x}_{t-\kappa+1}+\cdots+\Delta \PP^S {x}_{t})\otimes \PP^N {x}_{t}.  
	\end{equation}
	Using the summation by parts, 
    Assumptions \ref{assum1} and \ref{assum1c}, 
    and the fact that $\|T^{-1/2}\PP^Nx_t\| = O_p(1)$, the following can be shown: for $j=1,\ldots,\kappa$,
	\begin{equation*}
		-\frac{1}{T}\sum_{t=\kappa+1}^T \Delta \PP^S {x}_{t-\kappa+j} \otimes \PP^N{x}_t =   \frac{1}{T}\sum_{t=\kappa+2}^T \PP^S {x}_{t-\kappa+j-1} \otimes \PP^N \Delta {x}_t + o_p(1)  \to_{\pp} \mathbb{E}[u_{t-\kappa+j-1}^S \otimes u_t^N]. 
	\end{equation*}
	Since $\mathbb{E}[u_{t-\kappa+j-1}^S \otimes u_t^N] = \mathbb{E}[u_{t}^S \otimes u_{t+\kappa-j+1}^N]$ due to stationarity (see \eqref{eqreg1}), we find that
	\begin{equation}\label{eqpf001}
		- \frac{1}{T}\sum_{t=\kappa+1}^T  (\Delta \PP^S x_{t-\kappa+1}+\Delta \PP^S x_{t-\kappa+2}+\cdots+\Delta \PP^S x_{t})\otimes \PP^N x_{t} \to_{\pp}  \sum_{j=1}^{\kappa} \mathbb{E}[u_{t}^S \otimes u_{t+\kappa-j+1}^N]. 
	\end{equation}
	We have $T^{-1}\sum_{t=\kappa+1}^{T} e_{t-\kappa} = O_p(T^{-1/2})$ under Assumption \ref{assum1}, and $\sup_{1 \leq t\leq T}\|\PP^N {x}_{t}\| = O_p(T^{-1/2})$ as well (see e.g. \citealp{berkes2013weak,NSS}). We thus find that 
	\begin{equation}\label{eqpf001a}
		\frac{1}{T}\sum_{t=\kappa+1}^T \PP^S  e_{t-\kappa} \otimes \PP^N {x}_{t} = O_p(1).
	\end{equation}
	Note that the operator in \eqref{eqpf001a} is equal to $T^{-1}\sum_{t=\kappa+1}^T e_{t-\kappa} \otimes \PP^N {x}_{t-\kappa-1} +  T^{-1}\sum_{t=\kappa+1}^T e_{t-\kappa} \otimes (\Delta \PP^N {x}_{t-\kappa}+\cdots+\Delta \PP^N {x}_{t})$, 
	which is not generally negligible unless $e_t=0$ for $t\geq 1$ under our assumptions.   
	One may deduce from the proof of Theorem 3.1 of \cite{seo2020functional} that $T^{-1}\sum_{t=1}^T \PP^S x_t \otimes \PP^N x_t \to_{\pp} V_{1,0} =_d \int dW^S \otimes W^N + \sum_{j \geq 0}\mathbb{E}[u^S_t \otimes u^N_{t-j}]$. Combining this result with \eqref{eqpf000}, \eqref{eqpf001} and \eqref{eqpf001a}, we find that 
	\begin{align} \label{eqpf001aa}
		\PP^N\widehat{C}_\kappa \PP^S - \frac{1}{T}\sum_{t=\kappa+1}^T \PP^S e_{t-\kappa} \otimes \PP^N {x}_{t} \to_{\pp} V_{1,\kappa} =_d  \int dW^S \otimes W^N + \sum_{j \geq -\kappa}\mathbb{E}[u^S_t \otimes u^N_{t-j}]. 
	\end{align}
	Let ${\YYYY}_T$ be defined as in \eqref{equpsilon}, i.e., ${\YYYY}_T =  (T^{-2}\PP^N  \widehat{D}_\kappa \PP^N)^\dag (T^{-1} \PP^N  \widehat{C}_\kappa^\ast \PP^N)(T^{-1}\sum_{t=\kappa+1}^T \PP^S e_{t-\kappa} \otimes \PP^N {x}_{t}).$ From \eqref{eqadd02}, \eqref{eqaddpf01} and \eqref{eqpf001aa}, we find that
\begin{align}
&\left(T^{-2}\PP^N  \widehat{D}_\kappa \PP^N\right)^\dag T^{-1} \PP^N  \widehat{D}_\kappa \PP^S - \YYYY_T \notag  \\&= \left(T^{-2}\PP^N  \widehat{D}_\kappa \PP^N\right)^\dag  \left(T^{-1}\PP^N\widehat{C}_\kappa^\ast \PP^N \right)\left(\PP^N\widehat{C}_\kappa \PP^S- \frac{1}{T}\sum_{t=\kappa+1}^T \PP^S e_{t-\kappa} \otimes \PP^N {x}_{t}\right) +  o_p(1) \to_{\pp} \mathcal A_\kappa, \label{eqpf003}
\end{align}
where $\mathcal A_{\kappa}=_d(V_1^\ast V_1)^\dag V_1^\ast  V_{1,\kappa}$. Since  $\|\PP^N\widehat{\PP}^S_\kappa  \widehat{\Lambda}\|_{\opnorm} = O_p(1)$, we find from \eqref{eq002} and \eqref{eqadd02} that
	\begin{equation}\label{eqpf002}
		T \PP^N\widehat{\PP}^S_\kappa = - \left\{\left(T^{-2}\PP^N  \widehat{D}_\kappa \PP^N\right)^\dag T^{-1} \PP^N  \widehat{D}_\kappa \PP^S  - \YYYY_T \right\} - \YYYY_T +o_p(1). 
	\end{equation}
	Using similar arguments used in the proof of Claim 3 (of Theorem 3.1) of \cite{seo2020functional}, it can be shown that $	T \PP^N\widehat{\PP}^S_\kappa = - T(\PP^N\widehat{\PP}^S_\kappa)^\ast + o_p(1)$. From \eqref{eq001}, we find that $T (\widehat{\PP}^N_\kappa - \PP^N)  = - T \PP^N\widehat{\PP}^S_\kappa  - T (\PP^N\widehat{\PP}^S_\kappa)^\ast + o_p(1)$. It is then deduced from \eqref{eqpf003}, \eqref{eqpf002} and similar arguments used in the proof of Theorem 3.1 of  \cite{seo2020functional} that $T (\widehat{\PP}^N_\kappa - \PP^N) - \YYYY_T - \YYYY_T^\ast \to_{\pp}  \mathcal A_{\kappa} + \mathcal A_{\kappa}^\ast$ as desired. 

The limiting behavior of $T (\widehat{\PP}_\kappa^{S} -\PP^S)$ is deduced from that $T (\widehat{\PP}_\kappa^{N}   - \PP^N) = -T (\widehat{\PP}_\kappa^{S} -\PP^S)$.
\end{proofs}

\subsection{Proof of the results in Section \ref{sec_proposed_estimator2}}
Subsequently, we provide our proof of the desired result, focusing on the case where $\mathcal H_y = \mathcal H$, and hence $y_t$ and $u_t$ are function-valued, with $f$ understood as a bounded linear operator on $\mathcal H$. The other case, where $\mathcal H_y=\mathbb{R}$, is, as may be expected, simpler and requires only a moderate modification.
\begin{proofs}[Proof of Theorem \ref{thm1}]
	We first note that $T^{-1}\sum_{t=1}^T \tilde{x}_{t-\kappa} \otimes f (\tilde{x}_t) = f  \widehat{C}_\kappa^\ast$, and hence
	\begin{equation*}
		\widehat{f^N_{\kappa}}  = \left(\frac{1}{T}\sum_{t=1}^T \tilde{x}_{t-\kappa} \otimes (f (\tilde{x}_t)  + \tilde{u}_t)  \right) \widehat{C}_\kappa (\widehat{D}_\kappa)_{\KK}^{-1} \widehat{\PP}^N_\kappa  
		= f \widehat{\PP}_\kappa^N  + \left(\frac{1}{T}\sum_{t=1}^T \tilde{x}_{t-\kappa} \otimes \tilde{u}_t \right) \widehat{C}_\kappa (\widehat{D}_\kappa)_{\KK}^{-1} \widehat{\PP}^N_{\kappa}.
	\end{equation*}
	Observe that $\widehat{C}_\kappa (\widehat{D}_\kappa)_{\KK}^{-1} \widehat{\PP}^N_\kappa= \widehat{C}_\kappa \widehat{\PP}^N_\kappa  (\widehat{D}_\kappa)_{\KK}^{-1} \widehat{\PP}^N_\kappa = \widehat{\QQ}^N_\kappa  \widehat{C}_\kappa \widehat{\PP}^N_\kappa (\widehat{D}_\kappa)_{\KK}^{-1} \widehat{\PP}^N_\kappa$. Using the fact that $\widehat{\PP}_\kappa^N$ and $\widehat{\QQ}^N_\kappa$ are orthogonal projections (and thus idempotent), we find that 
	\begin{equation} \label{pfeq04}
		T(	\widehat{f^N_{\kappa}} -f{\PP}^N ) = Tf(\widehat{\PP}_\kappa^N - \PP^N) +  \left(\frac{1}{T}\sum_{t=1}^T   \widehat{\QQ}^N_{\kappa} \tilde x_{t-\kappa} \otimes \tilde u_t \right)\frac{\widehat{\QQ}^N_{\kappa}  \widehat{C}_\kappa \widehat{\PP}^N_{\kappa} }{T} \left(T^2 \widehat{\PP}^N_{\kappa} ({\widehat{D}_\kappa})_{\KK}^{-1} \widehat{\PP}^N_{\kappa}\right).
	\end{equation}	 
	From a slight modification of our proof of Theorem~\ref{thm0}, it can be shown that $\|\widehat{\QQ}_\kappa^S-\PP^S\|_{\opnorm} = O_p(T^{-1})$. We also note that $T^2 \widehat{\PP}^N_{\kappa} (\widehat{D}_\kappa)_{\KK}^{-1} \widehat{\PP}^N_{\kappa} = T^2 \sum_{j=1}^{d_N} \llambda^{-1}\edex{j}{\widehat{D}_{\kappa}^N}\PPi\edex{j}{\widehat{D}_{\kappa}^N}$, and since $T^{-2} \widehat{\PP}^N_{\kappa} \widehat{D}_\kappa \widehat{\PP}^N_{\kappa} \to_p V_1^\ast V_1$ (where $V_1=_d\int W^N \otimes W^N$), we have, for $j=1,\ldots,d_N$, $T^{-2}\llambda\edex{j}{\widehat{D}_{\kappa}^N} \to_p \llambda\edex{j}{V_1^\ast V_1}$, which are distinct almost surely, and also $\PPi\edex{j}{\widehat{D}_{\kappa}^N} \to_p \PPi\edex{j}{A^\ast A}$. Combining these results with the arguments used to establish (S6.7) of \cite{seo2020functional} and the limiting behavior of $\PP^N\widehat{C}_{\kappa}\PP^N$ discussed in the proof of Theorem~\ref{thm0}, we  find that 
	\begin{equation}\label{pfeq04a}
        	T^{-1}{\widehat{\QQ}^N_{\kappa}  \widehat{C}_\kappa \widehat{\PP}^N_{\kappa} } \left(T^2 \widehat{\PP}^N_{\kappa} ({\widehat{D}_\kappa})_{\KK}^{-1} \widehat{\PP}^N_{\kappa}\right) \to_{\pp}  V_1^\dag =_d \left(\int W^N \otimes W^N\right)^\dag.  
	\end{equation}
	Moreover, using Assumptions \ref{assum1} and \ref{assum1c}, and the fact that $\tilde{u}_t= u_t -f(e_t)$, 
	we first find that 
	\begin{equation}\label{eqaddaa1}
		\frac{1}{T}\sum_{t=1}^T   \widehat{\QQ}^N_{\kappa} \tilde x_{t-\kappa} \otimes \tilde u_t  = \frac{1}{T}\sum_{t=1}^T   {\PP}^N  x_{t-\kappa} \otimes  u_t  - \frac{1}{T}\sum_{t=1}^T   {\PP}^N  x_{t-\kappa} \otimes  f(e_t)   + o_p(1),
	\end{equation}
	where we use the employed conditions that $T^{-1}\sum_{t=1}^T  e_{t-\kappa} \otimes  u_t = o_p(1)$ and $T^{-1}\sum_{t=1}^T  e_{t-\kappa} \otimes  e_t=o_p(1)$. Letting ${\YYY}_T =T^{-1}\sum_{t=1}^T   {\PP}^N  x_{t-\kappa} \otimes  f(e_t)$, which is $O_p(1)$, we find from \eqref{eqaddaa1} that  
	\begin{align}
		\frac{1}{T}\sum_{t=1}^T   \widehat{\QQ}^N_{\kappa} \tilde x_{t-\kappa} \otimes \tilde u_t  + {\YYY}_T &= \frac{1}{T}\sum_{t=1}^T   {\PP}^N  x_{t} \otimes  u_t - \frac{1}{T}\sum_{t=1}^T   (\Delta {\PP}^N  x_{t-\kappa+1}+\ldots + \Delta {\PP}^N  x_{t}) \otimes  u_t \to_{\pp}  V_{2,\kappa},  \label{eqpf002a}
	\end{align}
	where $V_{2,\kappa}=_d \int W^N \otimes d W^u - \sum_{j \geq \kappa}\mathbb{E}[u_{t-j}^N \otimes u_{t}]$ and the convergence is deduced from arguments similar to those used in our proof of Theorem \ref{thm0} for the limiting behavior of $\PP^N\widehat{C}_{\kappa} \PP^S$, together with the fact that $T^{-1} \sum_{t=1}^T u_{t-j}^N \otimes u_t \to_p \mathbb{E}[ u_{t-j}^N \otimes u_t]$ under the employed assumptions. Note also that, from Theorem \ref{thm0}, the following can be deduced:
	\begin{equation}\label{eqaddaa2}
		Tf(\widehat{\PP}_\kappa^N - \PP^N) - f(\Upsilon_T)= f(T(\widehat{\PP}_\kappa^N - \PP^N) - \Upsilon_T) \to_{\pp} f(\mathcal A_\kappa^\ast + \mathcal A_\kappa).
	\end{equation}
	From 	\eqref{pfeq04}--\eqref{eqaddaa2}, we find that 
	\begin{align}
		T(	\widehat{f^N_{\kappa}} -f{\PP}^N ) - f(\Upsilon_T) + \YYY_T{\widehat{\QQ}^N_{\kappa}  \widehat{C}_\kappa \widehat{\PP}^N_{\kappa} } ({\widehat{\PP}^N_{\kappa} \widehat{D}_\kappa \widehat{\PP}^N_{\kappa}})_{\KK}^{-1} \to_{\pp}  f(\mathcal A_{\kappa} + \mathcal A_{\kappa}^\ast ) + V_{2,\kappa} V_1^\dag,
	\end{align}
	which proves Theorem \ref{thm1}; more specifically, when $\{u_t\}_{t\geq1}$ is a martingale difference with respect to $\mathfrak F_t$, as assumed in Theorem \ref{thm1}, we have $\sum_{j \geq \kappa}\mathbb{E}[u_{t-j}^N \otimes u_{t}] = 0$, and hence obtain the desired result. 
	
	Since $	\widehat{f^N_{\kappa}}-f\widehat{\PP}^N_{\kappa} = 	(\sum_{t=1}^T   \widehat{\QQ}^N_{\kappa} \tilde x_{t-\kappa} \otimes \tilde u_t ){\widehat{\QQ}^N_{\kappa}  \widehat{C}_\kappa \widehat{\PP}^N_{\kappa} } ({\widehat{\PP}^N_{\kappa} \widehat{D}_\kappa \widehat{\PP}^N_{\kappa}})_{\KK}^{-1}$, the following is also deduced from the above arguments:
	\begin{equation}\label{eqaddresult1}
		\widehat{f^N_{\kappa}}-f\widehat{\PP}^N_{\kappa} = O_p(T^{-1}),
	\end{equation}
	which will be used in our proof of Theorem \ref{thm2}. 
\end{proofs}

\begin{proofs}[Proof of Theorem \ref{thm2}] 
	We let $(\widehat{D}_{\kappa}^S)^{-1}_{\KK}$ denote $(\widehat{D}_{\kappa})^{-1}_{\KK} \widehat{\PP}_{\kappa}^S$ (see \eqref{eqaddnotation}).
	Noting the facts that 
	$\widehat{C}_\kappa (\widehat{D}_\kappa^S)_{\KK}^{-1} \widehat{\PP}^S_{\kappa}= \widehat{C}_\kappa \widehat{\PP}^S_{\kappa}  (\widehat{D}_\kappa^S)_{\KK}^{-1} \widehat{\PP}^S_\kappa = \widehat{\QQ}^S_{\kappa}  \widehat{C}_\kappa \widehat{\PP}^S_{\kappa}  (\widehat{D}_\kappa^S)_{\KK}^{-1} \widehat{\PP}^S_{\kappa}$, $\widehat{\QQ}^S_{\kappa}$ is idempotent, $\tilde{x}_t = O_p(T^{1/2})$ 
	and $\widehat{f^N_{\kappa}} - f^N = O_p(T^{-1})$ (see Theorem \ref{thm1}), we write $	\widehat{f^S_{\kappa}}$ as follows:
	\begin{align}
		\widehat{f^S_{\kappa}}  &= \left(\frac{1}{T}\sum_{t=1}^T \tilde{x}_{t-\kappa} \otimes (f (\tilde{x}_t)  + \tilde{u}_t -\widehat{f^N_{\kappa}}  (\tilde{x}_t))  \right) \widehat{C}_\kappa (\widehat{D}_\kappa^S)_{\KK}^{-1} \widehat{\PP}_\kappa^{S} \notag \\
		&= \left(\frac{1}{T}\sum_{t=1}^T \tilde{x}_{t-\kappa} \otimes (f^S (\tilde{x}_t)+ \tilde{u}_t + \widehat{\mathcal W}_t)\right) \widehat{C}_\kappa (\widehat{D}_\kappa^S)_{\KK}^{-1} \widehat{\PP}_\kappa^{S} \notag \\ 
		&= f^S  \widehat{\PP}_\kappa^{\KK_S} + \left(\frac{1}{T}\sum_{t=1}^T \widehat{\QQ}^{S}_{\kappa} \tilde{x}_{t-\kappa} \otimes (\tilde{u}_t + \widehat{\mathcal W}_t)\right) \widehat{\QQ}_\kappa^{S} \widehat{C}_\kappa \widehat{\PP}_\kappa^{\KK_S} (\widehat{D}_\kappa)_\KK^{-1} \widehat{\PP}_\kappa^{\KK_S}, \label{eqthm01}
	\end{align}
	where $\widehat{\mathcal W}_t = f^N(\tilde{x}_t)-\widehat{f^N_{\kappa}}(\tilde{x}_t)$.
	Observe that 
	\begin{align}
		&\widehat{\QQ}_\kappa^{S} \widehat{C}_\kappa \widehat{\PP}_\kappa^{\KK_S} (\widehat{D}_\kappa)_\KK^{-1} \widehat{\PP}_\kappa^{\KK_S} \notag \\ &= \left( \sum_{j=d_N+1}^{\KK} {\sqrt{\llambda\edex{j}{\widehat{D}_{\kappa}}}} \varv\edex{j}{\widehat{D}_{\kappa}} \otimes  \widehat{\QQ}_\kappa^{S} \varv\edex{j}{\widehat{E}_{\kappa}}\right)  \left( \sum_{j=d_N+1}^{K} \frac{1}{\llambda\edex{j}{\widehat{D}_{\kappa}}} \varv\edex{j}{\widehat{D}_{\kappa}} \otimes  \varv\edex{j}{\widehat{D}_{\kappa}}\right)\notag \\
		&= \left( \sum_{j=d_N+1}^{\KK} \frac{1}{\sqrt{\llambda\edex{j}{\widehat{D}_{\kappa}}}} \varv\edex{j}{\widehat{D}_{\kappa}} \otimes  \widehat{\QQ}_\kappa^{S} \varv\edex{j}{\widehat{E}_{\kappa}}\right)   =   O_p(\alpha^{-1/2}). \label{eqthm02}
	\end{align}
	Since $\|\widehat{\QQ}_\kappa^S-\PP^S\|_{\opnorm} = O_p(T^{-1})$ (see our proof of Theorem \ref{thm1}) and $\|\widehat{\mathcal W}_t\| \leq \|f^N-\widehat{f^N_{\kappa}}\|_{\opnorm}\|\tilde{x}_t\| = O_p(T^{-1/2})$ uniformly in $t$,  $\|T^{-1}\sum_{t=1}^T \widehat{\QQ}^{S}_{\kappa} \tilde{x}_{t-\kappa} \otimes (\tilde{u}_t + \widehat{\mathcal W}_t) \|_{\opnorm}= O_p(T^{-1/2})$ under Assumptions \ref{assum1} and \ref{assum1c}. 
	From \eqref{eqthm01}, \eqref{eqthm02} and nearly identical arguments used in the proof of Theorem 1 of \cite{seong2021functional}, we find that  $\|\widehat{f^S_{\kappa}} - f^S\|_{\opnorm} \to_p 0$  as long as $T^{-1/2} \sum_{j=1}^{\KK_S} \ttau_{j}(D_{\kappa}^S) \to_p 0$, which is implied by the employed condition that $T^{-1/2} \alpha^{-1/2}\sum_{j=1}^{\KK_S} \ttau_{j}(D_{\kappa}^S) \to_p 0$. 
	
	We next show \eqref{eqthm2}. From \eqref{eqthm01}, we have
	\begin{align}
		&\sqrt{T/\theta_{\KK_S}(\zeta)}(\widehat{f^S_{\kappa}}(\zeta)-	f^S\widehat{\PP}_\kappa^{\KK_S}(\zeta)) =    \left(\frac{1}{\sqrt{T \theta_{\KK_S}(\zeta)}}\sum_{t=1}^T  \widehat{\QQ}^S_{\kappa} \tilde{x}_{t-\kappa} \otimes (\tilde{u}_t + \widehat{\mathcal W}_t) \right) \widehat{\QQ}^S_{\kappa}  \widehat{C}_\kappa \widehat{\PP}^S_{\kappa} (\widehat{D}_\kappa)_\KK^{-1} \widehat{\PP}^S_{\kappa}(\zeta). \label{eqaddadd01}
	\end{align}	 
	We first show that \eqref{eqaddadd01} reduces to 
	\begin{align} \label{eqlater00}
		\sqrt{T/\theta_{\KK_S}(\zeta)}(\widehat{f^S_{\kappa}}(\zeta)-	f^S\widehat{\PP}_\kappa^{\KK_S}(\zeta)) =	\left(\frac{1}{\sqrt{T \theta_{\KK_S}(\zeta)}}\sum_{t=1}^T  \widehat{\QQ}^S_{\kappa} \tilde{x}_{t-\kappa} \otimes \tilde{u}_t \right) \widehat{\QQ}^S_{\kappa}  \widehat{C}_\kappa \widehat{\PP}^S_{\kappa} (\widehat{D}_\kappa)_\KK^{-1} \widehat{\PP}^S_{\kappa}(\zeta) + o_p(1).
	\end{align}	 
	To see this, we observe that 
	\begin{align}
		&\left\|\left(\frac{1}{\sqrt{T \theta_{\KK_S}(\zeta)}}\sum_{t=1}^T  \widehat{\QQ}^S_{\kappa} \tilde{x}_{t-\kappa} \otimes \widehat{\mathcal W}_t \right) \widehat{\QQ}^S_{\kappa}  \widehat{C}_\kappa \widehat{\PP}^S_{\kappa} (\widehat{D}_\kappa)_\KK^{-1} \widehat{\PP}^S_{\kappa}(\zeta)\right\| \notag \\ &\leq 		  \left\|f^N-\widehat{f^N_{\kappa}}\right\|_{\opnorm}\left\|\frac{1}{\sqrt{T \theta_{\KK_S}(\zeta)}}\sum_{t=1}^T  \widehat{\QQ}^S_{\kappa} \tilde{x}_{t-\kappa} \otimes \tilde{x}_t\right\|_{\opnorm} \left\|\widehat{\QQ}^S_{\kappa}  \widehat{C}_\kappa \widehat{\PP}^S_{\kappa} (\widehat{D}_\kappa)_\KK^{-1} \widehat{\PP}^S_{\kappa}(\zeta) \right\|\notag \\ &= O_p\left(\frac{1}{\sqrt{T \theta_{\KK_S}(\zeta)}}\right) O_p\left(\frac{1}{T}\sum_{t=1}^T  \PP^S \tilde{x}_{t-\kappa} \otimes \tilde{x}_t + \frac{1}{T}\sum_{t=1}^T   (\widehat{\QQ}^S_{\kappa} -\PP^S) \tilde{x}_{t-\kappa} \otimes \tilde{x}_t\right) O_p \left(\alpha^{-1/2}\right), \label{eqaadad1} 
	\end{align}	  
	where the second equality follows from (a) $\|f^N-\widehat{f^N_{\kappa}}\|_{\opnorm} = O_p(T^{-1})$ and (b) $\|\widehat{\QQ}^S_{\kappa}  \widehat{C}_\kappa \widehat{\PP}^S_{\kappa} (\widehat{D}_\kappa)_\KK^{-1} \widehat{\PP}^S_{\kappa}\|_{\opnorm} \leq O_p(1/\sqrt{\llambda\edex{\KK}{\widehat{D}_{\kappa}}})\leq O_p(\alpha^{-1/2})$ (see \eqref{eqthm02}). We also find from the proof of Theorem 3.1 of \cite{seo2020functional} that $T^{-2}\sum_{t=1}^T  \tilde{x}_{t-\kappa} \otimes \tilde{x}_t = O_p(1)$ and also $T^{-1}\sum_{t=1}^T  \PP^S \tilde{x}_{t-\kappa} \otimes \tilde{x}_t = O_p(1)$. These results, together with the fact that $\|\widehat{\QQ}^S_{\kappa}-\PP^S\|_{\opnorm} = O_p(T^{-1})$, imply that the middle term in the right hand side of \eqref{eqaadad1} is $O_p(1)$. We will show later that there is an estimator $\widehat{\theta}_{\KK_S}$ of ${\theta}_{\KK_S}$ such that  $|\widehat{\theta}_{\KK_S}(\zeta) - \theta_{\KK_S}(\zeta)|= o_p(1)$ and $\widehat{\theta}_{\KK_S}(\zeta) = O_p(\alpha^{-1})$. This implies that $\theta_{\KK_S}(\zeta) = O_p(\alpha^{-1})$. Combining all these results, we find that 
	\begin{align*}
		\left\|\left(\frac{1}{\sqrt{T \theta_{\KK_S}(\zeta)}}\sum_{t=1}^T  \widehat{\QQ}^S_{\kappa} \tilde{x}_{t-\kappa} \otimes \widehat{\mathcal W}_t \right) \widehat{\QQ}^S_{\kappa}  \widehat{C}_\kappa \widehat{\PP}^S_{\kappa} (\widehat{D}_\kappa)_\KK^{-1} \widehat{\PP}^S_{\kappa}(\zeta)\right\| = O_p(1/\sqrt{T\alpha^2}) = o_p(1),
	\end{align*}
	under our assumption that $T\alpha^2 \to \infty$. Thus  \eqref{eqlater00} is established. 
	
	We next focus on the limiting behavior of the term appearing in the right hand side of \eqref{eqlater00}. 	Note that  $\widehat{\PP}_\kappa^S-\PP^S = O_p(T^{-1})$ and  $\widehat{\QQ}_\kappa^S-\PP^S = O_p(T^{-1})$, from which 	it is not difficult to show that $\| \widehat{\QQ}^S_{\kappa}  \widehat{C}_\kappa \widehat{\PP}^S_{\kappa}- {C}_\kappa^S \|_{\opnorm} = \| \widehat{C}_\kappa \widehat{\PP}^S_{\kappa}- {C}_\kappa^S \|_{\opnorm} = O_p(T^{-1/2})$ (and thus $\|\widehat{\PP}^S_{\kappa}  \widehat{D}_\kappa \widehat{\PP}^S_{\kappa} - D_{\kappa}^S\|_{\opnorm} = O_p(T^{-1/2})$ as well). 	Moreover, if $\llambda\edex{1}{D_{\kappa}^S}>\llambda\edex{2}{D_{\kappa}^S}>\cdots > 0$ and  
	$T^{-1/2}\alpha^{-1/2}\sum_{j=1}^{\KK_S}\ttau\edex{j}{D_{\kappa}^S} \to_p 0$ as assumed in  \eqref{eqcondition_a}, we may deduce from nearly identical arguments used in the proof of Theorem 2 of \cite{seong2021functional} that  $\|\widehat{\QQ}^S_{\kappa}  \widehat{C}_\kappa \widehat{\PP}^S_{\kappa}  (\widehat{D}_\kappa)_\KK^{-1} \widehat{\PP}^S_{\kappa}(\zeta) - \PP^S C_{\kappa}^S(D_\kappa^S)_{\KK_S}^{-1}(\zeta)\|\to_p0$ (see (S2.4)-(S2.6) in their paper). 
	Combining all these results, we may rewrite \eqref{eqaddadd01} (or \eqref{eqlater00}) as follows, ignoring asymptotically negligible terms:
	\begin{align} \label{eqlater0}
		\sqrt{T/\theta_{\KK_S}(\zeta)}(\widehat{f^S_{\kappa}}(\zeta)-	f^S\widehat{\PP}_\kappa^{\KK_S}(\zeta)) =      \left(\frac{1}{\sqrt{T \theta_{\KK_S}(\zeta)}}\sum_{t=1}^T {\PP}^{S} \tilde{x}_{t-\kappa} \otimes \tilde{u}_t \right)C_{\kappa}^S(D_\kappa^S)_{\KK_S}^{-1}(\zeta) + o_p(1).
	\end{align}	
	Let $\zeta_t = [ \PP^{S}  \tilde{x}_{t-\kappa} \otimes \tilde{u}_t]  C_{\kappa}^S(D_\kappa^S)_{\KK_S}^{-1}(\zeta) = \langle \PP^{S}  \tilde{x}_{t-\kappa} ,  C_{\kappa}^S(D_{\kappa}^S)_{\KK_S}^{-1} (\zeta) \rangle\tilde{u}_t.$ Then, we have 
	\begin{equation*}
		\mathbb{E}[\zeta_t\otimes \zeta_t] 
		=  \mathbb{E}[\langle \PP^{S}  \tilde{x}_{t-\kappa} ,  C_{\kappa}^S(D_{\kappa}^S)_{\KK_S}^{-1} (\zeta) \rangle^2 \tilde{u}_t \otimes \tilde{u}_t]. 
	\end{equation*}
	Because $u_t$ is a martingale difference with respect to $\mathfrak F_{t}$, the following is deduced:
	\begin{align}
		\mathbb{E}[\zeta_t\otimes \zeta_t] &=  \mathbb{E}[\langle  \PP^{S} \tilde x_{t-\kappa} ,  C_{\kappa}^S(D_{\kappa}^S)_{\KK_S}^{-1} (\zeta) \rangle^2 \tilde u_t \otimes \tilde u_t] = \langle C_{\kappa}^S(D_\kappa^S)_{\KK_S}^{-1} (\zeta),  \widetilde{C}_{0}^S C_{\kappa}^S(D_\kappa^S)_{\KK_S}^{-1} (\zeta) \rangle  \widetilde{C}_{u} \notag \\ 
		&= \langle \zeta,  (D_\kappa^S)_{\KK_S}^{-1}   (C_{\kappa}^S)^\ast \widetilde{C}_{0}^S C_{\kappa}^S(D_\kappa^S)_{\KK_S}^{-1} (\zeta) \rangle  \widetilde{C}_{u} = \theta_{\KK_S} (\zeta) \widetilde{C}_u.\label{eqlater1}
	\end{align}
	As in (S2.8) and (S2.9) of \cite{seong2021functional}, we may deduce the following from \eqref{eqlater0}, \eqref{eqlater1}, and Assumptions~\ref{assum1} and \ref{assum1c}:  
	\begin{equation}
		\frac{1}{\sqrt{T}} \sum_{t=1}^T \frac{\zeta_t}{\sqrt{\theta_{\KK_S} (\zeta)}} \to_d N(0, \widetilde{C}_{u}). \label{eqradd01}
	\end{equation}
	Combining \eqref{eqradd01} with \eqref{eqlater0}, we find that $	\sqrt{T/\theta_{\KK_S}(\zeta)}(\widehat{f^S_{\kappa}}(\zeta)-	f^S\widehat{\PP}_\kappa^{\KK_S}(\zeta)) \to_d N(0, \widetilde{C}_{u})$. In addition, since $	\sqrt{T/\theta_{\KK_S}(\zeta)}(\widehat{f^N_{\kappa}}(\zeta) -	{f}\widehat{\PP}_\kappa^N(\zeta)) = o_p(1)$ (see \eqref{eqaddresult1}), we deduce the following desired result:
	\begin{align}
		\sqrt{T/\theta_{\KK_S}(\zeta)}(\hat{f}_{\kappa}(\zeta) -	{f}\widehat{\PP}_\kappa^{\KK}(\zeta))   &=  	\sqrt{T/\theta_{\KK_S}(\zeta)}(\widehat{f^N_{\kappa}}(\zeta)+\widehat{f^S_{\kappa}}(\zeta)  -	{f}\widehat{\PP}_\kappa^{N}(\zeta)-{f}\widehat{\PP}_\kappa^{\KK_S}(\zeta))  \notag  \\ 
		&=\sqrt{T/\theta_{\KK_S}(\zeta)}(\widehat{f^S_{\kappa}}(\zeta)  -	{f}^S\widehat{\PP}_\kappa^{\KK_S}(\zeta)-	{f}\PP^N\widehat{\PP}_\kappa^{\KK_S}(\zeta))  + o_p(1)  
		\notag \\ &=\sqrt{T/\theta_{\KK_S}(\zeta)}(\widehat{f^S_{\kappa}}(\zeta)  -	{f}^S\widehat{\PP}_\kappa^{\KK_S}(\zeta)) + o_p(1)   \to_d N(0,\widetilde C_{u}), \label{eqconver1}
	\end{align}
	where, for the last convergence result, we used the fact that $\sqrt{T/\theta_{\KK_S}(\zeta)}{f}\PP^N\widehat{\PP}_\kappa^{\KK_S}(\zeta) = O_p(\sqrt{\theta_{\KK_S}(\zeta)/T}) = o_p(1)$ since $\PP^N\widehat{\PP}_\kappa^{\KK_S}=O_p(T^{-1})$ (see Theorem \ref{thm0}). 
	
	We now discuss the consistency and asymptotic normality of our estimators in the case where repetition of eigenvalues of $D_{\kappa}^S$ is allowed. Let $\PP_\kappa^{\KK_S} = \sum_{j=1}^{\KK_S} \PPi\edex{j}{D_{\kappa}^S}$. If the conditions in Remark \ref{remadd} hold, we then deduce from Lemmas 3.1-3.2  (see also the proof of Theorem 3.1) of \cite{REIMHERR201562} that 
	\begin{equation} \label{eqthm02a}
		\left\|\widehat{\PP}^S_{\kappa} (\widehat{D}_\kappa)_{\KK}^{-1} \widehat{\PP}^S_{\kappa} - 
		(D_{\kappa}^S)_{\KK_S}^{-1} \right\|_{\opnorm} = O_p\left( \frac{\KK_S^{1/2} \|\widehat{\PP}^S_{\kappa}  \widehat{D}_\kappa \widehat{\PP}^S_{\kappa} - D_{\kappa}^S\|_{\opnorm}} {\llambda\edex{\KK_S}{D_{\kappa}^S}(\llambda\edex{\KK_S}{D_{\kappa}^S}- \llambda\edex{\KK_S+1}{D_{\kappa}^S}}) \right) = o_p(1),
	\end{equation}
	and from the facts that $\widehat{\PP}_\kappa^{\KK_S}=\widehat{\PP}^S_{\kappa} (\widehat{D}_\kappa)_{\KK}^{-1} \widehat{\PP}^S_{\kappa}\widehat{\PP}^S_{\kappa} \widehat{D}_\kappa \widehat{\PP}^S_{\kappa}$, ${\PP}_\kappa^{\KK_S} = (D_{\kappa}^S)_{\KK_S}^{-1}D_{\kappa}^S$ and $\|\widehat{\PP}^S_{\kappa}  \widehat{D}_\kappa \widehat{\PP}^S_{\kappa} - D_{\kappa}^S\|_{\opnorm}=O_p(T^{-1/2})$,  
	we also find that 
	\begin{align}
		\|\widehat{\PP}_\kappa^{\KK_S} - {\PP}_\kappa^{\KK_S}\| = O_p\left( \left\|\widehat{\PP}^S_{\kappa} (\widehat{D}_\kappa)_{\KK}^{-1} \widehat{\PP}^S_{\kappa} - 
		(D_{\kappa}^S)_{\KK_S}^{-1} \right\|_{\opnorm}\right) = o_p(1). \label{eqthm03}
	\end{align}
	Combining \eqref{eqthm01}, \eqref{eqthm02}, \eqref{eqthm03} and the fact that $\|T^{-1}\sum_{t=1}^T \widehat{\QQ}^{S}_{\kappa} \tilde{x}_{t-\kappa} \otimes (\tilde{u}_t + \widehat{\mathcal W}_t)) \|_{\opnorm}= O_p(T^{-1/2})$, we find that $\|\widehat{f^S_{\kappa}}- f^S \PP^{\KK_S}_{\kappa}\|_{\opnorm} \to_p 0$. Thus, it only remains to show that $\|f^S - f^S{\PP}_\kappa^{\KK_S}\|_{\opnorm} \to_p 0$ to establish the consistency under the employed conditions. Note that $\|f^S - f^S{\PP}_\kappa^{\KK_S}\|_{\opnorm}^2 =\|f^S (I- {\PP}_\kappa^{\KK_S})\|_{\opnorm}^2\leq \sum_{j=\KK_S+1}^\infty \|f(\varv\edex{j}{D_{\kappa}^S})\|^2$ and $\sum_{j=1}^\infty \|f(\varv\edex{j}{D_{\kappa}^S})\|^2 < \infty$ (see Assumption \ref{assum3}). Since $\KK_S \to_p \infty$, $\sum_{j=\KK_S+1}^\infty \|f(\varv\edex{j}{D_{\kappa}^S})\|^2 \to_p 0$ and thus $\|f^S - f^S{\PP}_\kappa^{\KK_S}\|_{\opnorm} \to_p 0$ as desired.  To show the asymptotic normality result \eqref{eqthm2} holds under the conditions in Remark \ref{remadd},  we first observe that $\sqrt{T/\theta_{\KK_S}(\zeta)}(\widehat{f^S_{\kappa}}(\zeta)-	f^S\widehat{\PP}_\kappa^{\KK_S}(\zeta))$ can also be written as \eqref{eqlater00} in this case. We note 
	the following, which can be directly deduced from  \eqref{eqthm02a}: $\|\widehat{\QQ}^S_{\kappa}  \widehat{C}_\kappa \widehat{\PP}^S_{\kappa}  (\widehat{D}_\kappa)_\KK^{-1} \widehat{\PP}^S_{\kappa}(\zeta) - \PP^S C_{\kappa}^S(D_\kappa^S)_{\KK_S}^{-1}(\zeta)\|=o_p(1)$. Thus, under either \eqref{eqcondition_a} or the conditions in Remark \ref{remadd}, \eqref{eqlater0} holds. The rest of the proof is identical, and \eqref{eqthm2} follows directly from the previously used arguments; details are omitted.
\end{proofs}

\begin{proofs}[Proof of Theorem \ref{prop1}] 
	Observe that 
	\begin{align}
		\sqrt{T/\theta_{\KK_S}(\zeta)}(\hat{f}_{\kappa}(\zeta)-f(\zeta)) = \sqrt{T/\theta_{\KK_S}(\zeta)}(\hat{f}_{\kappa}(\zeta)-f\widehat{\PP}^{\KK}_\kappa(\zeta))  + \sqrt{T/\theta_{\KK_S}(\zeta)} f(\widehat{\PP}_\kappa^{\KK}-I)(\zeta).  \label{eqradd02}
	\end{align}
	Due to the result given in \eqref{eqthm2}, it suffices to show that the second term in the right hand side of \eqref{eqradd02} is $o_p(1)$.
	From Theorems \ref{thm1} and \ref{thm2} and the fact that $\widehat{\PP}_\kappa^{\KK_S}\PP^N = O_p(T^{-1})$, we find that 
	\begin{align}
		\sqrt{T/\theta_{\KK_S}(\zeta)} f(\widehat{\PP}_\kappa^{\KK}-I)(\zeta) &= \sqrt{T/\theta_{\KK_S}(\zeta)} (f\widehat{\PP}_\kappa^N(\zeta) + f\widehat{\PP}_\kappa^{\KK_S}(\zeta)  - f^N(\zeta) -f^S(\zeta)) \notag \\ &= \sqrt{T/\theta_{\KK_S}(\zeta)}(f\widehat{\PP}_\kappa^{\KK_S}(\zeta) -f^S(\zeta)) + o_p(1) \notag\\&= \sqrt{T/\theta_{\KK_S}(\zeta)}f(\widehat{\PP}_\kappa^{\KK_S}-I)\PP^S(\zeta) + o_p(1). \notag 
	\end{align}
    Let $\varv^s\edex{j}{D^S_{\kappa}} = \sgn\{\langle \varv\edex{j}{D^S_{\kappa}}, \varv\edex{j}{\widehat{D}_{\kappa}^S} \rangle\} \varv\edex{j}{D^S_{\kappa}}$.
	We note that  $\| \widehat{\QQ}^S_{\kappa}  \widehat{C}_\kappa \widehat{\PP}^S_{\kappa}- {C}_\kappa^S \|_{\opnorm} = \| \widehat{C}_\kappa \widehat{\PP}^S_{\kappa}- {C}_\kappa^S \|_{\opnorm} = O_p(T^{-1/2})$ and thus $\|\widehat{D}_{\kappa}^S-D^S_{\kappa}\|_{\opnorm}=\|\widehat{\PP}^S_{\kappa}\widehat{D}_{\kappa}\widehat{\PP}^S_{\kappa}-D^S_{\kappa}\|_{\opnorm} = O_p(T^{-1/2})$ (see our proof of Theorem \ref{thm2}). Note also that 
	\begin{align}
		T\mathbb{E}[\langle (\PP^S  \widehat{C}_\kappa \PP^S- {C}_\kappa^S)\varv^s\edex{j}{D^S_{\kappa}},\varv^s\edex{\ell}{E^S_{\kappa}} \rangle^2] &\leq \sum_{s=0}^T \mathbb{E}[\varpi_t(j,\ell)\varpi_{t-s}(j,\ell)] 
	\notag	\\&\leq O(1)\mathbb{E}[\langle \PP^Sx_t, \varv^s\edex{j}{D^S_{\kappa}} \rangle^2 \langle \PP^Sx_{t-\kappa}, \varv^s\edex{\ell}{E^S_{\kappa}} \rangle^2] \notag\\& \leq O(\llambda\edex{j}{D^S_{\kappa}}\llambda\edex{\ell}{D^S_{\kappa}}), \label{eqaddadd1}
	\end{align}
	where the last equality follows from the Cauchy-Schwarz inequality, stationarity of $\PP^Sx_t$ and Assumption \ref{assum2add}\ref{assum2a} and the fact that $\llambda\edex{j}{D^S_{\kappa}}=\llambda\edex{j}{E^S_{\kappa}}$.  From \eqref{eqaddadd1} and the facts that  $\widehat{\PP}^{S}_{\kappa} \to_p \PP^S$ and $\widehat{\QQ}^{S}_{\kappa} \to_p \PP^S$, we find that $\langle (\widehat{\QQ}^S_{\kappa}  \widehat{C}_\kappa \widehat{\PP}^S_{\kappa}- {C}_\kappa^S) \varv^s\edex{j}{D^S_{\kappa}}, \varv^s\edex{\ell}{E^S_{\kappa}} \rangle^2 = O_p(\llambda\edex{j}{D^S_{\kappa}}\llambda\edex{\ell}{D^S_{\kappa}})$. Using this result and the employed conditions, the following can be shown:
    \begin{align}
		\|\varv\edex{j}{\widehat{D}_{\kappa}^S}-\varv^s\edex{j}{D^S_{\kappa}}\|^2 &= O_p(j^2 T^{-1}), \label{ssequation1} \\
		\|f(\varv\edex{j}{\widehat{D}_{\kappa}^S}-\varv^s\edex{j}{D^S_{\kappa}})\|^2 &= O_p(T^{-1})O_p(j^{2-2\varsigma} + j^{\rho+2-2\varsigma}),\label{ssequation2} \\
		\langle \varv\edex{j}{\widehat{D}_{\kappa}^S}-\varv^s\edex{j}{D^S_{\kappa}}, \PP^S(\zeta) \rangle^2 &= O_p(T^{-1}) j^{-2\delta_{\zeta} +2} + O_p(T^{-1}) j^{-2\delta_\zeta +2 + \rho}. \label{ssequation3} 
	\end{align} 
	The derivation of these results follows by arguments parallel to those used in the proofs of Theorems 3 and 4 of \cite{seong2021functional}, using their Lemma S1 (see, in particular, equations (S2.15), (S2.16), and (S2.33) therein).  We also consider the following decomposition of $f(\widehat{\PP}_{\kappa}^{\KK_S}-I)\PP^S(\zeta)$ as in the proof of Theorem 4 of  \cite{seong2021functional}:
	\begin{equation*}
		f(\widehat{\PP}_{\kappa}^{\KK_S}-I)\PP^S(\zeta) = A_1+A_2+A_3+A_4,
	\end{equation*}
	where 
    \begin{align}
		&A_1 = \sum_{j=1}^{\KK_S} \langle \varv\edex{j}{\widehat{D}_{\kappa}^S}-\varv^s\edex{j}{D^S_{\kappa}}, \PP^S(\zeta) \rangle f(\varv\edex{j}{\widehat{D}_{\kappa}^S}-\varv^s\edex{j}{D^S_{\kappa}}),   \notag \\  	
		&A_2 = \sum_{j=1}^{\KK_S} \langle \varv^s\edex{j}{D^S_{\kappa}}, \PP^S(\zeta) \rangle f(\varv\edex{j}{\widehat{D}_{\kappa}^S}-\varv^s\edex{j}{D^S_{\kappa}}),  \notag \\
		&A_3 = \sum_{j=1}^{\KK_S}\langle \varv\edex{j}{\widehat{D}_{\kappa}^S}-\varv^s\edex{j}{D^S_{\kappa}}, \PP^S(\zeta) \rangle f(\varv^s\edex{j}{D^S_{\kappa}}), \notag \\
		&A_4 = f(\PP_{\kappa}^{\KK_S} - I) \PP^S(\zeta),\notag
	\end{align}
    and $\PP_\kappa^{\KK_S} = \sum_{j=1}^{\KK_S} \PPi\edex{j}{D_{\kappa}^S}$, as defined in our proof of Theorem \ref{thm2}.
	Using \eqref{ssequation1}-\eqref{ssequation3}, we find that
    \begin{align}
		\|A_1\| &\leq \sum_{j=1}^{\KK_S} |\langle \varv\edex{j}{\widehat{D}_{\kappa}^S}-\varv^s\edex{j}{D^S_{\kappa}}, \PP^S\zeta \rangle| \|f(\varv\edex{j}{\widehat{D}_{\kappa}^S}-\varv^s\edex{j}{D^S_{\kappa}})\|  = O_p(T^{-1}) \sum_{j=1}^{\KK_S} j^{\rho-\varsigma-\delta_{\zeta}+2}\notag \\
		& \leq O_p(T^{-1} \KK_S^{\rho/2}) \sum_{j=1}^{\KK_S} j^{\rho/2-\varsigma-\delta_{\zeta}+1}. \notag
	\end{align}
	Similarly, 
    \begin{align}	
    \|A_2\| \leq  \sum_{j=1}^{\KK_S} |\langle \varv^s\edex{j}{D^S_{\kappa}}, \PP^S(\zeta) \rangle| \|f(\varv\edex{j}{\widehat{D}_{\kappa}^S}-\varv^s\edex{j}{D^S_{\kappa}})\| = O_p(T^{-1/2}) \sum_{j=1}^{\KK_S} j^{\rho/2-\varsigma-\delta_{\zeta}+1} \notag
    \end{align}
	and
    \begin{align}
		\|A_3\| = \sum_{j=1}^{\KK_S}|\langle \varv\edex{j}{\widehat{D}_{\kappa}^S}-\varv^s\edex{j}{D^S_{\kappa}}, \PP^S(\zeta) \rangle| \|f(\varv^s\edex{j}{D^S_{\kappa}})\| = O_p(T^{-1/2}) \sum_{j=1}^{\KK_S} j^{\rho/2-\varsigma-\delta_{\zeta}+1}. \notag
	\end{align}
	Note that $\rho/2+ 2 < \varsigma+\delta_{\zeta}$, under which  $\sum_{j=1}^{\KK_S} j^{\rho/2-\varsigma-\delta_{\zeta}+1}$ is summable and thus $O_p(1)$. Under our assumptions, it can also be shown that $\KK_S \leq (1 + o_p(1))\alpha^{-1/\rho}$ (see (S2.11) of \citealp{seong2021functional}). This result, along with the fact that $\alpha^{-1}T^{-1} \to 0$, implies that $T^{-1/2}\KK_S^{\rho/2} \to_p 0$. Combining all these results, we find that $\sqrt{T}(\|A_1\|+\|A_2\|+\|A_3\|) = O_p(1)$. Moreover, note that 
	\begin{align}
		\|	A_4\|^2 = \|f(\PP_{\kappa}^{\KK_S} - I) (\zeta)\|^2=  \sum_{j=\KK_S+1}^\infty \|\langle \varv^s\edex{j}{D_{\kappa}^S},\zeta \rangle f(\varv^s\edex{j}{D_{\kappa}^S})|^2 \leq   \sum_{j=\KK_S+1}^\infty j^{-2\delta_{\zeta} - 2\varsigma} \leq \alpha^{(2\delta_{\zeta} + 2\varsigma-1)/\rho},   \notag
	\end{align}
	where the last inequality follows from that the Euler-Maclaurin summation formula for
	the Riemann zeta-function (see e.g., (5.6) of \citealp{ibukiyama2014euler}) and the fact that $\KK_S \leq (1 + o_p(1))\alpha^{-1/\rho}$.
	This implies that  $\sqrt{T}\|A_4\| \leq  O(\sqrt{T \alpha^{(2\delta_{\zeta}+2\varsigma-1)/\rho}}) = O_p(1)$. 
	
	We have shown that $\sqrt{T/\theta_{\KK_S}(\zeta)}(\|A_1\|+\|A_2\|+\|A_3\| + \|A_4\|) = o_p(1)$ (since $\theta_{\KK_S}(\zeta) \to_p \infty$), from which the desired result immediately follows. 
\end{proofs}

\subsection{Proofs of the results in Section \ref{ap_sec_sup}}  \label{ap_sec_sup_proof}

\begin{proofs}[Proof of Corollary \ref{cor1}]
	Note that 
	\begin{equation*}
		\widehat{\theta}_{\KK_S}(\zeta) = \langle \zeta,  (\widehat{D}_\kappa^S)_{\KK_S}^{-1}   (\widehat C_{\kappa}^S)^\ast \widehat{C}_{0}^S \widehat C_{\kappa}^S(\widehat D_\kappa^S)_{\KK_S}^{-1} (\zeta) \rangle  = \langle   \widehat C_{\kappa}^S(\widehat{D}_\kappa^S)_{\KK_S}^{-1}  (\zeta),  \widehat{C}_{0}^S \widehat C_{\kappa}^S(\widehat D_\kappa^S)_{\KK_S}^{-1} (\zeta) \rangle.
	\end{equation*}
	Under either of \eqref{eqcondition_a} or the conditions in Remark \ref{remadd},  we have $\|\widehat{\QQ}^S_{\kappa}  \widehat{C}_\kappa \widehat{\PP}^S_{\kappa}  (\widehat{D}_\kappa)_\KK^{-1} \widehat{\PP}^S_{\kappa}(\zeta)  - \PP^S C_{\kappa}^S(D_\kappa^S)_{\KK_S}^{-1}(\zeta)\|=o_p(1)$. Moreover, note that  
	$\widehat C_{\kappa}^S(\widehat D_\kappa^S)_{\KK_S}^{-1} = \widehat{\QQ}^S_{\kappa} \widehat C_{\kappa}^S\widehat{\PP}^S_{\kappa} (\widehat{D}_\kappa^S)_{\KK_S}^{-1}\widehat{\PP}^S_{\kappa}$, $\widehat{\PP}^S_{\kappa}-\PP^S = O_p(T^{-1})$ 
	and $\|\widehat{C}_{0}^S-\tilde{C}_{0}^S\|_{\opnorm} \to_p 0$ holds under Assumptions \ref{assum1} and \ref{assum1c}. Therefore,  $\|\widehat C_{\kappa}^S(\widehat{D}_\kappa^S)_{\KK_S}^{-1}  (\zeta) - \PP^S C_{\kappa}^S(D_\kappa^S)_{\KK_S}^{-1}(\zeta)\|=o_p(1)$  and  $\|\widehat{C}_{0}^S  \widehat C_{\kappa}^S(\widehat{D}_\kappa^S)_{\KK_S}^{-1}  (\zeta) - \tilde{C}_{0}^S C_{\kappa}^S(D_\kappa^S)_{\KK_S}^{-1}(\zeta)\|=o_p(1)$. These results imply that $\widehat{\theta}_{\KK_S}(\zeta)/{\theta}_{\KK_S}(\zeta) \to_p 1$ because $\PP^S C_{\kappa}^S=C_{\kappa}^S$ and $\theta_{\KK_S}(\zeta)$ can be written as	\begin{equation}
		\theta_{\KK_S}(\zeta) = \langle C_{\kappa}^S (D_\kappa^S)_{\KK_S}^{-1}(\zeta) ,     \widetilde{C}_{0}^S C_{\kappa}^S(D_\kappa^S)_{\KK_S}^{-1} (\zeta) \rangle. \notag \label{eqtheta}
	\end{equation}
   Then the desired result \ref{cor1a} follows immediately. Moreover, by the consistency result $\|\widehat{f}_{\kappa}-f\|_{\opnorm} \to_p 0$ (Theorems \ref{thm1} and \ref{thm2}), we obtain $\|\widehat{C}_u-\widetilde{C}_u\|_{\opnorm} \to_p 0$, which establishes \ref{cor1b}.
\end{proofs}

\begin{proofs}[Proof of Corollary \ref{thm2app}]
	The proof is a straightforward adaptation of the existing proofs of Theorems \ref{thm0}–\ref{prop1}.
	Under Assumption \ref{assumapp}\ref{assumapp1}, with moderate modifications of the arguments used in our proofs of Theorems \ref{thm0} and \ref{thm1}, as well as those in \cite{seo2020functional} concerning the FPCA of cointegrated functional time series with deterministic components, the following result can be readily deduced:
	\begin{equation*}
		\|\widehat{\PP}_{c,\kappa}^N - \PP^N\|_{\opnorm} = O_p(T^{-1}), \quad 	\|\widehat{\PP}_{c,\kappa}^S - \PP^S\|_{\opnorm} = O_p(T^{-1})\quad \text{and} \quad \|\widehat{f}_{c,\kappa}^N - f^N\|_{\opnorm} = O_p(T^{-1}).
	\end{equation*}
	Moreover, from similar arguments used in our proof of Theorem \ref{thm2}, 
	the following--similar to \eqref{eqlater0}--can be deduced under Assumptions \ref{assumapp}\ref{assumapp1}–\ref{assumapp2}:
	\begin{equation*} 
		\sqrt{T/\theta_{c,\KK_S}(\zeta)}(\widehat{f}^S_{c,\kappa}(\zeta)-	f^S\widehat{\PP}_{c,\kappa}^{\KK_S}(\zeta)) =      \left(\frac{1}{\sqrt{T \theta_{c,\KK_S}(\zeta)}}\sum_{t=1}^T  \tilde{x}_{c,t-\kappa}^{S} \otimes \tilde{u}_{t} \right)C_{c,\kappa}^S(D_\kappa^S)_{c,\KK_S}^{-1}(\zeta) + o_p(1),
	\end{equation*}	
	where $ \tilde{x}_{c,t-\kappa}^{S} = {\PP}^{S} \tilde{x}_{c,t-\kappa}- \mu_{x,S}$.
	As in our proof of Theorem \ref{thm2}, we define $\zeta_{c,t} = [ \tilde{x}_{c,t-\kappa}^{S} \otimes \tilde{u}_t]  C_{c,\kappa}^S(D_{c,\kappa}^S)_{\KK_S}^{-1}(\zeta) = \langle \tilde{x}_{c,t-\kappa}^{S},  C_{c,\kappa}^S(D_{c,\kappa}^S)_{\KK_S}^{-1} (\zeta) \rangle\tilde{u}_t$, and deduce that 	$\frac{1}{\sqrt{T}} \sum_{t=1}^T \frac{\zeta_{c,t}}{\sqrt{\theta_{c,\KK_S} (\zeta)}} \to_d N(0, \widetilde{C}_{u})$. Then, as in \eqref{eqconver1}, we find that
	\begin{align*}
		\sqrt{T/\theta_{c,\KK_S}(\zeta)}(\hat{f}_{c,\kappa}(\zeta) -	{f}\widehat{\PP}_{c,\kappa}^{\KK}(\zeta))   
		&=\sqrt{T/\theta_{c,\KK_S}(\zeta)}(\widehat{f^S_{\kappa}}(\zeta)  -	{f}^S\widehat{\PP}_{c,\kappa}^{\KK_S}(\zeta)) + o_p(1) \to_d N(0,\widetilde C_{u}). 
	\end{align*}
	The extension from the above to the desired result, under the additional requirement Assumption \ref{assumapp}\ref{assumapp3}, follows in a similar manner--as in the proof of Theorem \ref{prop1}--from the asymptotic results provided by \cite{seong2021functional}, which can be readily extended to the case with an intercept, as discussed in their paper. Accordingly, the remainder of the proof is omitted.
\end{proofs}

\begin{proofs}[Proof of Lemma \ref{lemapp1}]
	We note that 
	\begin{equation} \label{eqlem01}
		T^{-1} \widehat{C}_0  = \frac{1}{T^2}\sum_{t=1}^T x_t\otimes x_t + \frac{1}{T^2}\sum_{t=1}^T x_t\otimes e_t + \frac{1}{T^2}\sum_{t=1}^T e_t\otimes x_t   +  \frac{1}{T^2}\sum_{t=1}^T e_t\otimes e_t. 
	\end{equation}
In \eqref{eqlem01}, Lemma 3.1 of \cite{Chang2016152} together with Assumptions \ref{assum1} and \ref{assum1c} implies that the latter three terms are negligible for large $T$.  
	If we let $X_t = \sum_{s=1}^t x_s$ and $\mathfrak E_t = \sum_{s=1}^t e_s$, we find that 
	\begin{equation} \label{eqlem02}
		T^{-3} \widehat{\mathcal K}_0  = \frac{1}{T^4}\sum_{t=1}^T X_t\otimes X_t + \frac{1}{T^4}\sum_{t=1}^T X_t\otimes \mathfrak E_t + \frac{1}{T^4}\sum_{t=1}^T \mathfrak E_t\otimes X_t   +  \frac{1}{T^4}\sum_{t=1}^T \mathfrak E_t\otimes \mathfrak E_t. 
	\end{equation}
	As shown in the proof of Lemma 1 of \cite{NSS}, $\|T^{-3/2}X_t\|_{\opnorm} = O_p(1)$ and  $\|T^{-1/2}\mathfrak E_t\|_{\opnorm} = O_p(1)$. This implies that the latter three terms in \eqref{eqlem02} are all negligible for large $T$.
\end{proofs}

\begin{proofs}[Proof of Proposition \ref{propapp1}]
	As shown by \cite{Chang2016152}, $\sum_{j=1}^{d_0}\PPi\edex{j}{\widehat{C}_0}$ converges to the orthogonal projection onto $\mathcal H^N$ and $\sum_{j=d_0+1}^{\ell}\PPi\edex{j}{\widehat{C}_0}$ converges to an orthogonal projection onto a subspace of $\mathcal H^S$. We then find that $(\sum_{j=d_0+1}^{\ell}\PPi\edex{j}{\widehat{C}_0}) \widehat{C}_0 (\sum_{j=d_0+1}^{\ell}\PPi\edex{j}{\widehat{C}_0})$ converges to a well defined nonrandom operator which is positive definite on $\ran (\sum_{j=1}^{\ell-d_0} \PPi\edex{j}{\widetilde{C}_0^S})$ which is included in $\mathcal H^S$. Using this result with our Lemma \ref{lemapp1} and the asymptotic results given by  \cite{NSS} (see Theorem 1 and Remark 4 in their paper), the desired result follows.   
\end{proofs}



\bigskip
\section{Functional time series of the regional growth rate}\label{AP_WGRP}
\subsection{Spatial distribution of the regional growth rate}\label{AP_WGRP_Dist}
\noindent This section describes the procedure of spatial disaggregation to obtain the spatial distribution of the regional growth rates. We use the real Gross Regional Product (GRP) data of \cite{leonie_wenz_2023_7573249} for 1960–2019 and the real GDP in millions of 2021 international dollars (converted using Purchasing Power Parities) from the Conference Board Total Economy Database (TED) for 1950–2019.\footnote{The Conference Board Total Economy Database™ (April 2022) — Output, Labor and Labor Productivity, 1950–2022, downloaded from https://www.conference-board.org/data/economydatabase/total-economy-database-productivity
 on April 13, 2023.}

\begin{table}[t]
	\centering
	\scriptsize
	\caption{Country list and number of grids for each country}
	\begin{tabular}{rrrrrrrr}
		\hline
		\multicolumn{8}{c}{Full-sample Available Country List (1950-2019): 59 Nations \& 1384 Grids} \\
		\hline
		Albania & Argentina & Australia & Austria & Belgium & Bolivia & Brazil & Bulgaria \\
		12    & 17    & 8     & 9     & 3     & 9     & 27    & 28 \\
		\hline
		Canada & Chile & China & Colombia & Denmark & Ecuador & Egypt & Ethiopia \\
		14    & 15    & 31    & 32    & 5     & 24    & 27    & 3 \\
		\hline
		Finland & France & Germany & Greece & Guatemala & Hungary & India & Indonesia \\
		5     & 13    & 16    & 7     & 21    & 20    & 33    & 32 \\
		\hline
		Iran & Ireland & Italy & Japan & Kenya & Malaysia & Mexico & Morocco \\
		31    & 26    & 20    & 47    & 48    & 15    & 32    & 24 \\
		\hline
		Mozambique & Netherlands & New Zealand & Nigeria & Norway & Pakistan & Paraguay & Peru \\
		11    & 12    & 15    & 22    & 19    & 4     & 17    & 24 \\
		\hline
		Philippines	& Poland & Portugal	& Romania & Russia & South Africa & South Korea & Spain \\
		17    & 16    & 25    & 42    & 79    & 9     & 17    & 18 \\
		\hline
		Sri Lanka & Sweden & Switzerland & Tanzania & Thailand & Turkey & UAE & UK \\
		9     & 21    & 26    & 25    & 77    & 81    & 7     & 4 \\
		\hline
		USA   & Uruguay & Vietnam &       &       &       &       &  \\
		51    & 19    & 63    &       &       &       &       &  \\
		\hline
		\multicolumn{8}{c}{Sub-sample Available Country List (1970-2019): 17 Nations \& 212 Grids} \\
		\hline
		Azerbaijan & Belarus & Uzbekistan & Croatia & Czech Republic & Estonia & Georgia & Kazakhstan \\
		10    & 6     & 14    & 21    & 14    & 15    & 11    & 14 \\
		\hline
		Kyrgyzstan & Latvia & Lithuania & Macedonia & Serbia & Slovakia & Slovenia & Ukraine \\
		9     & 5     & 10    & 8     & 25    & 8     & 12    & 27 \\
		\hline
		\multicolumn{2}{l}{Bosnia\&Herzegovina}    &       &       &       &       &       &  \\
		3   &       &       &       &       &       &       &  \\
		\hline
		\multicolumn{8}{c}{Omitted Country List due to Data Unavailability: 7 Nations \& 65 Grids} \\
		\hline
		Bahamas & Honduras & Laos  & Mongolia & Nepal & Panama & \multicolumn{2}{l}{Netherlands Antilles} \\
		3     & 18    & 2     & 22    & 7     &  10    & 3    &  \\
		\hline
	\end{tabular}%
	\label{tab:Country_list}%
\end{table}%

\begin{figure}[t]
	\begin{center}
		\includegraphics[height=0.3\textwidth, width=0.8\textwidth]{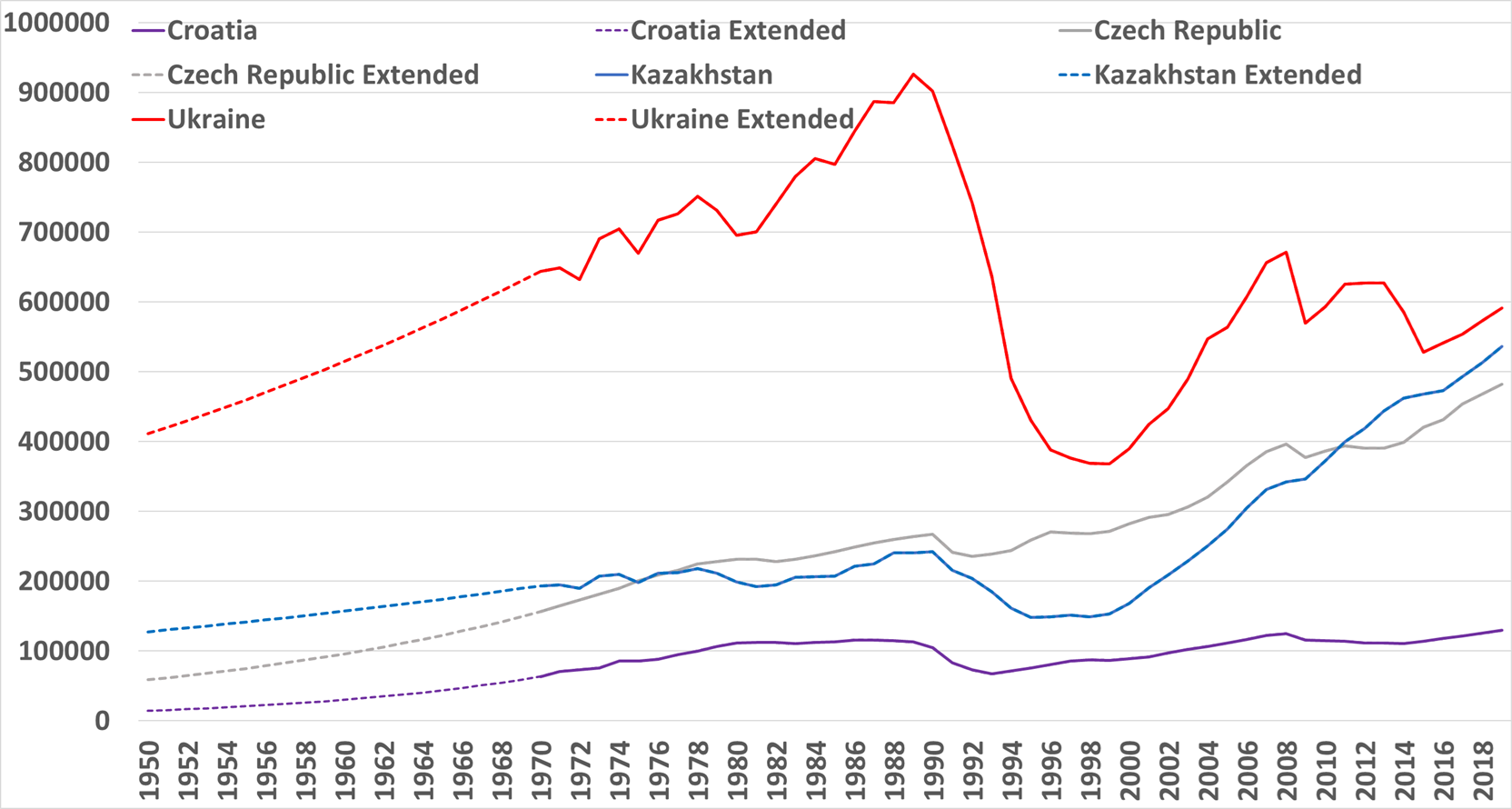}
		\caption{Extended TED-based real GDP level for four selected countries}
		\label{Fig:Extend_TED}
	\end{center}
\end{figure}

\indent We begin by aligning regional real growth rates from \cite{leonie_wenz_2023_7573249} with country-level real GDP data from the TED. The set of countries included in the analysis is provided in Table \ref{tab:Country_list}. For 17 countries with GDP data available only from 1970 onward—primarily in Eastern Europe—we extend the TED series back to 1950 by extrapolating with the average annual growth rate observed between 1970 and 1975. Figure \ref{Fig:Extend_TED} illustrates the extrapolated TED-based real GDP levels for four selected countries. Next, we construct regional income shares by aggregating regional product per capita with regional population data from \cite{leonie_wenz_2023_7573249}, normalized within each country–year. This procedure is feasible only when complete information on both per capita product and population is available for all regions in a given country–year; otherwise, the observations for that country–year are excluded.\footnote{We use the regional product per capita level (variable name: grp\_pc\_usd\_2015) rather than the regional product level to calculate the regional income shares because population data at the regional level are much less available for calculating the GDP at the country level. For the target variable, similarly, we consider the regional growth rate instead of the regional product per capita growth rate due to the same data limitations and endogeneity of population (climate change-induced mortality rate).} 
To address missing values, we apply interpolation and extrapolation procedures. When fewer than five regional observations are missing within a country–year, we implement limited cross-sectional (horizontal) interpolation, primarily at the beginning or end of a missing sequence. This method exploits the historical ratio of the omitted region’s income to the average income of other regions in the same country. When both endpoints of a missing time span are available, we use log-linear interpolation to impute the intermediate values.

\indent Having constructed the maximum possible coverage of regional income shares—and thereby GRP levels in a balanced panel (rectangular) form—we extend the series from 1951 to 2019 using two approaches, depending on sample length. Because regional income shares are relatively stable and tend to evolve gradually over time rather than exhibiting abrupt fluctuations, these smooth extrapolation methods are particularly appropriate. For regions with fewer than 15 available years, we extrapolate backward using the average ratio of the previous five years and forward using the average of the most recent five years (25 countries, 403 regions). For regions with 15 or more years of data, we fit the Nelson–Siegel two-factor model of \cite{diebold2006forecasting} to capture the gradual, nonlinear dynamics of income shares (56 countries, 1,238 regions).

\begin{figure}[t]
	\begin{center}
            \includegraphics[height=0.4\textwidth, width=0.99\textwidth]{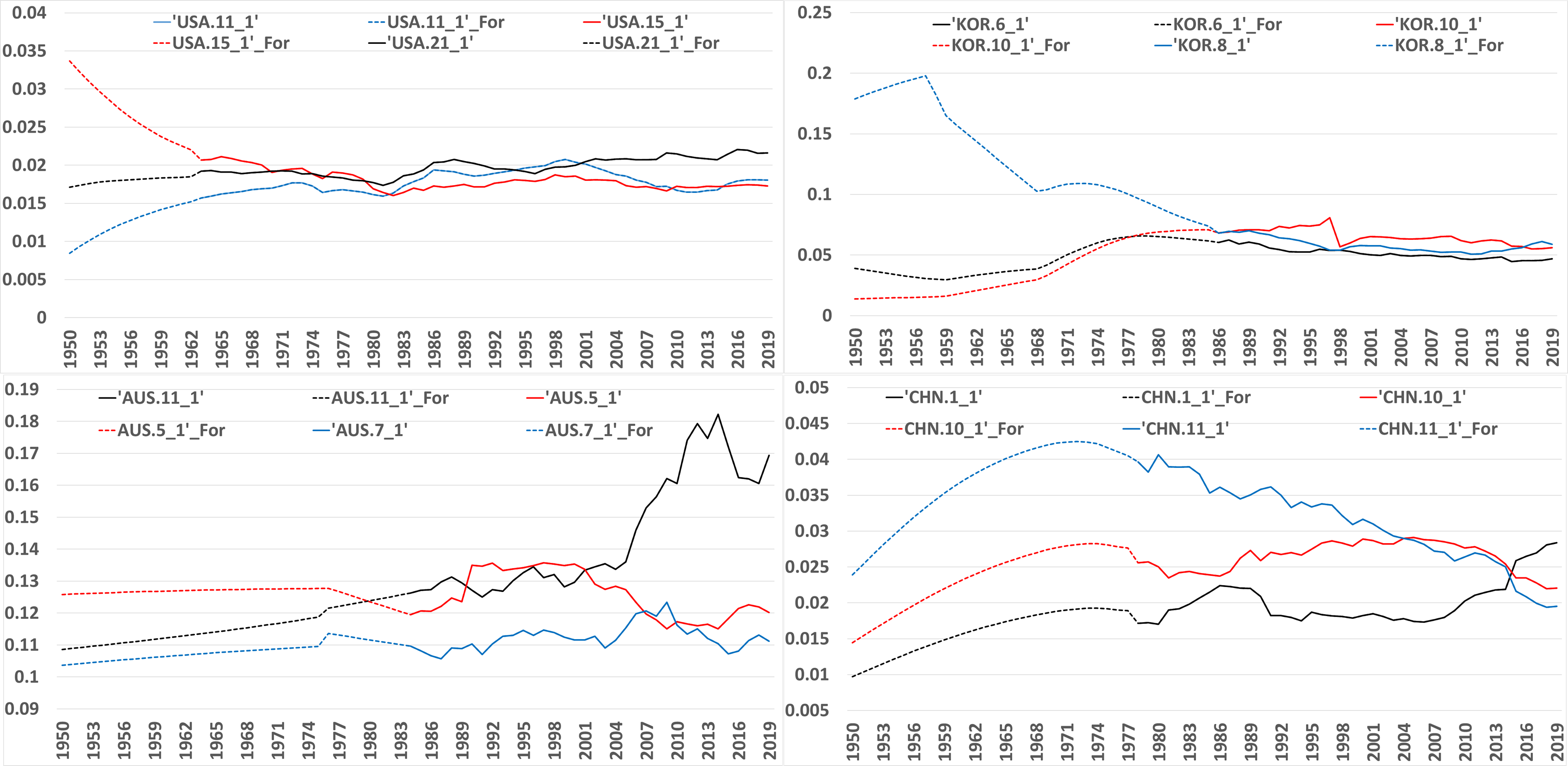}
		\caption{Extended regional income shares based on the Nelson-Siegel nonlinear function for four selected countries (U.S., South Korea, China, and Australia)}
		\label{Fig:Extend_ratio}
	\end{center}
\end{figure}

\indent Because small-sample settings render quadratic trend models inefficient under least squares estimation, we impose a parsimonious nonlinear structure that improves estimator precision while remaining consistent with the gradual evolution of regional income ratios. This strategy, analogous to the Nelson–Siegel approach originally applied to U.S. bond yields, offers an effective means of addressing the issue. After renormalizing the sum of income shares to unity for each country–year, the reconstructed GRP series yield a balanced panel. Taking first differences of the logarithm of GRP, the final dataset covers 1951–2019 and consists of 1,576 regional growth observations. Figure \ref{Fig:Extend_ratio} displays the extrapolated income share dynamics for four selected countries (the U.S., South Korea, China, and Australia).

\begin{figure}[t]
	\begin{center}
		\includegraphics[height=0.3\textwidth, width=0.4\textwidth]{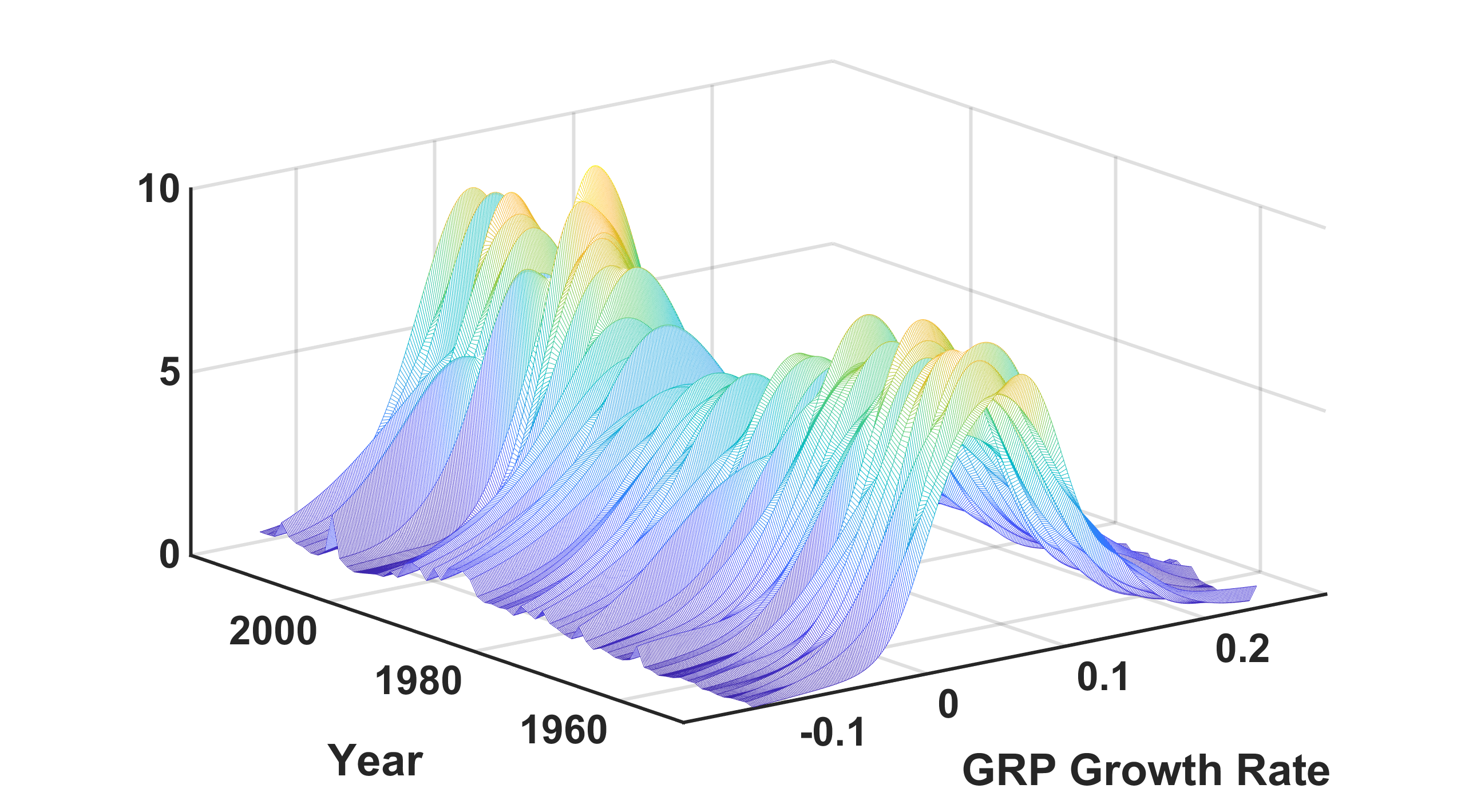}
		\includegraphics[height=0.35\textwidth, width=0.55\textwidth]{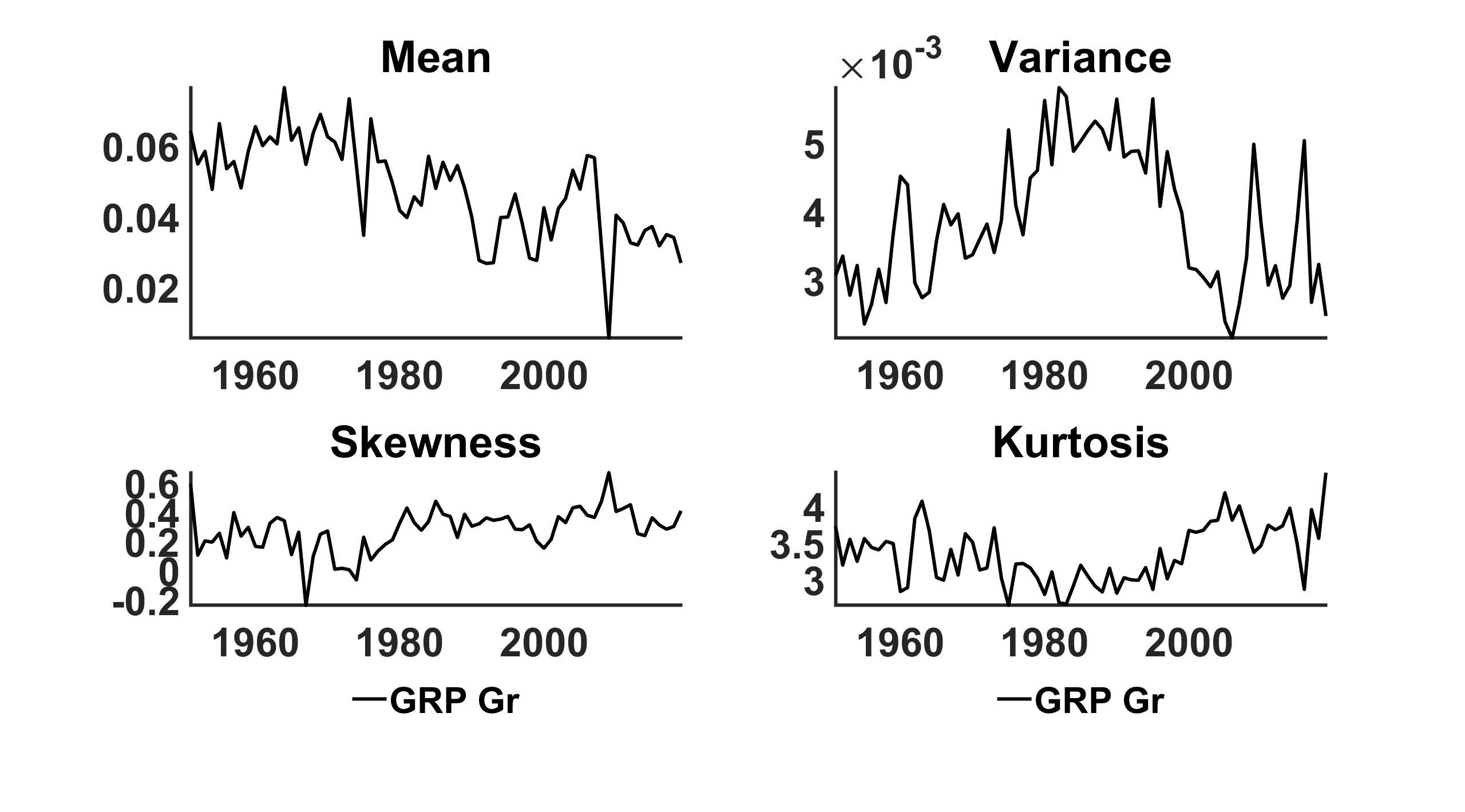}
		\caption{The generated spatial densities of the regional growth rate (left), and the first four central moments of the spatial densities of the regional growth rate (middle and right).}
		\label{Fig:GRP_Dist_Comp0}
	\end{center}
\end{figure}

\indent Figure \ref{Fig:GRP_Dist_Comp0} shows the time series of spatial distributions of regional growth rates, computed over the central 95\% of the total probability mass using a nonparametric kernel estimation method.\footnote{Because the CLR transform requires strictly positive densities, we evaluate moments on the central 95\% mass to avoid zero-density grids at the tails.} The cross-sectional mean has trended downward over time. The cross-sectional variance rises until the early 1980s and then declines—consistent with the subsequent Great Moderation—albeit with temporary spikes around global downturns (e.g., the late 2000s). Since the mid-1980s, both skewness and kurtosis have increased, indicating a more right-skewed and leptokurtic distribution: while average growth has slowed, a subset of regions has experienced very high growth, thickening the right tail and increasing overall tail weight.

\begin{figure}[htbp]
	\begin{center}
		\includegraphics[height=0.35\textwidth, width=0.95\textwidth]{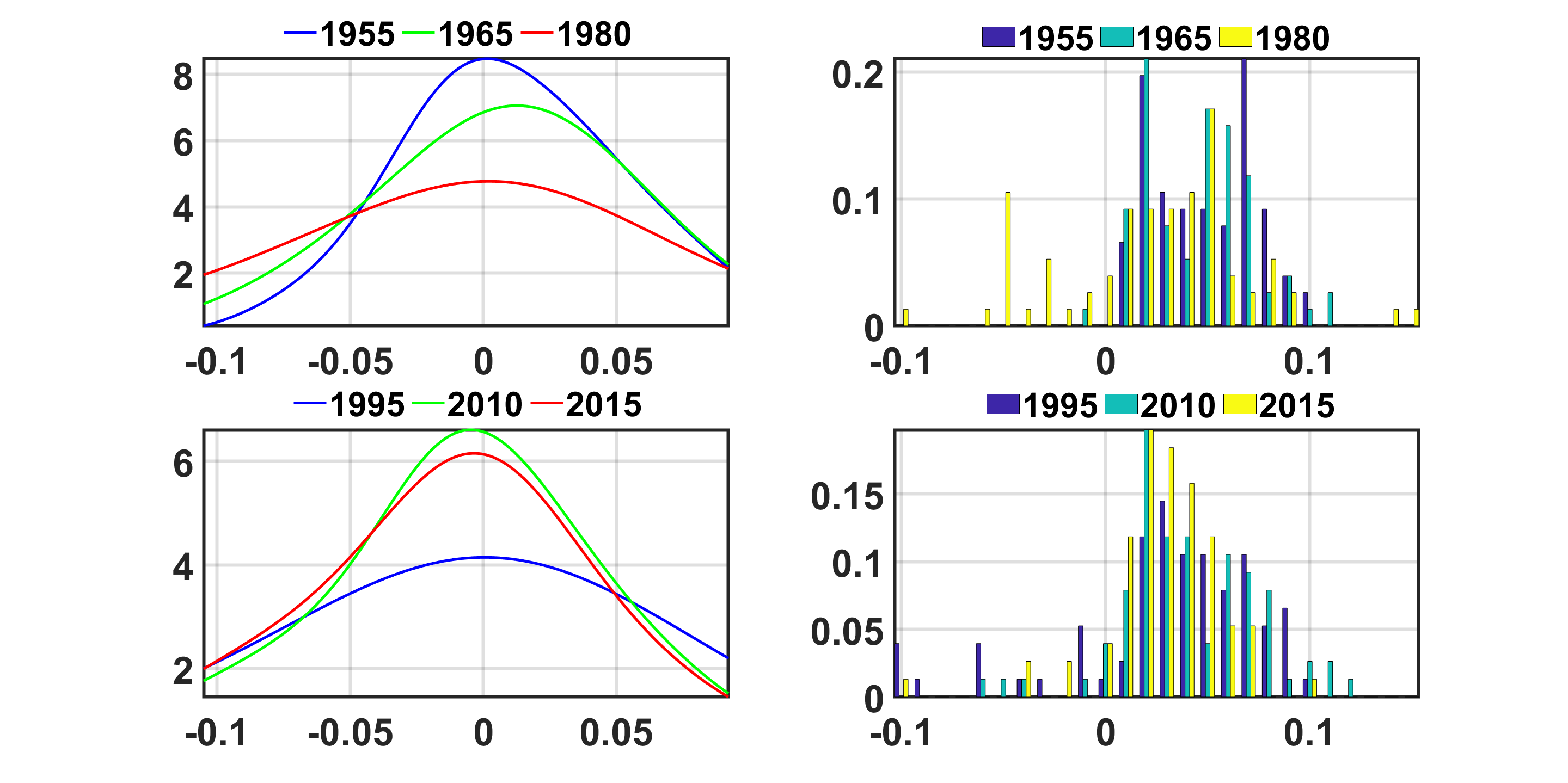}
		\caption{Estimated spatial densities of regional growth rate (left) and spatial histogram of TED-based world GDP growth rate (right) at selected years}
		\label{Fig:GRP_Comp2}
	\end{center}
\end{figure}

\indent Figure \ref{Fig:GRP_Comp2} compares kernel-based estimates of the cross-sectional density of regional growth rates (left panels) with simple cross-sectional histograms of TED-based country GDP growth (right panels) at selected years. The kernel densities are computed from regional growth rates constructed using the extrapolated regional income ratios described above, whereas the histograms are computed directly from TED growth rates and therefore do not rely on that extrapolation. This comparison serves as a validation check of our interpolation, extrapolation, and imputation procedures. Across years, the two distributions exhibit very similar shapes, indicating that the methods used to address data sparsity do not materially distort the nonparametric density estimates.

\subsection{Temperature-related regional growth rate}\label{AP_WGRP_Temp}
\noindent In this section, we outline the procedure for isolating the temperature-related component of regional growth rates. Based on the panel data structure spanning regions over time, a significant body of literature has investigated the effect of climate variables on regional economic activity. The linear panel regression model with two-way fixed effects has been employed to estimate the short-run response function, in the sense that the regional fixed effect, region-specific time trend, and time fixed effect would remove the permanent change, the gradual change, and the common trend in regional economic growth, respectively (\citealp{darwin1999impact}; \citealp{schlenker2010crop}; \citealp{dell2014we}; \citealp{carleton2016social}; \citealp{kolstad2020estimating}). Based on the Ramsey-type growth model and building upon the methodological literature, more specifically, \cite{kalkuhl2020impact} attempt to identify the short-run (immediate and transitory) and long-run (permanent) economic effects with the annual panel model, the long-difference model, and the cross-sectional model. The annual panel model is exploited to identify the short-run relationship, while the long-difference model for GRP growth rates and the cross-sectional model for the average level of the logarithmic GRP at a decade scale are exploited to identify the permanent impact of climate change. 

\indent Although the linear panel regression with two-way fixed effects would only capture the short-run relationship from the mean perspective, it fails to account for the fact that climate change could permanently affect not only the mean of regional economic growth but also the higher-order moments of its distribution. Moreover, the cross-sectional or panel regression model that is linear in the slope parameter would produce misleading information and large uncertainty. That is, the estimated slope coefficients of the linear regression model may underestimate the negative effects of climate change for developing countries, but overestimate those effects for developed countries.

\indent To estimate the marginal effects of climate variables while addressing potential collinearity with fixed effects and structural trends, we employ a two-step regression strategy. In the first stage, we remove both region-specific and year-specific effects from the regional growth rate to control for unobserved heterogeneity across regions and over time. While the two-way fixed-effects specification is standard in panel data analysis, its use in this setting raises concerns. Because climate variables such as temperature anomalies tend to evolve similarly across regions, largely following global time trends, including year fixed effects absorbs much of their temporal variation and obscures their independent influence. To address this issue, we adopt an alternative specification that retains region-specific quadratic trends but omits year fixed effects. This specification controls for long-term structural differences across heterogeneous regions while preserving temporal variation that is more likely to reflect climate-induced changes. In the second stage, we regress the residuals from the baseline model on the climate variables to estimate their marginal effects.

\indent As such, we consider the annual panel fixed effect regression model with omitted temperature (\citealp{hsiang2013quantifying}; \citealp{burke2015global}; \citealp{kalkuhl2020impact}; \citealp{newell2021gdp}; \citealp{meierrieks2023temperature}), as given by
\begin{align}\label{panel_FE}
	\Delta Y_{i,t} = p_{i}(t) + \delta_{i} + \mu_{t} + u_{i,t},  
\end{align}
where $\Delta Y_{i,t}$ is the regional growth rate in the region $i$ at time $t$, $p_{i}(t)$ is the region-specific quadratic time trends for technological, institutional, and demographic changes in the region $i$, $\delta_{i}$ is the region fixed effect, and $\mu_{t}$ is the time fixed effect. 

\begin{figure}[t]
	\begin{center}
		\includegraphics[height=0.35\textwidth, width=0.45\textwidth]{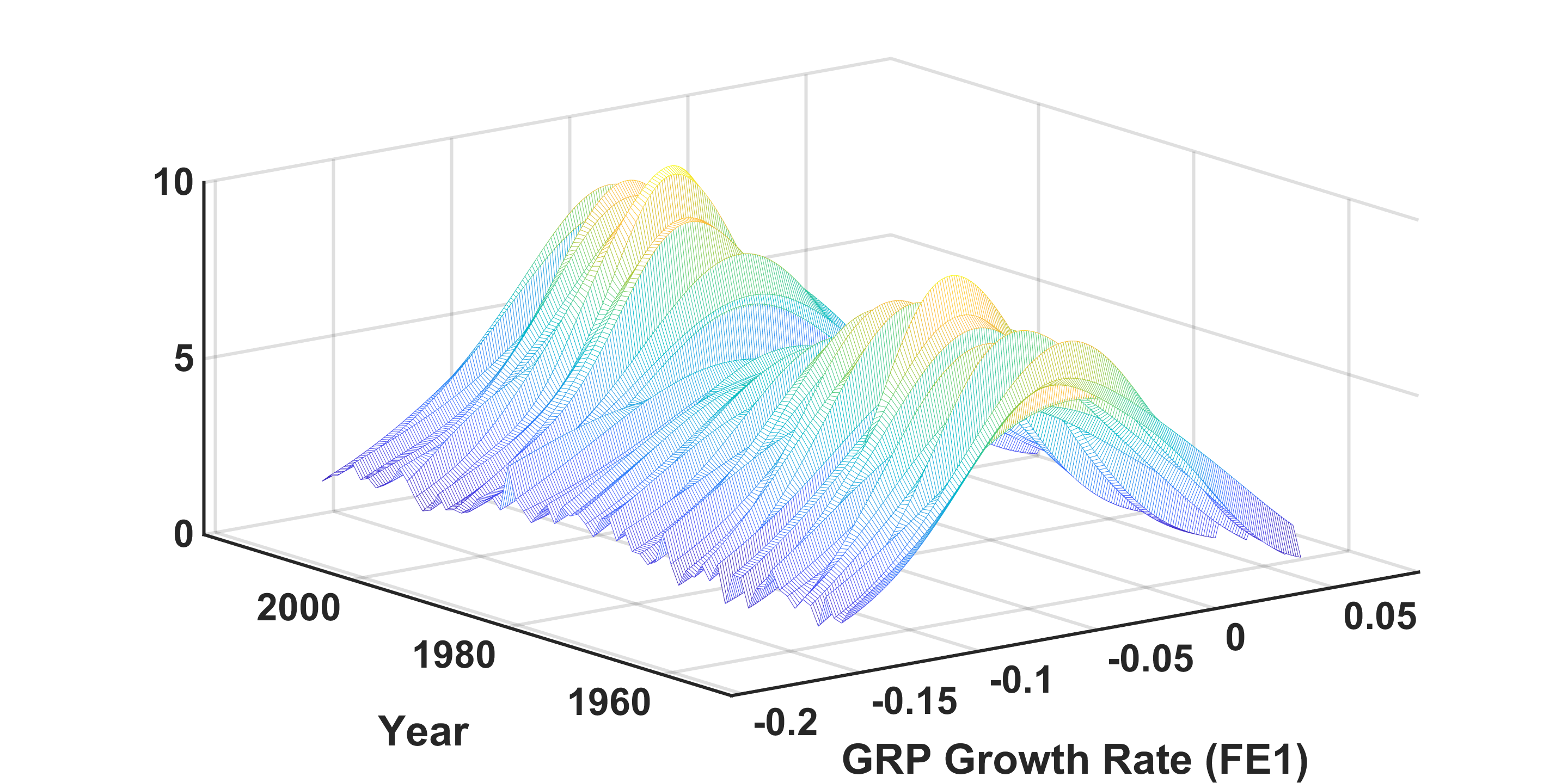} 
		\includegraphics[height=0.35\textwidth, width=0.45\textwidth]{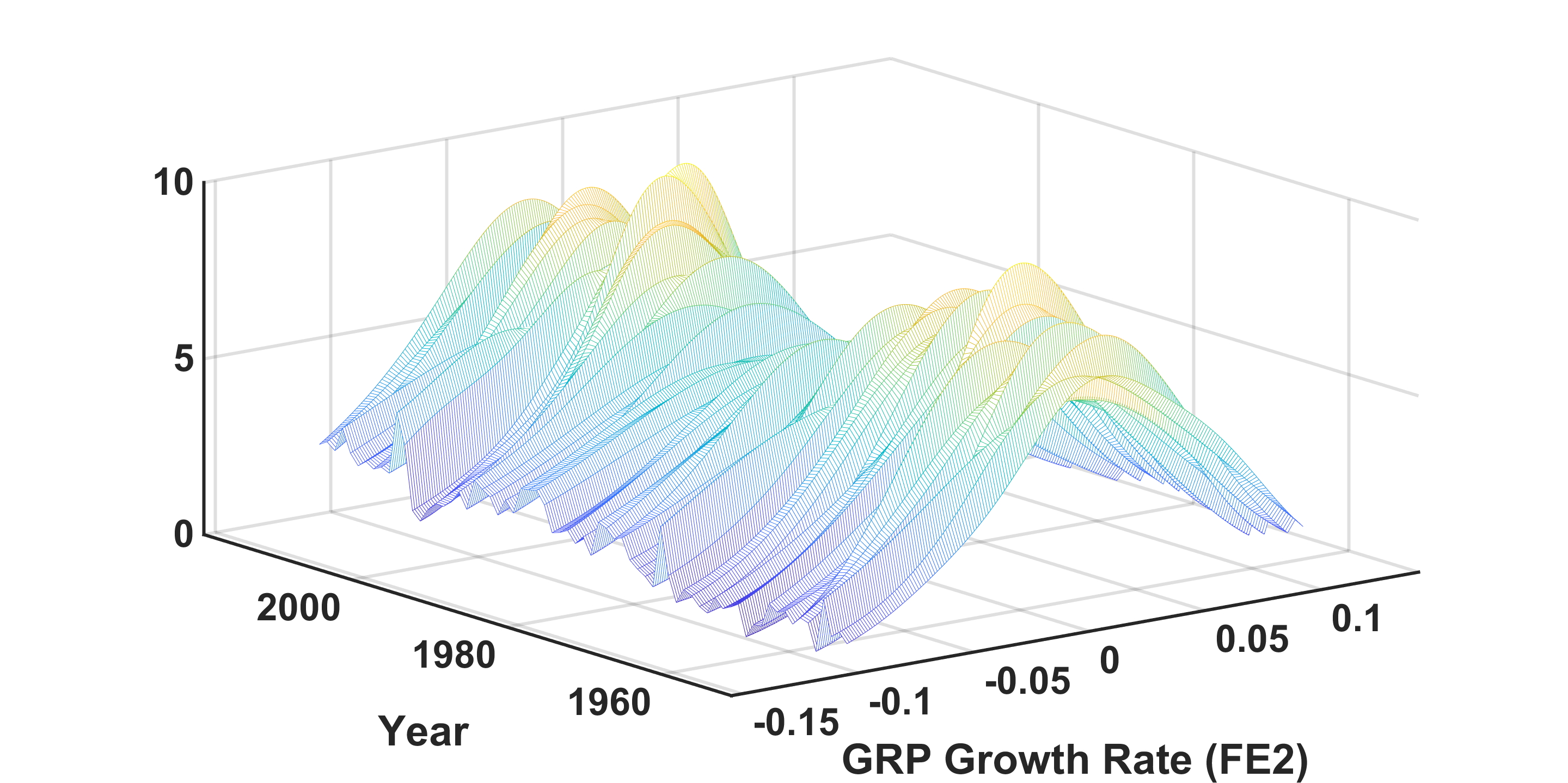} 
		\caption{The generated spatial densities of the temperature-related regional growth rate calculated by region-specific quadratic time trend and time-specific mean (FE1) (left), and those of the temperature-related regional growth rate calculated by region-specific quadratic time trend only (FE2) (right) from 1951 to 2019.}
		\label{Fig:GRP_Dist_Comp}
	\end{center}
\end{figure}

\indent We calculate deviations of regional growth rates from both region-specific quadratic time trends \((p_i(t) + \delta_i)\) and time-specific means \((\mu_t)\). The resulting stacked demeaned residuals, \(u_{i,t}\), represent the orthogonal component of regional growth, net of region- and time-specific income factors, and are less susceptible to the ``bad control'' problem \citep{angrist2009mostly}. We refer to this specification as FE1, where region-specific quadratic trends are included only if they are statistically significant at the 5\% level based on an F-test; otherwise, deviations are calculated using region-specific means \((\delta_i)\) alone.

\indent As an alternative specification to the earlier discussion, we also consider deviations based solely on region-specific quadratic time trends \((p_i(t) + \delta_i)\), which we denote as FE2. This model implicitly allows for the possibility that common time trends in regional growth may be partly driven by climate change. Assuming that exogenous income variation is adequately captured under either the FE1 or FE2 specification, this framework enables the identification of income responses to absolute temperature anomalies \citep{newell2021gdp}. Mapping the spatial distribution of the resulting residuals illustrates temperature-related variation in regional growth, capturing nonlinear responses to deviations in temperature anomalies relative to region- and time-specific baselines.

\begin{figure}[t]
	\begin{center}
		\includegraphics[height=0.45\textwidth, width=0.95\textwidth]{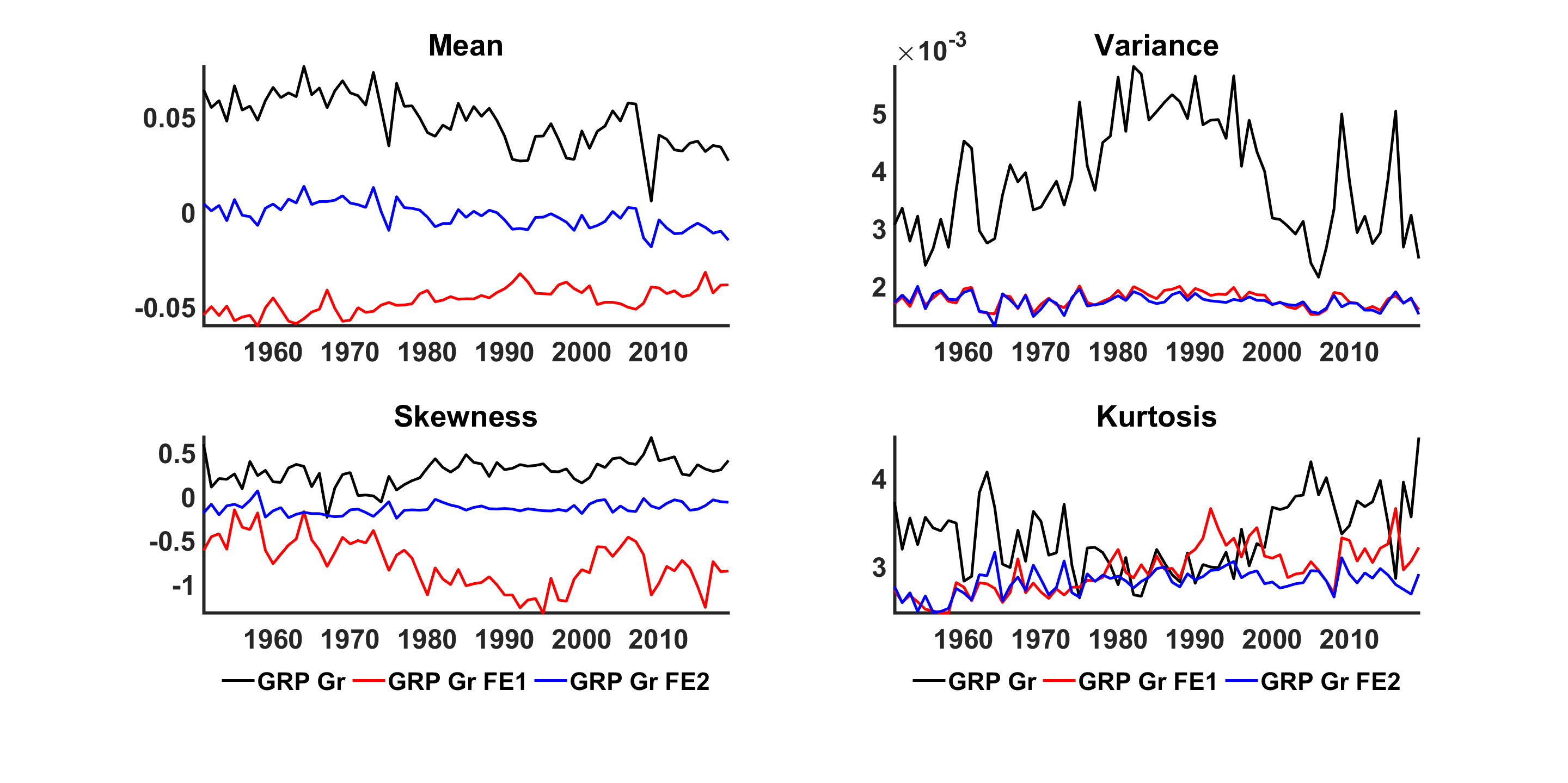}
		\caption{The comparison between the first four central moments of the spatial densities of the regional growth rate and those of the temperature-related regional growth rate using two-way panel fixed effect regression approaches from 1951 to 2019.}
		\label{Fig:GRP_Comp_Stat}
	\end{center}
\end{figure}

\indent For the nonparametric kernel density estimation, we restrict attention to the central 85\% of the total probability mass, thereby excluding anomalous behavior in the distributional tails and ensuring feasibility of the CLR transformation by avoiding zero-probability estimates. Figure \ref{Fig:GRP_Dist_Comp} illustrates the generated spatial densities of the temperature-related regional growth rate from 1951 to 2019. For more details, Figure \ref{Fig:GRP_Comp_Stat} compares the first four central moments—mean, variance, skewness, and kurtosis—of the spatial distributions of raw regional growth rates (GRP Gr) and those adjusted for income-related components using the FE1 and FE2 specifications (GRP Gr FE1 and FE2). Notably, the raw growth rates exhibit higher levels of variance, skewness, and kurtosis across time, indicating the presence of significant heterogeneity that may be attributable to persistent region-specific and time-specific income factors. When these factors are partially removed via FE1, which controls for both region and year fixed effects (or time-specific means), the moments shift considerably: the mean declines, and variance and skewness become more stable and compressed. However, the FE1 adjustment may over-correct by removing a substantial portion of the climate-related temporal variation, particularly due to the inclusion of year fixed effects.

\indent In contrast, the FE2 specification, which omits year fixed effects while retaining region-specific quadratic time trends, preserves more of the temporal variability associated with climate shocks. The central moments under FE2 are more moderate than the raw data, but less suppressed than in FE1—suggesting a more balanced removal of confounding income effects without eliminating key identifying variation related to climate. In particular, the reduced skewness and kurtosis under FE2, relative to the raw growth rates, imply that structural heterogeneity has been mitigated, while still retaining meaningful distributional shape and temporal dynamics. These results support the interpretation that the differences in statistical moments between the raw and adjusted growth rates primarily reflect the contribution of region- and time-specific income factors. Moreover, they suggest that FE2 provides a more appropriate adjustment framework for identifying the relationship between temperature anomalies and regional economic performance, especially when temporal climate trends are themselves of interest. Given the implausible temporal dynamics observed in the descriptive statistics under the FE1 specification, we proceed by focusing on the FE2-adjusted regional growth rate in our main analysis.

\newpage
\bibliography{FRIT_P1_Biblio}

\end{document}